\documentclass{article}

\pdfoutput=1

\usepackage[left=2cm, right=2cm, top=2cm]{geometry}
\usepackage{amsmath}
\usepackage{amsfonts}
\usepackage{jcappub}
\usepackage{subcaption}
\usepackage{hyperref} % for web links
\usepackage{mathtools}
\usepackage{amssymb}
\usepackage{multirow}
\usepackage{arydshln}
\usepackage[dvipsnames, table]{xcolor}
\usepackage{tikz}
\usepackage{colortbl}
\usepackage{pgf}
\usepackage{hhline}
\usepackage{soul}
\usetikzlibrary{calc}
\usepackage{zref-savepos}

\newcounter{NoTableEntry}
\renewcommand*{\theNoTableEntry}{NTE-\the\value{NoTableEntry}}

\newcommand*{\notableentry}{%
  \multicolumn{1}{@{}c@{}|}{%
    \stepcounter{NoTableEntry}%
    \vadjust pre{\zsavepos{\theNoTableEntry t}}% top
    \vadjust{\zsavepos{\theNoTableEntry b}}% bottom
    \zsavepos{\theNoTableEntry l}% left
    \hspace{0pt plus 1filll}%
    \zsavepos{\theNoTableEntry r}% right
    \tikz[overlay]{%
      \draw[black]
        let
          \n{llx}={\zposx{\theNoTableEntry l}sp-\zposx{\theNoTableEntry r}sp},
          \n{urx}={0},
          \n{lly}={\zposy{\theNoTableEntry b}sp-\zposy{\theNoTableEntry r}sp},
          \n{ury}={\zposy{\theNoTableEntry t}sp-\zposy{\theNoTableEntry r}sp}
        in
        (\n{llx}, \n{lly}) -- (\n{urx}, \n{ury})
        (\n{llx}, \n{ury}) -- (\n{urx}, \n{lly})
      ;
    }% 
  }%
}

\newcommand*{\opacity}{100}
\newcommand*{\minvaldBlim}{-88.72}
\newcommand*{\maxvaldBlim}{-30.45}
\newcommand*{\minvaldRlim}{-42.66}
\newcommand*{\maxvaldRlim}{-8.02}
\newcommand*{\minvalsigCal}{-65.12}
\newcommand*{\maxvalsigCal}{-16.10}

\newcommand{\fbseries}{\unskip\setBold\aftergroup\unsetBold\aftergroup\ignorespaces}
\makeatletter
\newcommand{\setBoldness}[1]{\def\fkbld@bold{#1}}
\makeatother

\definecolor{DarkRed}{rgb}{0.8, 0.2, 0.2}
\definecolor{DarkGreen}{rgb}{0.25, 0.75, 0.25}

\newcommand{\coldBlim}[1]{
            \pgfmathparse{int(round(100*((#1-\minvaldBlim)/(\maxvaldBlim-\minvaldBlim))))}
            \xdef\tempa{\pgfmathresult}
            \cellcolor{DarkGreen!\tempa!DarkRed!\opacity}#1
}
\newcommand{\coldRlim}[1]{
            \pgfmathparse{int(round(100*((#1-\minvaldRlim)/(\maxvaldRlim-\minvaldRlim))))}
            \xdef\tempa{\pgfmathresult}
            \cellcolor{DarkGreen!\tempa!DarkRed!\opacity}#1
}
\newcommand{\colsigCal}[1]{ 
            \pgfmathparse{int(round(100*((#1-\minvalsigCal)/(\maxvalsigCal-\minvalsigCal))))}
            \xdef\tempa{\pgfmathresult}
            \cellcolor{DarkGreen!\tempa!DarkRed!\opacity}#1
}

\definecolor{MediumBlue}{rgb}{0.5, 0.8, 1}
\definecolor{MediumRed}{rgb}{1, 0.7, 0.7}
\definecolor{MediumGreen}{rgb}{0.7, 0.9, 0.7}

\newcommand\dRlim{\delta R_{\rm lim}}
\newcommand\dBlim{\delta B_{\rm lim}}
\newcommand\sigmaCal{\sigma_{\rm Calib}}
\newcommand\dKlim{\delta K_{\rm lim}}

\title{\textbf{Impact of beam far side-lobe knowledge in the presence of foregrounds for \textit{LiteBIRD}}}%space-borne CMB experiment.}}

\author[1,2]{C.\,Leloup,}
\author[2]{G.\,Patanchon,}
\author[2]{J.\,Errard,}
\author[3,4]{C.\,Franceschet,}
\author[5]{J.\,E.\,Gudmundsson,}
\author[6]{S.\,Henrot-Versillé,}
\author[7]{H.\,Imada,}
\author[8]{H.\,Ishino,}
\author[1]{T.\,Matsumura,}
\author[9,10,11]{G.\,Puglisi,}
% ...
\affiliation[1]{Kavli Institute for the Physics and Mathematics of the Universe (Kavli IPMU, WPI), UTIAS, The University of Tokyo, Kashiwa, Chiba 277-8583, Japan}
\affiliation[2]{Université de Paris, CNRS, Astroparticule et Cosmologie, F-75013 Paris, France}
\affiliation[3]{Dipartimento di Fisica, Universita' degli Studi di Milano, Via Celoria 16 - 20133, Milano, Italy}
\affiliation[4]{INFN Sezione di Milano, Via Celoria 16 - 20133, Milano, Italy}
\affiliation[5]{The Oskar Klein Centre, Department of Physics, Stockholm University, SE-106 91 Stockholm, Sweden}
\affiliation[6]{Université Paris-Saclay, CNRS/IN2P3, IJCLab, 91405 Orsay, France}
\affiliation[7]{National Astronomical Observatory of Japan, Mitaka, Tokyo 181-8588, Japan}
\affiliation[8]{Okayama University, Department of Physics, Okayama 700-8530, Japan}
\affiliation[9]{Dipartimento di Fisica e Astronomia, Universitá degli Studi di Catania, Via S. Sofia,64, 95123, Catania, Italy}
\affiliation[10]{INAF, Osservatorio Astrofisico di Catania, via S.Sofia 78, I-95123 Catania, Italy}
\affiliation[11]{INFN, Sezione di Catania, via S.Sofia 64, I-95123, Catania, Italy}
% ...

\emailAdd{clement.leloup@ipmu.jp}

\abstract{
We present a study of the impact of a beam far side-lobe lack of knowledge on the measurement of the Cosmic Microwave Background $B$-mode signal at large scale. Beam far side-lobes induce a mismatch in the transfer function of Galactic foregrounds between the dipole and higher multipoles which degrades the performances of component separation methods. This leads
to foreground residuals in the CMB map. It is expected to be one of the main source of systematic effects in future CMB polarization observations. Thus, it becomes crucial for all-sky survey missions to take into account the interplays between beam systematic effects and all the data analysis steps. \textit{LiteBIRD} is the ISAS/JAXA second strategic large-class satellite mission and is dedicated to target the measurement of CMB primordial $B$ modes by reaching a sensitivity on the tensor-to-scalar ratio $r$ of $\sigma \left( r \right) \leq 10^{-3}$ assuming $r=0$. The primary goal of this paper is to provide the methodology and develop the framework to carry out the end-to-end study of beam far side-lobe effects for a space-borne CMB experiment. We introduce uncertainties in the beam model, and propagate the beam effects through all the steps of the analysis pipeline, most importantly including component separation, up to the cosmological results in the form of a bias $\delta r$. As a demonstration of our framework, we derive requirements on the calibration and modeling for the \textit{LiteBIRD}'s beams under given assumptions on design, simulation, component separation method and allocated error budget. In particular, we assume a parametric method of component separation with no mitigation of the far side-lobes effect at any stage of the analysis pipeline. %\clem{This makes the analysis particularly sensitive to an error on frequency maps normalization induced by the large-scale contamination from the Galactic plane seen by side-lobes. We expect our requirements to be relaxed by including dedicated mitigation techniques or by using less sensitive foreground cleaning methods, e.g. harmonic ILC.}

We show that $\delta r$ is mostly due to the integrated fractional power difference between the estimated beams and the true beams in the far side-lobes region, with little dependence on the actual shape of the beams, for low enough $\delta r$.
Under our set of assumptions, in particular considering the specific foreground cleaning method we used, we find that the integrated fractional power in the far side-lobes should be known at the level of $\sim 10^{-4}$
, to achieve the required limit on the bias $\delta r < 1.9 \times 10^{-5}$.
The framework and tools developed for this study can be easily adapted to provide requirements under different design, data analysis frameworks and for other future space-borne experiments, such as PICO or CMB-Bharat. We further discuss the limitations of this framework and potential extensions to circumvent them.
}


\begin{document}

%\today
\maketitle
\flushbottom

\def\matriximg{%
  \begin{matrix}
    \\
    \ \cdots \ \  \\
    \\
   \end{matrix}
}

\section{Introduction}
\label{section:Introduction}

Observations of the Cosmic Microwave Background (CMB) radiation have played a crucial role in establishing the concordance model of cosmology in the past 50 years. In particular, data from a sequence of space missions (COBE \cite{Bennett:1996ce}, WMAP \cite{Hinshaw_2013} and Planck \cite{Planck:2018nkj}) significantly improved our knowledge of the history of the Universe and its constituents. However, we have yet to probe the imprint of primordial gravitational waves in the curl component of the CMB polarized signal, the so called $B$ modes, which would constitute strong evidence of the hypothetical inflationary period \cite{Guth:1980zm,Linde:1981mu,Kamionkowski:1996ks,Zaldarriaga:1996xe,Chowdhury:2019otk}. CMB polarization is sourced by scalar, produced by primordial density fluctuations, and tensor perturbations, whose primordial contribution comes exclusively from gravitational waves in the early Universe. The relative amplitude between tensor and scalar modes is captured by the tensor-to-scalar ratio parameter $r$. A precise measurement of $r$ would allow us to shed new light on the physics of the early Universe and constrain, in particular, the multitude of inflation models. Currently, the best constraint on the tensor-to-scalar ratio is $r<0.032$ (95\% C.L. interval) using a combination of data from the Planck mission and the BICEP/Keck experiment \cite{Tristram:2021tvh}.

One of the main challenges in the precision primordial B-mode search is to distinguish between primordial $B$ modes from the inflationary period, and residuals from foreground polarized emissions of our own Galaxy as well as from instrumental systematic effects.
%The primordial $B$ modes and the Galactic polarized emissions are expected to have a larger signal at large angular scale.
A standard method to differentiate the sources is to observe the sky over a broad frequency range and make use of the fact that CMB and Galactic polarized emissions have a different spectral behaviour.

Imperfect knowledge of the optical response of a telescope, its so-called beam pattern, 
is one of the key systematic effects to be understood in order to properly process the large angular scale signal. In particular, the far side-lobes region at large angle can be very challenging to model and to measure. A number of studies have been carried out to understand the beam systematic effects propagating to observations of past CMB experiments and to evaluate the potential impacts on the scientific outcomes (e.g. in Planck \cite{Franco:2002hx,Burigana:2003ae,Huffenberger}). In the context of ground based experiments, e.g. the Atacama Cosmology Telescope or the Simons Observatory, Gallardo et al.\cite{Gallardo:2018zfa} addressed the systematic effects in the beam parameters including side-lobe pick-up.

\textit{LiteBIRD} \cite{Litebird-2019,LiteBIRD:2022cnt} is the second ISAS/JAXA strategic large-class mission. It will conduct a full-sky survey and measure precisely the polarization anisotropies of the CMB, with a combined sensitivity including statistical errors, foreground residuals and systematic uncertainties on the tensor-to-scalar ratio of $\sigma \left( r \right) \leq 0.001$, assuming $r=0$. \textit{LiteBIRD} will observe the sky in 15 frequency bands from 34 to 448~GHz, with an effective polarization sensitivity of 2.2~$\mu$K-arcmin and angular resolution ranging from 71 to 18~arcmin, allowing access to multipoles in the range $2 \leq \ell \leq 200 $, which will provide unique power to distinguish primordial $B$ modes from the foreground and gravitational weak-lensing $B$ modes. To achieve such a challenging scientific requirement, we need to evaluate the impact of instrumental systematic effects and impose strict requirements on their control. For a reliable estimation of these effects, one has to bridge the science goal and the instrumental specifications, which requires the implementation of various steps, e.g. instrument modeling, sky modeling, and component separation. While a number of studies address some of these steps, to our knowledge none made this bridge fully end-to-end \cite{Planck-Beams,Hivon:2016qyw,Lungu:2021slc,Gallardo:2018zfa,Huffenberger,Mitra:2010rt}.
The following work proposes to set up a general framework to study beam systematic effects from the instrumental beam simulations all the way up to their impact on the tensor-to-scalar ratio, which we applied to the particular case of \textit{LiteBIRD} \cite{LiteBIRD:2022cnt}. This allows us to evaluate the required knowledge of instrumental beam to achieve the scientific goal of \textit{LiteBIRD}, in the current experimental context as a first step towards future, more refined study cases. This framework makes use of computational approximations, physical assumptions and arbitrary choices that can impact these requirements. Most notably, the results depend on assumptions on the optical design, approximations for the convolution of the sky with the beams, and the choice of component separation method as well as the allocated error budget for this systematic effect. However, these assumptions can be changed and refined with minimal modifications to the analysis pipeline.

The paper is organized as follows. In Section~\ref{section:Overview of LiteBIRD} we describe \textit{LiteBIRD}'s instrumental characteristics relevant for this study. The analysis procedure, detailed in Section~\ref{section:Methodology}, is divided into two different approaches depending on the region of the beam under study. The first region, closer to the beam axis, will be accessible to measurements during calibration on the ground and in flight and its knowledge will therefore be impacted by measurement uncertainties. By assessing their impact on cosmological results, compared with the scientific goal of \textit{LiteBIRD}, we can set requirements on the accuracy of these calibration measurements. On the other hand, the second region, further from the beam axis, will be out of reach for precise direct measurements and will rely mostly on modeling combined with indirect measurements. We want to investigate the impact of these modeling uncertainties and the ideal location of the transition between the two regions in order to meet the requirements. These results are presented in Section~\ref{section:Results}, and their interpretations and limitations are given in Section~\ref{section:Discussions}.

\section{Overview of \textit{LiteBIRD}}
\label{section:Overview of LiteBIRD}

\begin{figure}[!htb]
\begin{center}
\includegraphics[width=0.8\textwidth]{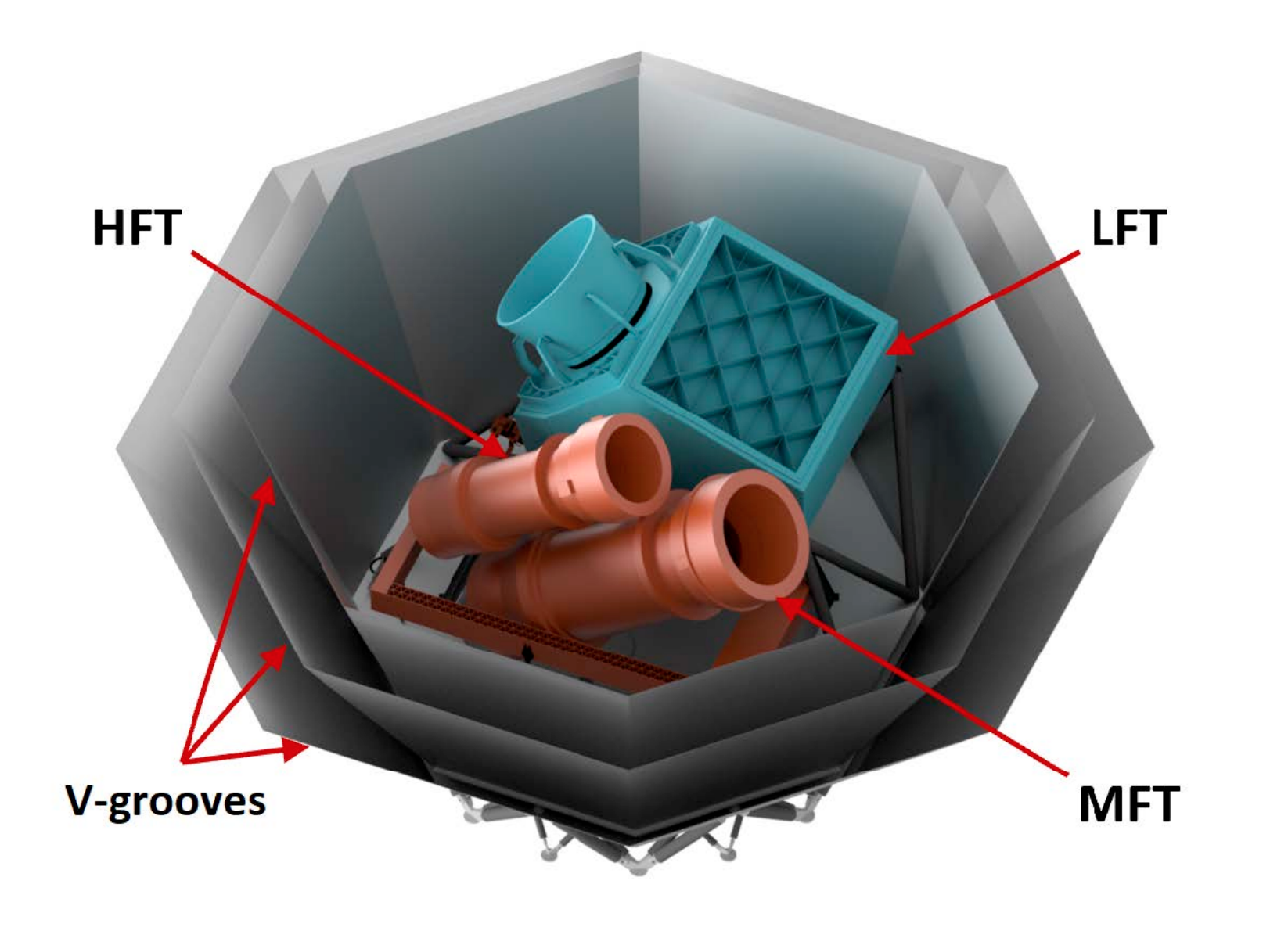}
\caption{Integration of the three telescopes, the LFT, the MFT and the HFT, to the payload of \textit{LiteBIRD}. The LFT follows a crossed-Dragone design while the MFT and HFT, mounted on the same mechanical structure, are fully refractive telescopes.}
\label{fig:LFT-MFT-HFT}
\end{center}
\end{figure}

%\subsection{LiteBIRD}
The \textit{LiteBIRD} satellite includes three telescopes at low, medium, and high frequencies (LFT \cite{LiteBIRD:2020tzb}, MFT and HFT \cite{LiteBIRD:2020zfx}). 
\begin{table}[!htb]
\begin{center}
\begin{tabular}{|c|c|c|c|c|c|}
\hline
& \vtop{\setbox0\hbox{\strut (GHz)}\hbox to\wd0{\hss\strut $\nu$\hss}\copy0} & \vtop{\setbox0\hbox{\strut $\Delta \nu$ ($\Delta \nu / \nu$)} \copy0\hbox to\wd0{\hss\strut (GHz) \hss}} & \vtop{\setbox0\hbox{\strut Beam size} \copy0\hbox to\wd0{\hss\strut (arcmin) \hss}} & \vtop{\setbox0\hbox{\strut Number of} \copy0\hbox to\wd0{\hss\strut bolometers \hss}} & \vtop{\setbox0\hbox{\strut ($\mu$K-arcmin)}\hbox to\wd0{\hss\strut Sensitivity \hss}\copy0} \\
\hline \hline
& 40 & 12 (0.30) & 70.5 & 48 & 37.42 \\ \cline{2-6}
& 50 & 15 (0.30) & 58.5 & 24 & 33.46 \\ \cline{2-6}
& 60 & 14 (0.23) & 51.1 & 48 & 21.31 \\ \cline{2-6}
& \multirow{2}*{68} & \multirow{2}*{16 (0.23)} & 41.6 & 144 & 19.91 \\ \cdashline{4-6}
& & & 47.1 & 24 & 31.77 \\ \cline{2-6}
& \multirow{2}*{78} & \multirow{2}*{18 (0.23)} & 36.9 & 144 & 15.55 \\ \cdashline{4-6}
& & & 43.8 & 48 & 19.13 \\ \cline{2-6}
& \multirow{2}*{89} & \multirow{2}*{20 (0.23)} & 33.0 & 144 & 12.28 \\ \cdashline{4-6}
& & & 41.5 & 24 & 28.77 \\ \cline{2-6}
& 100 & 23 (0.23) & 30.2 & 144 & 10.34 \\ \cline{2-6}
& 119 & 36 (0.30) & 26.3 & 144 & 7.69 \\ \cline{2-6}
\multirow{-12}{*}{\rotatebox[origin=c]{90}{LFT}} & 140 & 42 (0.30) & 23.7 & 144 & 7.25 \\ \hline \hline
\multirow{5}*{\rotatebox[origin=c]{90}{MFT}}
 & 100 & 23 (0.23) & 37.8 & 366 & 8.48 \\ \cline{2-6}
 & 119 & 36 (0.30) & 33.6 & 488 & 5.70 \\ \cline{2-6}
 & 140 & 42 (0.30) & 30.8 & 366 & 6.38 \\ \cline{2-6}
 & 166 & 50 (0.30) & 28.9 & 488 & 5.57 \\ \cline{2-6}
 & 195 & 59 (0.30) & 28.0 & 366 & 7.05 \\ \hline \hline
\multirow{5}*{\rotatebox[origin=c]{90}{HFT}}
 & 195 & 59 (0.30) & 28.6 & 254 & 10.50 \\ \cline{2-6} 
 & 235 & 71 (0.30) & 24.7 & 254 & 10.79 \\ \cline{2-6}
 & 280 & 84 (0.30) & 22.5 & 254 & 13.80 \\ \cline{2-6}
 & 337 & 101 (0.30) & 20.9 & 254 & 21.95 \\ \cline{2-6}
 & 402 & 92 (0.23) & 17.9 & 338 & 47.45 \\ \hline \hline
Total & & & & 4508 & 2.16 \\ \hline
\end{tabular}
\caption{\textit{LiteBIRD} specifications in its 22 frequency channels, from \cite{LiteBIRD:2020khw}. From left to right the columns are: the telescope covering the band, the band center frequency in GHz, the bandwidth in GHz and its ratio to the central frequency, the main beam FWHM in arcmin, the number of bolometers for each channel and the sensitivity in $\mu$K-arcmin.}
\label{tab:LB instrument}
\end{center}
\end{table}
With an aperture diameter of 400~mm and an angular resolution ranging from 71 to 24~arcmin, the LFT includes nine frequency bands, three of them redundant making twelve channels within the LFT, distributed from the lower bound of the lowest frequency channel at 34~GHz to the upper bound of the highest frequency channel at 161~GHz, in order to cover the spectral domains of both CMB and low frequency Galactic emission. Its optical design follows a crossed-Dragone configuration, with a rotating Half-Wave Plate (HWP) as its first optical component. The LFT focal plane is made of tri-chroic lens-coupled Transition Edge Sensors (TES) detectors cooled down to 100~mK. On the other hand, the MFT and HFT,
%of \textit{LiteBIRD}
spanning from 89 to 448~GHz, consist of two fully refractive telescopes held on a single mechanical structure, and so are part of a common system called the MHFT. The frequency bands of the MFT range from 89 to 224~GHz, and the HFT from 166 to 448~GHz, with an angular resolution between 18 and 38~arcmin. Therefore, \textit{LiteBIRD} is composed of 15 frequency bands, which partially overlap each other. As a result, the total of 22 frequency channels are distributed from 34 to 448~GHz. Their main characteristics (bands, beam sizes, sensitivites, etc.) are detailed in Table~\ref{tab:LB instrument}. Figure~\ref{fig:LFT-MFT-HFT} shows the integration of the LFT, MFT and HFT on the \textit{LiteBIRD} satellite~\cite{LiteBIRD:2020tzb,LiteBIRD:2020zfx}.

The \textit{LiteBIRD} main scientific requirement $\sigma \left( r \right) < 0.001$, assuming $r=0$, is very challenging from the instrumental point of view. It requires an unprecedented sensitivity at the largest scales and an extreme control of systematic effects. In particular, a good characterization of the beams is of utmost importance as this has a major impact on the quality of observed data. Following the successful NASA WMAP and ESA Planck experiences, the telescope's main-beam response will be calibrated using the planets~\cite{WMAP-Beams,Huffenberger,Planck-Beams,Planck:2015wtm}. Although some information about the near and far side-lobes can be obtained in flight using planets and brighter objects such as the Moon, such methods face strong limitations. Therefore, most of the knowledge of the side-lobes response typically rely on a mathematical model validated by the telescope characterization on the ground prior to the launch. Note that we call side-lobes the region of the beam pattern at angles $\gtrsim 5^{\circ}$ away from the beam axis given the optical system of \textit{LiteBIRD}. The side-lobes characterization of a cryogenically cooled telescope at a millimeter-wave range is known to be challenging. A modeling based performance forecast is also computationally expensive. As a result, it is essential to study the needed accuracy of calibration measurements of the beam side-lobes at an early phase of the project to plan effectively the calibration strategy of \textit{LiteBIRD}.

\begin{figure}[!htb]
\begin{center}
\includegraphics[width=0.37\textwidth]{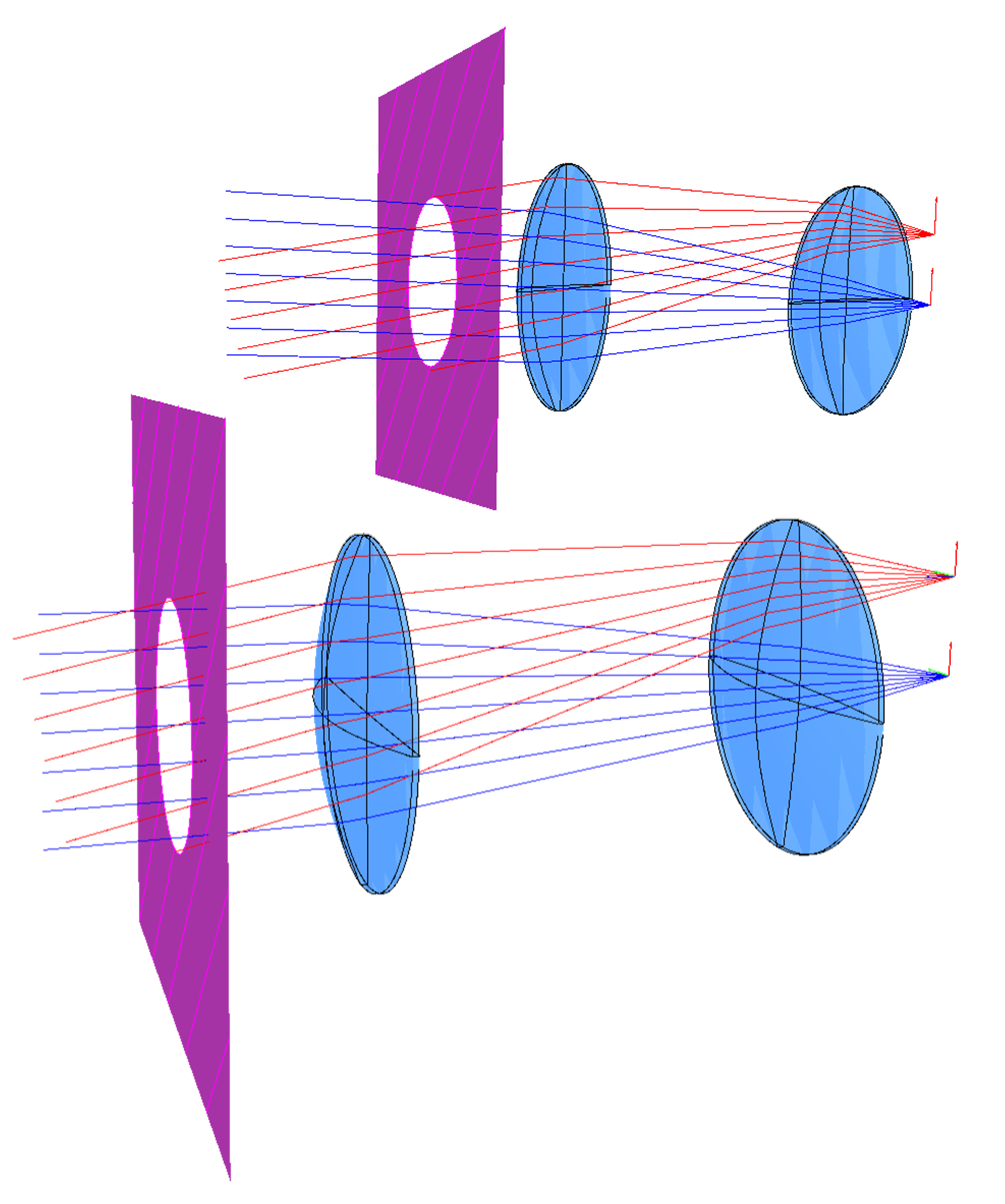}
\includegraphics[width=0.58\textwidth]{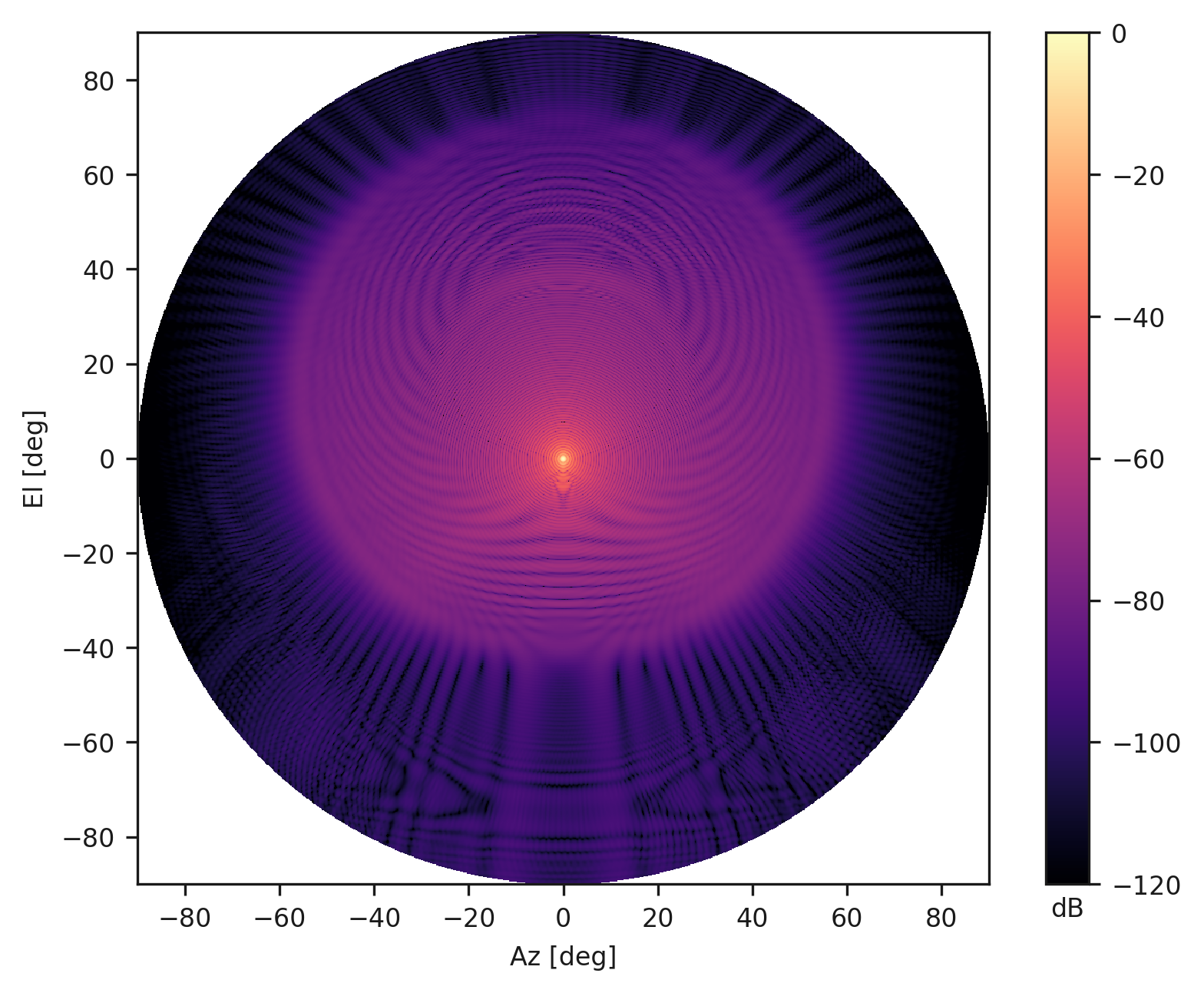}
\caption{Left: Ray diagrams of MFT (bottom) and HFT (top) models, as implemented into \texttt{GRASP}. The on-axis and edge pixel fields (14 deg) are the blue and red rays, respectively. Right: A full $2\pi$ far field beam map (normalized to peak amplitude) for a 100 GHz detector located at the edge of the focal plane.}
\label{fig:mhft_model}
\end{center}
\end{figure}

The current \textit{LiteBIRD} beam model is based on simulations of the beam response of individual detectors at the center frequency of the frequency bands with \texttt{GRASP} \cite{GRASP}, a software tool based on several electromagnetic methods (Physical Optics and Physical Theory of Diffraction, Method of Moments, etc.) accounting for the optical elements expected for each camera deployed in. Given the level of complexity, \texttt{GRASP} simulations have been performed on a smaller set of detectors on the focal planes at this stage, leaving their complete description for future work. For a given pixel in the focal plane, if the \texttt{GRASP} simulation is not available, we consider the closest one (in terms of distance from the axis) and rotate the beam map accordingly to the detector location. Our current \texttt{GRASP} models of the LFT and MHFT telescopes include most of the elements of the nominal optics design, which directly contribute to the response in the main beam region. However, they also include approximations, so they are far from being exhaustive in predicting the radiation pattern at larger angles, in the far side-lobes region, over the whole 4$\pi$ solid angle. This is particularly true for the LFT model \cite{LiteBIRD:2020tzb}. The MHFT model (see Fig.~\ref{fig:mhft_model}) includes the nominal optical elements of the two refractive telescopes, from the focal plane to the aperture towards the sky: the beam former (a lenselet coupled to sinuous antenna for MFT and a spline-profiled horn for HFT), the two Ultra High Molecular Weight Polyethylene (UHMW-PE) dielectric lenses, the aperture stop of the telescope and a perfectly absorbing fore-baffle. A half-wave plate will be used at the vicinity of the aperture, but is not included in the current beam modeling simulations because of the complexity of its integration. These elements do not directly contribute to most of the asymmetries in the far side-lobes region of the beam, which are mainly due to the impact of the large mechanical structures of the satellite on the beam (V-grooves, structural elements, etc.) that are not taken into account, but include asymmetries for off-center detectors in the focal plane. A more realistic impact of asymmetries is left for future work.

\section{Methodology}
\label{section:Methodology}

\subsection{Simulation of the effect of imperfect beam knowledge}
\label{subsection:Simulation of the effect of imperfect beam knowledge}

\begin{figure}[!htb]
\begin{center}
\includegraphics[width=\textwidth]{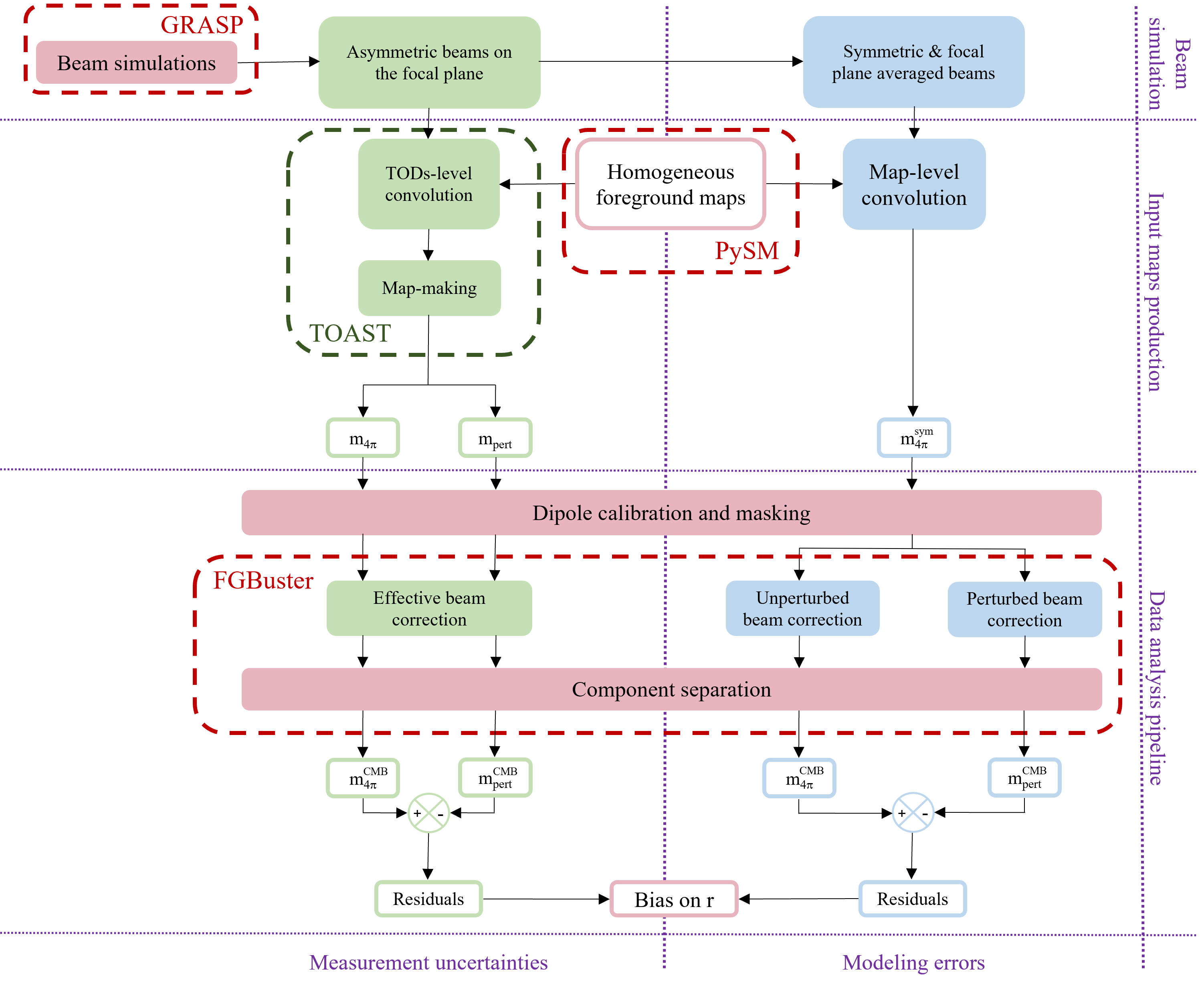}
\caption{Block diagram summarizing the main analysis steps described in Section \ref{section:Methodology} to estimate the bias on the tensor-to-scalar ratio, defined as the measured $r$ assuming its true value to be $r_{\rm true}=0$, from an uncertainty on the beam shape. In all cases, the perturbed maps ($\bold{m_{\rm pert}}$) include a convolution with perturbed beams while unperturbed maps ($\bold{m_{4\pi}}$) are convolved with a reference unperturbed beam. The comparison of the reconstructed CMB maps in these two cases are interpreted as systematic residuals and used to compute $\delta r$. Sections~\ref{subsubsection:Perturbation Case} and \ref{subsubsection:Modeling Case} describe two different methods, using different definitions of perturbed and unperturbed maps, the former (in green) being more realistic while the latter (in blue) is faster and more flexible. The steps in red are common to both methods. We will see in Section~\ref{subsubsection:Comparison of the Detailed Method and the Axisymmetric Method} that the two methods give compatible results.}
\label{fig:block diagram}
\end{center}
\end{figure}

In order to study errors arising from an inacurate knowledge of the beam patterns, we have to simulate the effect of uncertainties on these shapes. We use two distinct approaches to study two different cases. The first approach consists in introducing a localized perturbation of the beam amplitude to account for either statistical or systematic measurement uncertainties during calibration. The second approach relies entirely on a model to estimate the beam shape for angles larger than an angle $\theta_{\rm lim}$, this would correspond to the case where known systematic effects during calibration prevent us from measuring accurately the beam shape at very large angle. In the following, we will refer to the former as the Perturbation Case and to the latter as the Modeling Case. We treat these two approaches separately in Sections~\ref{subsubsection:Perturbation Case} and \ref{subsubsection:Modeling Case} respectively, and the general procedure is schematically described in Figure~\ref{fig:block diagram}.

\subsubsection{Perturbation Case}
\label{subsubsection:Perturbation Case}

\subsubsection*{Beam convolution of the sky} 

We use \texttt{PySM}~\cite{Thorne:2016ifb} to simulate the Galactic microwave emission accounting for thermal dust, synchrotron, Anomalous Microwave Emission (AME), free-free, respectively the \texttt{d0,s0,a1,f1} models \footnote{These refer to models of dust, synchrotron, AME and free-free foreground emissions respectively, with homogeneous spectral energy densities (SEDs) over the sky. AME and free-free are not polarized but are needed in the dipole calibration step where temperature maps are used, see Section~\ref{subsubsection:Perturbation Case}.}. We further include the emission of radio sources both in intensity and polarization following the modeling described in \cite{Li:2021ial,Puglisi:2017lpn}. As the goal of this work is to assess the residuals from systematic uncertainties in terms of CMB $B$ modes, we do not include CMB emission nor other systematic effects to better single out the effect. We do not include noise in the simulations, but due to the component separation treatment described in Section~\ref{subsection:Component separation}, the impact of noise is taken into account later in the analysis process. Before beam convolution, the signal is integrated across the \textit{LiteBIRD} frequency band (assumed to be a top-hat for all the detectors). Because the beams themselves are simulated only at the center frequency of each channel, this corresponds to assuming the beams to have no frequency dependence, or alternatively the beam at the central frequency to be representative of the band-averaged beam. This will need to be checked in future, more precise studies including the frequency dependence of the beam patterns.
%, therefore treating the simulated beams as the band-averaged beam.
We then use the \texttt{TOAST} \cite{TOAST} software to generate the Time-Ordered Data (TOD) with the nominal scanning strategy of \textit{LiteBIRD}, accounting for a portion of the focal plane detectors: 50\% for LFT, 80\% for MFT and 90\% for HFT. The beam convolution is done by the {\tt{conviqt}} algorithm \cite{prezeau2010} implemented in \texttt{TOAST}.

As mentioned earlier, we have not taken the HWP into account at the step of the beam modeling. For experiments employing a continuously spinning HWP, the presence of these extra-optical components needs to be correctly accounted for at the step of beam convolution in addition to the conventional beam convolution. A methodology to include HWP rotation with beam convolution by adding an extra-dimension in the data-cube was proposed in \cite{Duivenvoorden:2018zdp,duivenvordeen2021}. Given the required resolution and number of detectors of current and forthcoming CMB experiments, this represents a limiting factor that makes the convolutions  almost unfeasible. We thus present here an approximated procedure able to ensure convolution of polarized signal with the realistic beams simulated with \texttt{GRASP} in presence of a spinning HWP. The convolution procedure relies on the assumption that the beam employed for the polarized signals, i.e. $b^Q$ and $b^U$, is assumed to be the same as the unpolarized one, $b^{I}= \tilde{I}$. By following the notations adopted by \citep{Duivenvoorden:2018zdp}, we indicate beam maps in terms of the \emph{Stokes} parameters $b^{\mu} = (\tilde{I}, \tilde{Q},\tilde{U},\tilde{V})$:
\begin{eqnarray}
    b^I &=& (\tilde{I}, 0 , 0, 0)  \nonumber\\
     b^Q &=& (0, \tilde{I},  0, 0)  \label{eq:beamIQU}\\
      b^U &=& (0,0,\tilde{I},  0) \nonumber,
\end{eqnarray}
with $\tilde{I}$ being the total intensity component of the \texttt{GRASP} simulated beam. Once we construct the three beam maps $b^I, b^Q,b^U$, we use the spherical harmonic (SH) transform to expand each map into  $b^X_{\ell m} $ with $X=T,E,B$. Based on our assumption \eqref{eq:beamIQU} that there is no $I \rightarrow P$ leakage in our simulations, we have $b^{I, E} _{\ell m}=b^{I, B} _{\ell m} = b^{Q, T} _{\ell m}= b^{U, T} _{\ell m}=0$.

The transformation in harmonic domain speeds up the convolution step as it becomes a simple product of the $b^X_{\ell m} $ with the SH coefficients of the input sky $a_{\ell m}$. We perform 3 separate convolutions: 
\begin{enumerate}
\item multiplication in harmonic space of the unpolarized $a^I_{\ell m}$ by the $b^{I}_{\ell m}$ beam;
\item multiplication in harmonic space of the polarized component $_{- 2}a^P_{\ell m}$ by the $_{- 2}b^{P}_{\ell m}$ beam;
\item multiplication in harmonic space of the polarized component $_{2}a^P_{\ell m}$ by the $_{2}b^{P}_{\ell m}$ beam;
\end{enumerate}
where $a^I_{\ell m}$ and $_{\pm 2}a^P_{\ell m}$ stands for the spherical harmonic coefficients of total intensity and both the $E$-mode and $B-$mode components, respectively. The final convolved TOD is thus \cite{Duivenvoorden:2018zdp}:

\begin{equation}
    \small
    d_t \propto \sum _{\ell m} \sqrt{\frac{4\pi}{2\ell +1 }} \left[  b^{I*}_{\ell s} a _{\ell m}^I   +\frac{1}{2}  \left( {}_{-2}b^{P*}_{\ell s} {}_{-2} a^{P}_{\ell m}\, {\rm e}^{-4i\phi_t}  +{}_{2}b^{P*}_{\ell s} {}_{2} a^{P}_{\ell m}{\rm e}^{4i\phi_t}\right)\right] \, {\rm e}^{-is\psi_t} {}_{s}Y_{\ell m } (\Omega_{t}), \label{eq:TODmodel}
\end{equation}
with $\psi_{t}$ being the  orientation angle  of the detectors at time $t$ and $\phi_{t}$ the angle of the HWP at time $t$. Using Eq.~\eqref{eq:TODmodel}, we can efficiently separate the data sampling and the convolution into two steps \citep{prezeau2010}, by evaluating the inverse spherical harmonics over the sphere for each azimuthal mode $s$ of the beam. Once the maps are computed, we use the pointing and phase information to sample the TOD $d_t$. It was shown in \citep{Duivenvoorden:2018zdp} that $s_{max} \ll \ell_{max}$ thanks to the azimuthal band-limit of the simulated beam so that each $s$ mode can be independently treated and recursively estimated via $e^{i(s+1)\psi} = e^{i s \psi}e^{i\psi}$.

Using this framework, we produce several series of convolutions by applying different apodization functions to the nominal beam patterns to consider the side-lobes contribution only:
\begin{equation}
    A \left( \theta, \theta_{\rm cut} \right) = \left\{ \begin{array}{ll}
         0 & \text{if } \theta < \theta_{\rm apo} \\
         \frac{1}{2} \left( 1 - \mathrm{cos} \left( \frac{\left( 2\theta - \theta_{\rm cut} \right) \pi}{\theta_{\rm cut}} \right) \right) & \text{if } \theta_{\rm apo} < \theta < \theta_{\rm cut} \\
         1 & \text{if } \theta_{\rm cut} < \theta,
    \end{array} \right.  \label{eq:apodization function}
\end{equation}
where $\theta_{\rm apo} = \theta_{\rm cut}/2$ is the angle at which we start the apodization and $\theta$ is defined in $\left[ 0^{\circ}, 180^{\circ} \right]$. This axisymmetric apodization is applied to the beam pattern to extract the contribution of the side-lobes starting from the angle $\theta_{\rm cut}$. In this way, we extracted the side-lobes starting from 3 different angular distances ($\theta_{\rm cut} = 5^{\circ},\, 10^{\circ},\, 15^{\circ}$), producing $\mathbf{m} \left( \theta > \theta_{\rm cut} \right) = \mathbf{m_{\rm cut}}$ maps, and  the full 4$\pi$ beam, producing $\mathbf{m_{4\pi}}$ maps. Note that this is an abusive notation as \texttt{GRASP} produces beam maps only up to $\theta = \pi/2$, so $\mathbf{m_{4\pi}}$ is convolved by a $2\pi$ beam, rigorously speaking, and the beam is assumed to vanish for $\theta > \pi/2$. Nevertheless, we will continue using it in the following. The three apodization functions used in the following are illustrated in Figure \ref{fig:ring window functions}. These three angular ranges are chosen to probe several levels of side-lobe pick-up given the specifics of the \textit{LiteBIRD} optics and the experimental constraints to characterize at high significance level the beam pattern.

We want to emphasize that even though all the specific features related to co-polar components are obtained by \texttt{GRASP} simulations, they are lost from the use of the transformations in Eq.~\eqref{eq:beamIQU}. Thus, they are not encoded in the convolution and the beam non-idealities (e.g. side-lobe pick up) for the co-polar component are assumed to be similar to the ones from $\tilde{I}$. Moreover, the results presented here are obtained with an approximated convolution method as the cross-polar component of the beam, albeit small, is totally neglected in the beam decomposition Eq.~\eqref{eq:beamIQU}. This is mainly supported by the fact that linear polarization detectors are designed to have a minimal cross-polar response, and thus instrumental beams are often approximated in the literature as just co-polar. The derivation of requirements with an implementation properly accounting for cross-polar components of the beam is left for future work.

\subsubsection*{Perturbed beam maps}

The study presented here relies on the production of simulated maps including the generation of sky emissions, map-making and realistic beam convolution in various settings. Given the complexity and time consumption of such simulations, we keep their production to the bare minimum and simulate the effect of an imperfect beam knowledge directly at the map level. Therefore, we combine the previously defined sets of maps $\mathbf{m_{\rm cut}}$ and $\mathbf{m_{4\pi}}$ to produce maps convolved with perturbed beams. By producing the map difference $\bold{m} \left( \theta > \theta_{\rm inf} \right)-\bold{m} \left( \theta > \theta_{\rm sup} \right) \equiv \bold{m_{inf}}-\bold{m_{sup}}$, we are left with maps that correspond to a convolution of the sky by the beams in the angular region between $\theta_{\rm inf}$ and $\theta_{\rm sup}$, i.e. in an angular window function defined as:
\begin{equation}
    W \left( \theta \right) \equiv A \left( \theta, \theta_{\rm inf} \right) - A \left( \theta, \theta_{\rm sup} \right) \label{eq:window function}.
\end{equation}

\begin{figure}[!htb]
\begin{center}
\includegraphics[width=\textwidth]{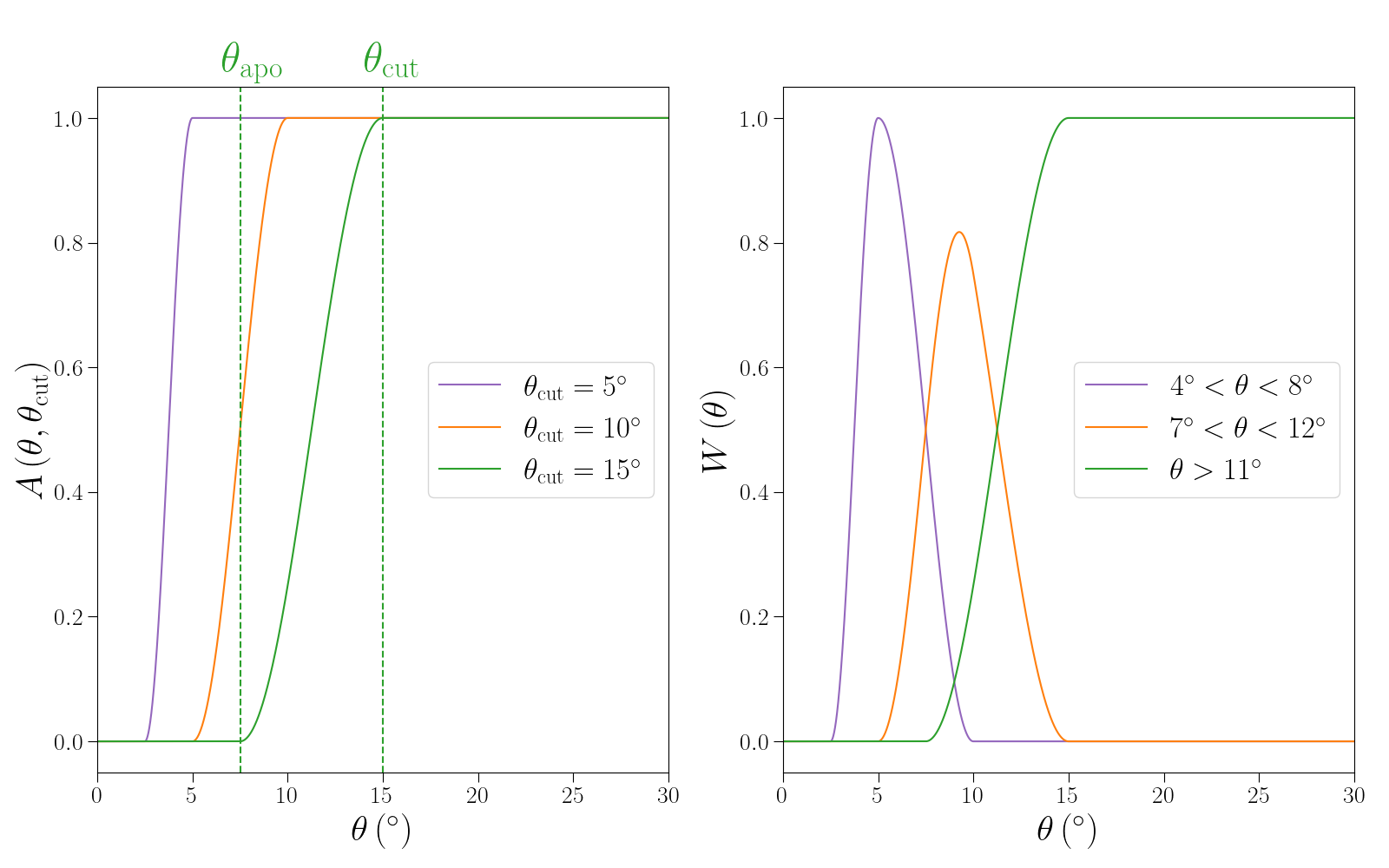}
\caption{(Left) Apodization functions as defined in Eq.~\eqref{eq:apodization function} for three values of $\theta_{\rm cut}$: $5^{\circ}$, $10^{\circ}$ and $15^{\circ}$. The definition of $\theta_{\rm apo}$ and $\theta_{\rm cut}$ are emphasized for the case $\theta_{\rm cut}=15^{\circ}$. (Right) Beam window functions defined as in Eq.~\eqref{eq:window function}, for three sets of $\left[ \theta_{\rm inf}, \theta_{\rm sup} \right]$: [$5^{\circ}$, $10^{\circ}$], [$10^{\circ}$, $15^{\circ}$] and [$15^{\circ}$, $180^{\circ}$]. The corresponding angular ranges at half maximum are: [$4^{\circ}$, $8^{\circ}$], [$7^{\circ}$, $12^{\circ}$] and [$11^{\circ}$, $180^{\circ}$].}
\label{fig:ring window functions}
\end{center}
\end{figure}

Throughout this paper, we consider three angular ranges $\left[ \theta_{\rm inf}, \theta_{\rm sup} \right]$: [$5^{\circ}$, $10^{\circ}$], [$10^{\circ}$, $15^{\circ}$] and [$15^{\circ}$, $180^{\circ}$]. To be more representative, we describe the angular range spanned by these windows in terms of the angles at half maximum, which are respectively: [$4^{\circ}$, $8^{\circ}$], [$7^{\circ}$, $12^{\circ}$] and [$11^{\circ}$, $180^{\circ}$]. The window functions corresponding to these angular ranges can be seen in Figure~\ref{fig:ring window functions}.

\begin{figure}[!htb]
\begin{center}
\includegraphics[width=\textwidth]{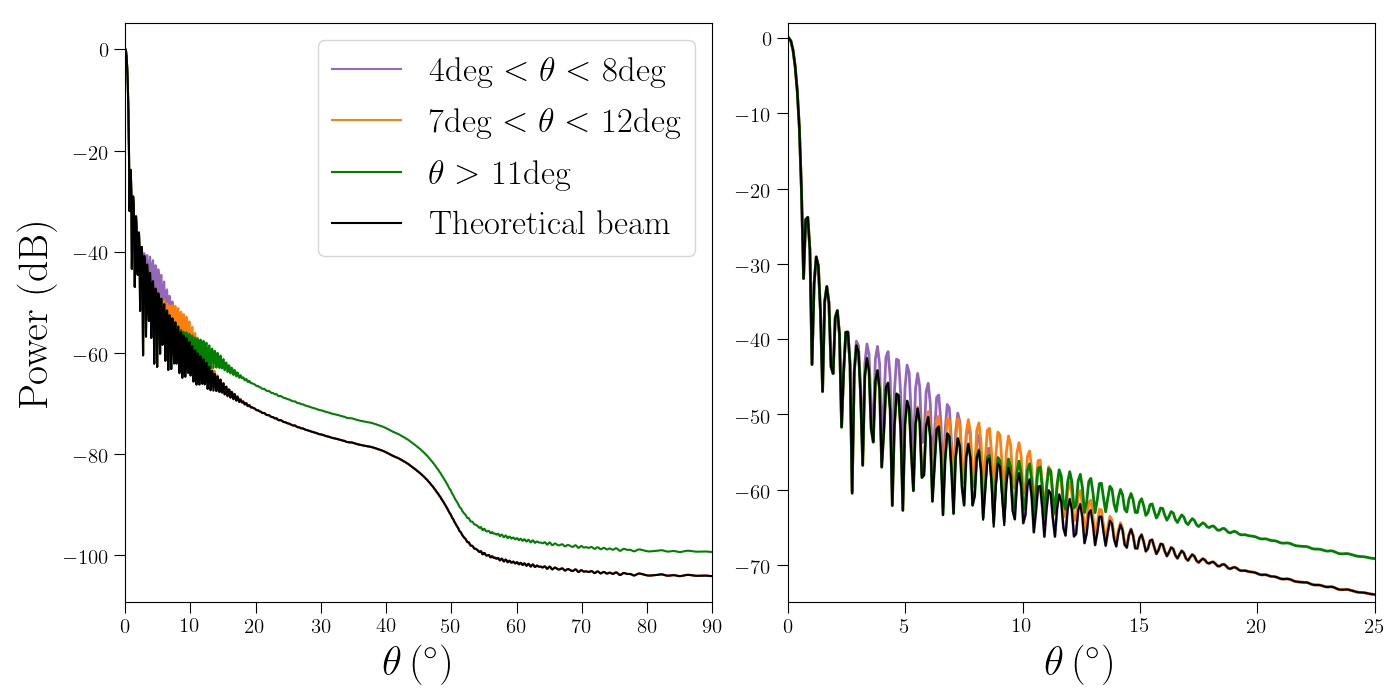}
\caption{Examples of perturbed beams used in this analysis in \textit{LiteBIRD}'s 100~GHz channel, in the three angular windows [4deg, 8deg] (purple), [7deg, 12deg] (orange) and [11deg, 180deg] (green), compared to the nominal beam (black) used in the convolution of the reference $\bold{m_{4\pi}}$ maps, for a perturbation amplitude $\alpha = 2.0$. The left panel shows the whole beam profile while the right panel is zoomed on the $\left[ 0^{\circ}, 25^{\circ} \right]$ angular range.}
\label{fig:perturbed beams}
\end{center}
\end{figure}

We see that, given these three combinations of maps we can select parts of the beam on rings at different angular radial distances from the beam axis, with some small overlap between the three cases. This way, it is possible to generate maps convolved by a perturbed beam, where the perturbation is localized in these annular rings and the shape of the perturbation is that of the nominal beam. These maps are produced using the following combination:
\begin{equation}
    \bold{m_{pert}} \left( \alpha \right) = \bold{m_{4\pi}} + \alpha \left( \bold{m_{inf}}-\bold{m_{sup}} \right), \label{eq:map combination}
\end{equation}
where $\alpha$ is an arbitrary parameter that drives the amplitude of the beam perturbation, that will be referred to simply as the perturbation amplitude in the following. Note that, because the beam amplitude is a positive quantity, we must have $\alpha \geq -W_{\rm max}^{-1}$, where $W_{\rm max}$ is the maximum of the window function over the angular range. An example of the perturbed beam profile (averaged over the detectors and symmetrized, see Section~\ref{subsubsection:Modeling Case}) by which $\bold{m_{pert}} \left( \alpha \right)$ at 100~GHz in the LFT is convolved, with $\alpha= 2.0$, is given in Figure~\ref{fig:perturbed beams}.

\subsubsection*{Dipole calibration}

Since component separation methods are based on the relative comparison of the amplitude of observed maps in the different frequency channels, these maps must be properly normalized beforehand. In particular, the combination of maps described by Eq.~\eqref{eq:map combination} modifies the map normalization compared to that of $\bold{m_{4\pi}}$. Therefore, we need to renormalize these maps before applying component separation. The renormalization is performed by matching the temperature dipoles $C_{\ell = 1}^{TT}$ of the perturbed maps with those of $\bold{m_{4 \pi}}$. This is equivalent to a perfect calibration of the maps on the dipole, in other words the perturbation of the beam we introduce has no impact on the dipole calibration. This is justified because the procedure of gain calibration on the dipole is entirely decoupled from the beam calibration. In the present work, we want to isolate the effect of imperfect beam knowledge from other systematic effects so the study of a potential coupling with dipole miscalibration, as well as with other systematic effects, is left for future work. For a study of the effect of errors on gain calibration in the context of \textit{LiteBIRD}, see \cite{Ghigna:2020wat}.

Using this normalization convention, the maps convolved by a perturbed beam with a normalization factor $\mu$, hereafter the ``perturbed maps", are defined by:
\begin{equation}
    \bold{m_{pert}} \left( \alpha \right) = \mu \left[ \bold{m_{4\pi}} + \alpha \left( \bold{m_{inf}}-\bold{m_{sup}} \right) \right], \label{eq:normalized map combination}
\end{equation}
After a little algebra, we find the analytical expression for $\mu$ to be determined by the dipoles of the relevant maps previously defined:
\begin{equation}
    \mu = \sqrt{\frac{C_{\ell = 1}^{4\pi}}{C_{\ell = 1}^{4\pi} + 2\alpha \left( C_{\ell = 1}^{4\pi-\rm inf} - C_{\ell = 1}^{4\pi-\rm sup} \right) + \alpha^{2} \left( C_{\ell = 1}^{\rm inf} - 2C_{\ell = 1}^{\rm inf-sup} + C_{\ell = 1}^{\rm sup} \right)}},
\end{equation}
where $C_{\ell = 1}^{X}$ is the value of the temperature power-spectrum of $X$ at $\ell=1$ ($4\pi$ standing for $\bold{m_{4\pi}}$, inf for $\bold{m_{inf}}$ and sup for $\bold{m_{sup}}$) and $C_{\ell = 1}^{X-Y}$ is the value of the temperature cross-spectrum of $X$ and $Y$.

Once all these preparation steps are performed, the resulting maps are effectively convolved with the following perturbed beams:
\begin{equation}
    B_{\rm pert}^{\nu}(\theta,\phi) = \mu_{\nu} (1 + \alpha_{\nu} W(\theta))\, B_0^{\nu}((\theta,\phi), \label{eq:perturbed beam}
\end{equation}
where $B_{0}^{\nu}$ corresponds to the simulated \texttt{GRASP} beams used to convolve the sky maps in the frequency channel $\nu$.

\subsubsection*{Beam correction and masking}
\label{subsubsection:Beam correction and masking}

The observed maps are convolved with the beams. In order to follow the steps of the real analysis pipeline, we deconvolve the observed data in harmonic domain using transfer functions $b_{\ell}^{\nu}$ in each frequency band to obtain beam corrected harmonic coefficients $\bold{a}_{\ell m, \rm corr}^{\nu}$. In the real analysis, these transfer functions would be computed from the beam maps estimated from calibration and modeling. Note that beam asymmetries present in the input beam models are not taken into account in this step, but will not affect the final results as will be explained in Section~\ref{subsection:Component separation}. Indeed, we are interested in the effect of imperfect beam knowledge on the cosmological results so we will ultimately compare the contamination in the reconstructed CMB maps from the perturbed ($\bold{m_{pert}^{\nu}}$) and unperturbed cases ($\bold{m_{4\pi}^{\nu}}$). Therefore, as long as the same $b_{\ell, \rm eff}^{\nu}$ are used in the perturbed and unperturbed cases, our results will not depend on the shape of the effective beams. For this reason, we defer the details of the computation of these transfer functions to Appendix~\ref{appendix:Correction by the effective beam}.

The last step needed to produce the input data for the component separation step is to apply a mask to the Galactic plane where the foreground signal is very strong. This amounts to scaling each pixel's amplitude using weights $\bold{w} \in \left[ 0, 1 \right]$ where $\bold{w}$ is 0 in the masked region, and smoothly gets to 1 in the region used for the analysis. Therefore, we first need to translate back the corrected $\bold{a}_{\ell m, \rm corr}$ into pixel amplitudes $\bold{m_{pert}^{corr}}$ before applying the mask weights:
\begin{equation}
    \bold{m_{pert}^{masked}} = \bold{w} \cdot \bold{m_{pert}^{corr}}.
\end{equation}

This step, in itself, will introduce mixing between the $Q$ and $U$ components leading to $E \rightarrow B$ leakage. However, this should not have a significant impact on residuals from beam far side-lobes systematic mismatch as will be explained in section \ref{subsection:Component separation}. In this study, we used the Planck HFI Galactic mask with $f_{sky} \sim 51 \%$ \cite{PLA}.

Finally, because we use component separation in the harmonic domain, we perform a last translation of maps into harmonic domain, where we are dealing with pseudo-$\bold{a}_{\ell m}$'s because of the masking.

\subsubsection{Modeling Case}
\label{subsubsection:Modeling Case}

We know that, in a realistic set-up, direct measurements of the beams will be very challenging in the region very far from the center of the beam,
%In such a region, measurements are very hard to make
and our main estimation of the beam amplitude would be from modeling. We want to study the impact of modeling errors in this region and its dependence in the angle $\theta_{\rm lim}$ from which measurements are absent, which constitutes our Modeling Case. We employ a different method, as opposed to the ``Detailed Method" presented in Section~\ref{subsubsection:Perturbation Case}, referred to as the ``Axisymmetric Method", to carry out this study. This is because the Detailed Method that relies on focal plane simulation and TOD level convolution is computationally heavy, which is not well suited for the Modeling Case as we need to explore a broader parameter space (of the order of $\sim10^4$ configurations) see Section~\ref{subsection:Requirements for the Modeling Case}. As a result, we employ in the Axisymmetric Method a faster and more flexible way of producing beam convolved maps based on an axisymmetric approximation.

It appears clear that the previously described methodology of local perturbation of the beam is also not very well suited for the study of the Modeling Case. We develop in this section a more adapted approach.

\subsubsection*{Beam convolved map production}

The approach developed in Section~\ref{subsubsection:Perturbation Case} is based on extensive simulations of realistic sky maps convolved with perturbed beams including simulations of the focal plane, scanning strategy effects, and taking some level of beam asymmetry into account. In the context of the current section, as we mentioned above, we need a more flexible and quicker procedure. Therefore, we adopt here another similar but simpler approach. The maps are convolved by the same beam in all studied cases, the assumed true beams $\overline{B_{0}^{\nu}}$ are taken from the simulated \texttt{GRASP} beams $B_{0}^{\nu}$ by averaging over the simulated detectors of the focal plane and symmetrizing them around the beam axis, i.e. averaging over the $\phi$ coordinate. On the other hand, they are corrected using the transfer functions corresponding to the assumed beam model of the analysis. In harmonic space, the $\bold{a}_{\ell m}$ to be used in component separation are defined as:
\begin{equation}
    \bold{a}_{\ell m} = \frac{b_{\ell}^{\rm GRASP}}{b_{\ell}^{\rm model}} \bold{a}_{\ell m}^{\rm sky},
\end{equation}
with $\bold{a}_{\ell m}^{sky}$ corresponding to simple band-passes integrated foreground maps assuming the spatially homogeneous model \texttt{d0s0} of \texttt{PySM} \cite{Thorne:2016ifb} and the $b_{\ell}$'s being the transfer functions of the averaged symmetrized \texttt{GRASP} beams and of the beam models. We also apply the same masking scheme. In the same way, we make sure to reproduce the effect of dipole calibration, assumed perfect, by re-scaling the model transfer function such that $b_{\ell = 1}^{\rm model} = b_{\ell = 1}^{\rm GRASP}$.

We will show in Section~\ref{subsubsection:Comparison of the Detailed Method and the Axisymmetric Method} that this method relying on the axisymmetric beam approximation and the more realistic one described in Section~\ref{subsubsection:Perturbation Case} gives, in fact, very comparable results in our settings. The interpretation of this correspondence will also be discussed in Section~\ref{section:Discussions}.

\subsubsection*{Beam modeling at large angle}

We assume the beam to be perfectly known in the central region, for angles lower than an arbitrary $\theta_{\rm lim}$. For angles $\theta > \theta_{\rm lim}$, we assume an empirical beam model to simulate the complete lack of information on the true beam shape in this region. As explained earlier, one of the goals of the present study is to estimate angles $\theta_{\rm lim}$ for which the induced error from this lack of information is compatible with the scientific goals of the mission. Two conservative empirical models would be a constant amplitude $B_{\rm lim} = B_{\rm true} \left( \theta_{\rm lim} \right)$, or otherwise to cut the beam for $\theta>\theta_{\rm lim}$. However, both these two cases are too simplistic and overly pessimistic. In addition, the results obtained by using this kind of empirical model would be very dependent on the assumed true beam used in the analysis.

\begin{figure}[!htb]
\begin{center}
\includegraphics[width=0.8\textwidth]{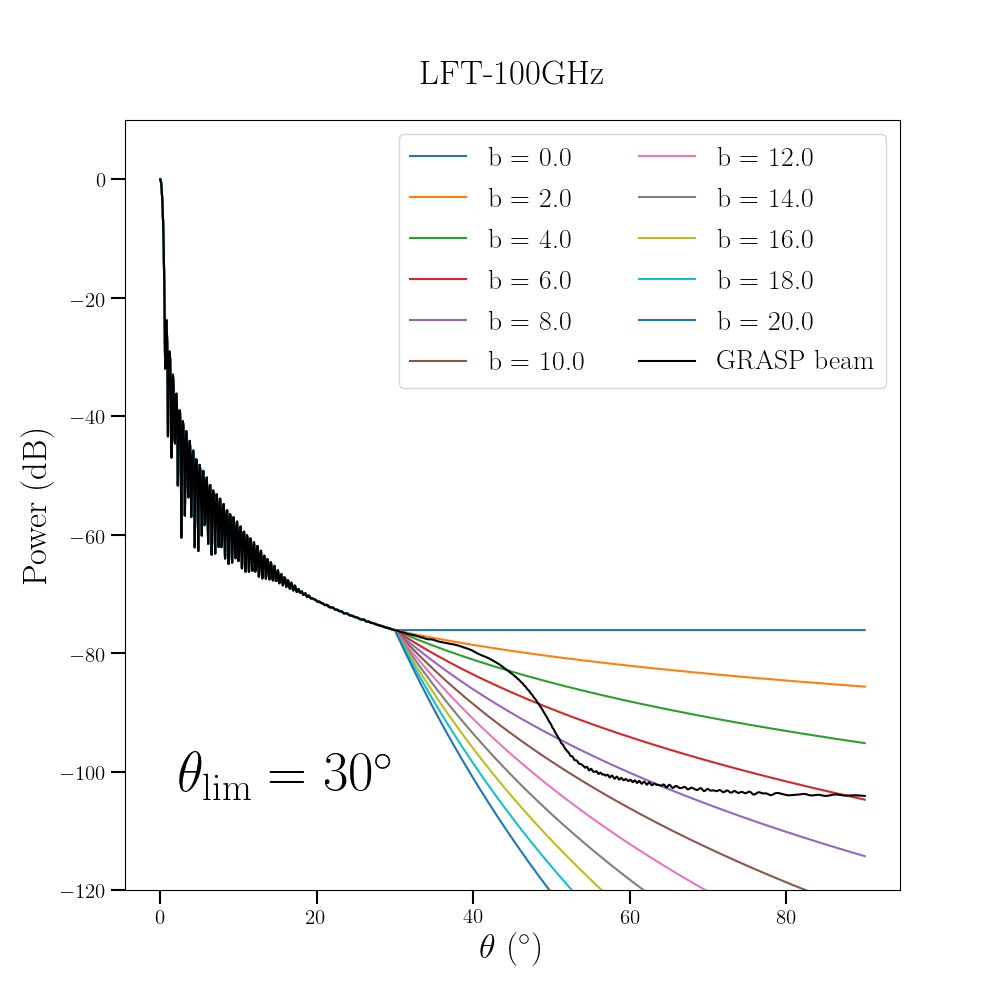}
\caption{Examples of beam models in \textit{LiteBIRD}'s LFT 100~GHz channel compared to the averaged and symmetrized \texttt{GRASP} beam used as true beam reference. The limit angle is taken to be $\theta_{lim} = 30^{\circ}$, and the arbitrary parameters of the power law model is taken to be $b \in \left[ 0, 20 \right]$.}
\label{fig:syst beam models}
\end{center}
\end{figure}

Therefore, we need a more complex model to capture the essential features, to be determined, of the true beam. The question of how complex the model should be is essential in this analysis since complexity is at the expense of flexibility and generality. With these considerations in mind, we keep the complexity low and consider simple power laws for $\theta > \theta_{\rm lim}$ with an arbitrary parameter $b \geq 0$:
\begin{equation}
        B_{\rm model}^{\nu} \left( \theta, \theta_{\rm lim}, b \right) = \left( 1 - A' \left(  \theta, \theta_{\rm lim} \right) \right) \overline{B_{0}^{\nu}} + A' \left(  \theta, \theta_{\rm lim} \right) B_{\rm lim}^{\nu} \left( \frac{\theta_{\rm lim}}{\theta} \right)^{b}. \label{eq:beam model}
\end{equation}
The function $A'$ is an apodization function, with a much sharper transition than in Eq.~\eqref{eq:apodization function}, because it needs to compensate the high amplitude of the power law $B_{\rm lim}^{\nu} \left( \frac{\theta_{\rm lim}}{\theta} \right)^{b}$ at low $\theta$ in a large range of possible $\theta_{\rm lim}$. It is taken to be the following logistic function:
\begin{equation}
    A' \left( \theta, \theta_{\rm lim} \right) = \frac{1}{1+e^{20 \left( \theta_{\rm lim} - \theta \right)/1^{\circ}}}.
\end{equation}
The case $b=0$ in Eq.~\eqref{eq:beam model} gives back the constant model and the case $b \rightarrow \infty$ leads to the cut beam. Therefore, this model includes the two most non-informative models, but also all the intermediate cases. Examples of such beam models are shown in Figure~\ref{fig:syst beam models} for several values of $b$ and $\theta_{\rm lim} = 30^{\circ}$, in \textit{LiteBIRD}'s LFT 100~GHz channel. We will see in Section \ref{subsection:Requirements for the Modeling Case} that these very simple models are enough to grasp the most important characteristic of the beams, i.e. the residual power in the far side-lobes.

\subsection{Component separation}
\label{subsection:Component separation}

\subsubsection{Spectral parameters estimation}
\label{subsubsection:Spectral parameters estimation}

The goal of component separation for foreground cleaning is to retrieve the CMB and possibly other components emission maps from the observed frequency maps $\bold{m^{\nu}}$. In particular, we are concerned with the contamination from polarized Galactic dust, dominant at high frequencies, and synchrotron, dominant at low frequencies, which are expected to have a significant impact on the scientific results if not properly cleaned \cite{Planck:2018yye,Krachmalnicoff:2015xkg,Errard:2015cxa}. %Taking the foreground cleaning step into account is very important in this analysis since the main effect of beam miscorrection is a mismatch in the transfer function of foregrounds between $\ell = 1$ (over which we calibrate) and higher multipoles %which affects the foreground cleaning procedure. the residual leakage of foreground emissions from the Galactic plane to other regions 
%due to convolution by the beam, not entirely accounted for and removed by foreground cleaning.
Taking the foreground cleaning step into account is very important in this analysis since the main effect of beam miscorrection is to shift the relative amplitude between frequency channels, changing the effective SEDs of the components for multipoles $\ell \geq 2$.
The component separation technique used in this analysis is based on the method described in \cite{Stompor:2008sf,Stompor:2016hhw} and implemented in the \texttt{FGBuster} package, with the notable difference that it operates in harmonic domain. This component separation method is based on a parametric approach where the foreground components spectral properties are modeled using analytic emission laws. The dust component is modeled by a modified black-body law with temperature $T_{d}$ and spectral parameter $\beta_{d}$, while the synchrotron emission is modeled by a simple power law with spectral parameter $\beta_{s}$. These three parameters are assumed constant over the sky and are fitted using the maximum-likelihood principle. Therefore, this component separation method is based on fitting the SEDs from the relative amplitudes of the frequency maps which are significantly altered for all $\ell \geq 2$ by the dipole calibration of maps with large leakage from the Galactic plane. This makes our approach to foreground cleaning particularly sensitive to the effect of a mismatch in the far side-lobes region.

The observed maps are assumed to be the result of combined emissions from the different sources, CMB and foreground components, together with instrumental noise. Therefore, the observed amplitudes $\bold{m}_{p}$ in a single pixel gathered in a vector of $n_{\nu}$ frequencies are assumed to be produced by $n_{c}$ components, and is modeled in pixel space as:
\begin{equation}
	\bold{m}_{p} = \bold{A} \left( \beta \right) \bold{s}_{p} + \bold{n}_{p}.
\end{equation}

In the above data model, $\bold{A}$ is a matrix that gives each components' amplitude scaling across frequency bands and depends on a set of foreground parameters that we collectively refer to as the spectral parameters $\beta$, $\bold{s}_{p}$ is a vector describing the amplitude of the different components in the pixel $p$, and $\bold{n}_{p}$ is the amplitude of instrumental noise in this pixel assuming white noise determined from the sensitivities in Table~\ref{tab:LB instrument}. These elements have the following dimensions:
\begin{equation}
	\bold{m}_{p}, \bold{n}_{p} = \left. \begin{bmatrix}
		\\
		\ \vdots \ \  \\
		\\
	\end{bmatrix} \right\} n_{\nu}
	\qquad
	\bold{A} = \left[ \vphantom{\matriximg} \right. \underbrace{\matriximg}_{\rule{0pt}{1.5ex} \textstyle n_{c}}\kern-\nulldelimiterspace\left.\left. \vphantom{\matriximg} \right]\right\}\kern-2\nulldelimiterspace \ n_{\nu}
	\qquad
	\bold{s}_{p} = \left. \begin{bmatrix}
		\\
		\ \vdots \ \  \\
		\\
	\end{bmatrix} \right\} n_{c}.
\end{equation}

Assuming that the noise is Gaussian with covariance matrix $\bold{N}_{p}$, the log-likelihood for the data given the model is:
\begin{equation}
	 \mathcal{S} \equiv -2\ \mathrm{ln} \ \mathcal{L} = \mathrm{const} + \sum_{p} \left( \bold{m}_{p} - \bold{A} \bold{s}_{p} \right)^{T} \bold{N}_{p}^{-1} \left( \bold{m}_{p} - \bold{A} \bold{s}_{p} \right).
	 \label{eq:likelihood}
\end{equation}

In the following, we drop the constant term as it does not impact the position of the maximum. As mentioned previously, the parameters over which we maximize the likelihood are the foreground spectral parameters $\beta$ and the amplitudes of the components $\bold{s}$. The vanishing partial derivatives of the likelihood at the maximum imply:
\begin{eqnarray}
	- \sum_{p} \left( \frac{\partial \bold{A}}{\partial \beta} \ \bold{s}_{p} \right)^{T} \bold{N}_{p}^{-1} \left( \bold{m}_{p} - \bold{A} \bold{s}_{p} \right) = 0 \\
	\bold{s}_{p} = \left( \bold{A}^{T}\bold{N}_{p}^{-1}\bold{A} \right)^{-1} \bold{A}^{T}\bold{N}_{p}^{-1} \bold{m}_{p}. \label{eq:max s}
\end{eqnarray}

Evaluating the likelihood Eq.~\eqref{eq:likelihood} at the maximum of $\bold{s}_{p}$, Eq.~\eqref{eq:max s} allows us to define the profile likelihood, or spectral likelihood, which depends only on the spectral parameters $\beta$:
\begin{equation}
	\mathcal{S}_{spec} = - \sum_{p} \bold{m}_{p}^{T}\bold{N}_{p}^{-1}\bold{A} \left( \bold{A}^{T}\bold{N}_{p}^{-1}\bold{A} \right)^{-1} \bold{A}^{T}\bold{N}_{p}^{-1} \bold{m}_{p}.
\end{equation}

As was pointed out at the end of Section~\ref{subsubsection:Beam correction and masking}, we work in harmonic domain instead of pixel domain. Therefore, we have to expand $\bold{m}_{p}$ and $\bold{N}_{p}$ on the basis of spherical harmonics. We use the spherical harmonic transform operator $\bold{Y}_{\ell m}$ and define the data harmonic coefficients $\bold{a}_{\ell m}$ as well as the noise covariance in harmonic domain $\mathcal{N}_{\ell_{1} m_{1}, \ell_{2} m_{2}}$ by the following relations:
\begin{equation}
     \bold{m}_{p} = \sum_{\ell, m} \bold{Y}_{\ell m, p}^{\dagger} \bold{a}_{\ell m} \qquad \mathrm{and} \qquad \bold{N}_{p}^{-1} = \sum_{\ell_{1}, \ell_{2}, m_{1}, m_{2}} \bold{Y}_{\ell_{1} m_{1}, p} \mathcal{N}_{\ell_{1} m_{1}, \ell_{2} m_{2}}^{-1} \bold{Y}_{\ell_{2} m_{2}, p}^{\dagger}.
\end{equation}

Given the orthogonality of spherical harmonics on the full sky, that we can use even though we only observe a limited patch of the sky by setting $\bold{m}_{p}$ outside the patch to be zero, the spectral likelihood becomes:
\begin{equation}
    \mathcal{S}_{spec} = - \sum \bold{a}_{\ell_{1} m_{1}}^{\dagger} 
    \mathcal{N}_{\ell_{1} m_{1}, L_{1}M_{1}}^{-1} \bold{A} \left( \bold{A}^{T}\mathcal{N}_{L_{1}M_{1}, L_{2}, M_{2}}^{-1}\bold{A} \right)^{-1} \bold{A}^{T}  \mathcal{N}_{L_{2}M_{2}, \ell_{2} m_{2}}^{-1} \bold{a}_{\ell_{2} m_{2}},
\end{equation}

where the sum runs over all harmonic indices. If we further assume $\bold{N}^{-1}$ to be diagonal in harmonic space, i.e. homogeneous noise, and to depend only on $\ell$, the spectral likelihood simplifies to:
\begin{equation}
    \mathcal{S}_{spec} = - \sum_{\ell, m} \bold{a}_{\ell m}^{\dagger} \mathcal{N}_{\ell}^{-1} \bold{A} \left( \bold{A}^{T} \mathcal{N}_{\ell}^{-1}\bold{A} \right)^{-1} \bold{A}^{T} \mathcal{N}_{\ell}^{-1} \bold{a}_{\ell m}. \label{eq:harmonic spectral likelihood}
\end{equation}

Since we are interested in forecasting the impact of beam shape uncertainty on component separation, we only work with the likelihood averaged over noise realizations. Moreover, as detailed in the previous sections, we apply a correction for the beam shape by deconvolving the maps using effective beams $b_{\ell}^{eff}$, therefore the noise covariance matrix is scaled by a factor $\left( b_{\ell}^{eff} \right)^{-2}$. The noise averaged spectral likelihood is given by:
\begin{equation}
    \left< \mathcal{S}_{spec} \right>_{noise} = - \sum_{\ell, m} \mathrm{Tr} \left[ \mathcal{N}_{\ell}^{-1} \bold{A} \left( \bold{A}^{T} \mathcal{N}_{\ell}^{-1} \bold{A} \right)^{-1} \bold{A}^{T} \mathcal{N}_{\ell}^{-1} \left( \bold{\hat{a}}_{\ell m} \bold{\hat{a}}_{\ell m}^{\dagger} + \mathcal{\hat{N}}_{\ell} \right) \right],
    \label{eq:averaged spectral likelihood}
\end{equation}
where $\bold{\hat{a}}_{\ell m}$ are the true (noiseless) observed multipoles, including only beam convolved foreground emissions, and $\hat{\mathcal{N}_{\ell}}$ is the true noise covariance. If we assume that $\hat{\mathcal{N}}_{\ell} = \mathcal{N}_{\ell}$, this likelihood reduces to the spectral likelihood Eq.~\eqref{eq:harmonic spectral likelihood} involving noiseless data, and it was shown in \cite{Stompor:2016hhw} that CMB data does not play any role on the likelihood maximization. This justifies the use of foreground only input maps in this analysis. Note that the likelihood is defined for each Stokes parameter independently, because we assume no leakage from T to P, or from E to B. Indeed, this effect is expected to be mostly mitigated by the HWP, and the residual leakage to be of second order compared to foreground residuals. Therefore, we consider only the CMB $B$ modes in this work.

\subsubsection{Impact on cosmological results}
\label{subsubsection:Impact on cosmological results}

By maximizing the spectral likelihood Eq.~\eqref{eq:averaged spectral likelihood}, we recover spectral parameters $\beta$. These are then used to clean foreground emissions from frequency maps and recover the CMB map. The estimate of the components is given by:
\begin{equation}
    \bold{\bar{s}}_{\ell m} = \left( \bold{A}^{T} \mathcal{N}_{\ell}^{-1} \bold{A} \right)^{-1} \bold{A}^{T} \mathcal{N}_{\ell}^{-1} \bold{\hat{a}}_{\ell m}.
\end{equation}

An error on the spectral indices will lead to foreground leakage into CMB maps and larger reconstructed CMB power spectra. The perturbation introduced to account for the uncertainty on the shape of beam far side-lobes can impact the measurement of the spectral parameters, therefore introducing systematic residuals on the CMB B-mode power spectrum and systematic bias $\delta r$ on the tensor-to-scalar ratio $r$.

Because we want to disentangle the impact of systematic error from beam shape mismatch and intrinsic systematic errors of the component separation procedure, we define the residuals to be the difference between the recovered CMB $B$-mode multipoles in a reference case $\bold{\bar{s}}_{\ell m}^{\rm ref}$ that assumes no beam perturbation and the recovered CMB $B$-mode multipoles in the perturbed case $\bold{\bar{s}}_{\ell m}$ (see \cite{LiteBIRD:2022cnt}):
\begin{equation}
    \delta \bold{s}_{\ell m} = \bold{\bar{s}}_{\ell m} - \bold{\bar{s}}_{\ell m}^{\rm ref}. \label{eq:residuals}
\end{equation}

In particular, because the reference case differs from the perturbed case only by the absence of the perturbation, this comparison removes the contribution to the residuals from a mismatch between the true foreground SEDs and the model used in the component separation, from the masking procedure and other effects which are present both in the reference and perturbed cases. This justifies the use of simplistic foreground models such as the \texttt{d0} and \texttt{s0} models from \texttt{PySM}.

The bias $\delta r$ is defined to be the measured value of $r$ assuming its true value to be $r=0$. Therefore, we can extract it from the power spectrum $C_{\ell}^{\rm res}$ of the residuals $\delta \bold{s}_{\ell m}$ by maximizing the following cosmological likelihood as a function of $r$:
\begin{equation}
-2 \mathrm{ln} \mathcal{L}_{cosmo} = f_{sky} \sum_{\ell} \left( 2\ell + 1 \right) \left( \mathrm{ln} C_{\ell}^{\rm th} \left( r \right) + \frac{C_{\ell}^{\rm th} \left( r=0 \right) + C_{\ell}^{\rm res}}{C_{\ell}^{\rm th} \left( r \right)} \right), \label{eq:cosmo likelihood}
\end{equation}
where $C_{\ell}^{th}$ is the theoretical CMB $B$-mode power spectrum that includes all expected contributions:
\begin{equation}
    C_{\ell}^{\rm th} = rC_{\ell}^{GW} + C_{\ell}^{\rm lens} + N_{\ell} + R_{\ell}^{\rm fore},
\end{equation}
where $C_{\ell}^{GW}$ is the primordial CMB BB power-spectrum for $r=1$, $C_{\ell}^{\rm lens}$ is the contribution from gravitational lensing, $N_{\ell}$ is the power-spectrum of the noise after component separation and $R_{\ell}^{\rm fore}$ is the expected power-spectrum of foreground residuals from statistical uncertainties propagated through component separation. The contributions $N_{\ell}$ and $R_{\ell}^{\rm fore}$ are estimated following \cite{LiteBIRD:2022cnt}.

\section{Results}
\label{section:Results}

The purpose of this section is to use the frameworks described in Section~\ref{section:Methodology} to build the bridge between scientific requirements for \textit{LiteBIRD} and requirements on the instrument knowledge for the specific case of beam far-sidelobes. Table~\ref{tab:Results organization} describe how the results for the various cases and methods are organized in the Section.

\begin{table}[!htb]
\begin{center}
\Large
\begin{tabular}{c|c|c|}
\cline{2-3}
& Perturbation Case &  Modeling Case \\ \hline
\multicolumn{1}{|c|}{Detailed Method} & \ref{subsection:Requirements for the Perturbation Case} & \notableentry \\ \hline
\multicolumn{1}{|c|}{Axisymmetric Method} & \ref{subsubsection:Comparison of the Detailed Method and the Axisymmetric Method} & \ref{subsubsection:Bias on the tensor-to-scalar ratio from beam mismodeling at large angle} \\ \hline
\end{tabular}
\caption{Organization of Section~\ref{section:Results} for the various cases. We remind the reader that the Perturbation Case corresponds to the study of measurement uncertainties while the Modeling Case corresponds to the study of modeling errors, and that the Detailed Method adopts a full simulation of the asymmetric beam, focal plane and scanning strategy while the Axisymmetric Method follows an axisymmetric map-based approach.}
\label{tab:Results organization}
\end{center}
\end{table}

\subsection{Requirements for the Perturbation Case}
\label{subsection:Requirements for the Perturbation Case}

\subsubsection{Bias on the tensor-to-scalar ratio from the beam perturbation}
\label{subsubsection:Standard method}

The goal of this section is to use the approach described in \ref{subsubsection:Perturbation Case} to evaluate the impact of an imperfect beam knowledge from calibration errors and to produce requirements on calibration measurements given the methodology assumptions of this work, detailed in Section~\ref{section:Methodology}.

\begin{figure}[!htb]
\begin{center}
\includegraphics[width=\textwidth]{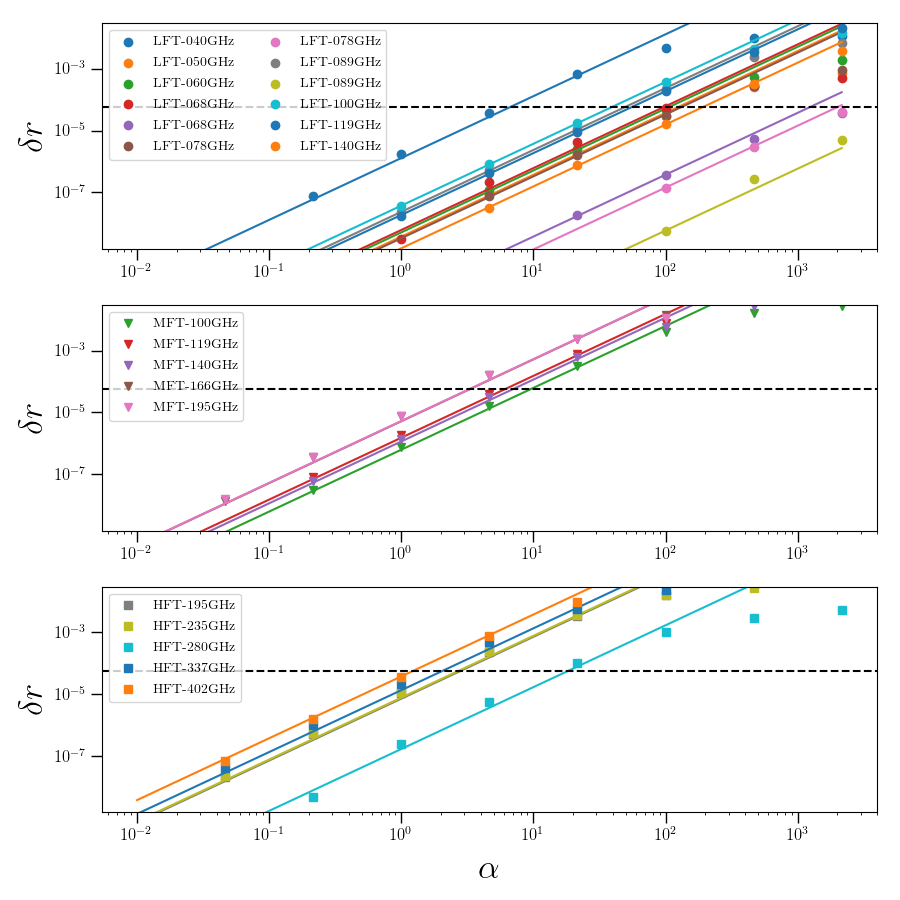}
\caption{Evolution of the bias on $r$ as a function of $\alpha$ in each frequency channel of the LFT (top), MFT (middle) and HFT (bottom). The solid lines correspond to the quadratic law $\delta r_{\nu} \propto \alpha_{\nu}^{2}$ that best fits the points at perturbation amplitudes $\alpha \leq 100$, where the quadratic scaling breaks down in frequency channels that induce a large bias. These results are obtained in the angular window at largest angle for illustration, the behaviour in the two other windows being similar.}
\label{fig:alpha versus deltar}
\end{center}
\end{figure}

To this aim, we estimate the bias on $r$ introduced by a perturbation of the beam in a single frequency channel and in a single angular window at a time, with a perturbation amplitude $\alpha^{\nu, W}$, following Eq.~\eqref{eq:normalized map combination} and the corresponding beam Eq.~\eqref{eq:perturbed beam}. The beams in other channels are left unperturbed such that the induced bias on the tensor-to-scalar ratio $\delta r^{\nu, W}$ comes only from the perturbed channel. From a given set of arbitrarily assigned error budgets per channel $\delta r_{\rm lim}^{\nu, W}$, we find the corresponding limit values $\alpha_{\rm lim}^{\nu, W}$ of the amplitudes $\alpha^{\nu, W}$ that induce a bias $\delta r^{\nu, W} = \delta r_{\rm lim}^{\nu, W}$. We found that the bias on $r$ scales as $\delta r \propto \alpha^{2}$, as can be seen in Figure \ref{fig:alpha versus deltar}. This is to be expected since $\alpha$ scales the amplitude of the perturbation, therefore the power-spectrum is scaled by $\alpha^{2}$.

If we assume the effect on each channel to be independent, which is a reasonable assumption for small biases as is the case in this study, $a_{\ell m}$ of the corresponding residuals would add in quadrature. So, because $\delta r$ is the amplitude of their variance the total bias is obtained by simply summing the biases from the different channels. Therefore, the total bias will correspond to the total error budget assigned to the beams far side-lobe systematic effect $\Delta r_{\rm FSL} = \sum_{\nu, W} \delta r_{\rm lim}^{\nu, W}$. We follow the total budget for beam far side-lobe systematic bias defined by the \textit{LiteBIRD} collaboration: $\Delta r_{\rm FSL} = 1.9\times 10^{-5}$ \cite{LiteBIRD:2022cnt}. Note that this corresponds to $\sim 3\%$ of the total mission systematic budget which is $0.001/\sqrt{3}$. We further assume, for now, that each channel have the same contribution to $\Delta r$. In our case, there are 22 frequency channels and 3 angular windows, so the allowed bias from each channel is assigned with equal weight and is simply $\delta r_{\rm lim}^{\nu, W} = \left( 1.9 / 66 \right) \times 10^{-5}$. Note that assuming different values of $\Delta r_{\rm FSL}$ or $\delta r_{\rm lim}^{\nu, W}$ would lead to different requirements.

\begin{figure}[!htb]
\begin{center}
\includegraphics[width=0.9\textwidth]{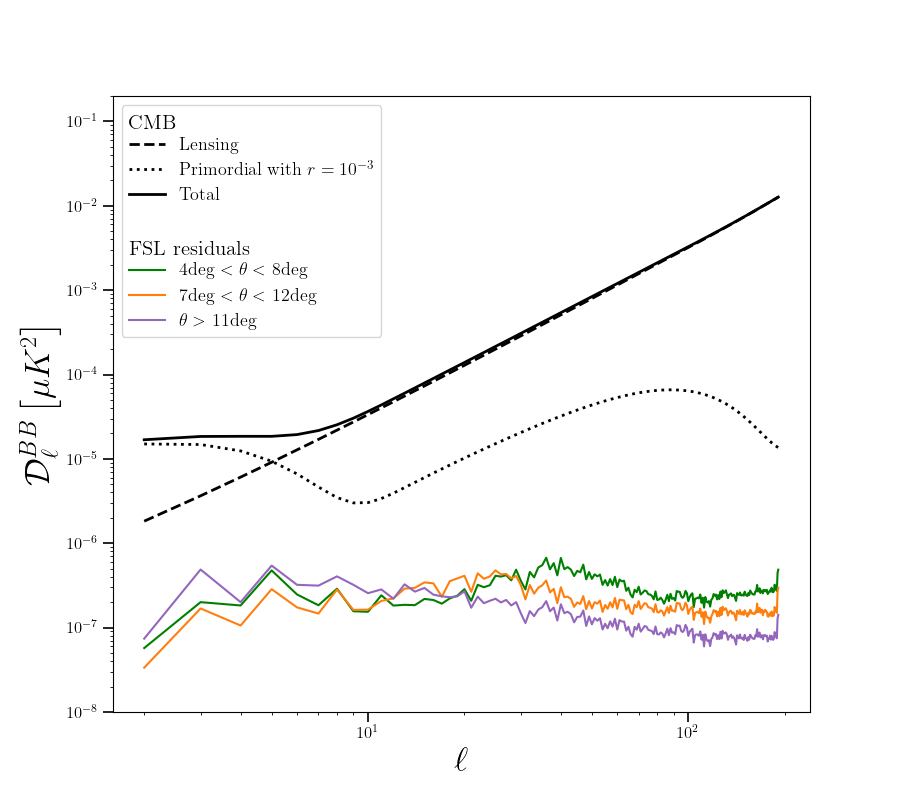}
\caption{Average of the residual $B$-mode power spectra from beam far side-lobes mismatch in the 3 angular windows, for 100 realizations of randomly produced sets of perturbation amplitudes $\alpha^{\nu, W}$ parameters following the requirements that $\alpha^{\nu, W} \in \left[ {\rm max} \left( -W_{\rm max}^{-1},-\sqrt{3}\alpha_{\rm lim}^{\nu, W} \right); \sqrt{3}\alpha_{\rm lim}^{\nu, W} \right]$ for all frequency channels in the given angular window. The corresponding total bias on $r$ is evaluated, combining the three windows and using \eqref{eq:cosmo likelihood}, to be $\Delta r = 2.08 \times 10^{-5}$.}
\label{fig:Residual Cls}
\end{center}
\end{figure}

In order to check that the correlation between the effect in the different frequency channels is limited, we estimated the total contribution $\Delta r$ with a hundred realizations of simultaneously varying the perturbation amplitudes $\alpha^{\nu, W}$ in all frequency channels at once but treating the angular windows separately, with a uniform distribution in the range $\left[ {\rm max} \left( -W_{\rm max}^{-1},-\sqrt{3}\alpha_{\rm lim}^{\nu, W} \right); \sqrt{3}\alpha_{\rm lim}^{\nu, W} \right]$. We use a uniform distribution instead of a Gaussian distribution with standard deviation $\alpha_{\rm lim}^{\nu, W}$ to avoid negative beam amplitudes. Nevertheless, we maintain the same variance as the Gaussian distribution by introducing a factor of $\sqrt{3}$ to the width of the uniform distribution. The average bias due to the far side-lobe systematic effect is found to be $\Delta r = 2.08 \times 10^{-5}$, in excess by less than $\sim 10\%$ as compared to the pre-defined bias budget defined above which we consider as validating our assumption of small correlations between the channels. Note that these results are slightly different as those presented in \cite{LiteBIRD:2022cnt} where it was assumed that $\alpha \geq -1$, which tends to over-estimate the average bias compared to the more precise limit used here. The average residual power spectra are shown in Figure~\ref{fig:Residual Cls}.

From this procedure, we derive requirements on the perturbation amplitudes $\alpha^{\nu, W}$, given $\delta r_{\rm lim}^{\nu, W}$, as an upper bound $\alpha_{\rm lim}^{\nu, W}$. However, though the $\alpha_{\rm lim}$ parameters are useful intermediate to evaluate the effect of imperfect beam knowledge, they are not directly accessible and we want to relate them to physical properties of the beams. We describe three such properties in the following sections.

\subsubsection{Beam perturbation power}

First, we introduce the parameter $\dRlim$ that quantifies the relative difference of power between the perturbed and unperturbed beams:
\begin{equation}
    \dRlim^{\nu, W} \equiv \frac{\int B_{\rm lim}^{\nu, W} \left( \theta \right) W \left( \theta \right) d\Omega}{\int \overline{B_{0}^{\nu}} \left( \theta \right) d\Omega}. \label{eq:dRlim}
\end{equation}

In this expression, we defined the fractional power difference between the perturbed and nominal beams $B_{\rm lim}^{\nu, W} \left( \theta \right) = \alpha_{\rm lim}^{\nu, W} \overline{B_{0}^{\nu}} \left( \theta \right)$ with the averaged beam normalized at the peak, i.e. $\overline{B_{0}^{\nu}} \left( \theta=0 \right) = 1$ as will always be the case in the following. This quantity clearly represents the fraction of power in the perturbation expressed in units of the total power of the beam, and is illustrated in Figure~\ref{fig:Rlim Blim} for the LFT 100~GHz channel. The $\dRlim^{\nu, W}$ values found in the 22 frequency channels and three angular windows are given in Table \ref{tab:dRlim requir}.

\begin{table}[!htb]
\begin{center}
{
\setlength{\tabcolsep}{0.5pt}
\begin{tabular}{|c|c|c|c|c|c|c|c|c|c|c|c|}
\hline
\multicolumn{12}{|c|}{\rule{0pt}{2.5ex} $\dRlim \left( \mathrm{dBi} \right)$} \\ \hline\hline
\rule{0pt}{2.5ex} \multirow{2}*{\vbox{\setbox0\hbox{\strut (GHz)}\hbox to\wd0{\hss\strut $\nu$\hss}\copy0}} & \multicolumn{11}{c|}{LFT} \\ \cline{2-12}
\rule{0pt}{2.5ex} & 40 & 50 & 60 & \multicolumn{2}{c|}{68} & \multicolumn{2}{c|}{78} & \multicolumn{2}{c|}{89} & 100 & 119 \\ \hline
\rule{0pt}{2.5ex} $4^{\circ} <\theta< 8^{\circ}$ & \coldRlim{-23.54} & \coldRlim{-13.45} & \coldRlim{-17.68} & \coldRlim{-13.27} & \coldRlim{-8.02} & \coldRlim{-18.36} & \coldRlim{-16.43} & \coldRlim{-23.14} & \coldRlim{-15.50} & \coldRlim{-25.61} & \coldRlim{-27.57} \\ \hline
\rule{0pt}{2.5ex} $7^{\circ} <\theta< 12^{\circ}$ & \coldRlim{-25.41} & \coldRlim{-15.51} & \coldRlim{-19.68} & \coldRlim{-15.75} & \coldRlim{-10.99} & \coldRlim{-19.21} & \coldRlim{-17.28} & \coldRlim{-24.13} & \coldRlim{-16.49} & \coldRlim{-26.59} & \coldRlim{-28.46} \\ \hline
\rule{0pt}{2.5ex} $11^{\circ} <\theta$ & \coldRlim{-27.49} & \coldRlim{-17.44} & \coldRlim{-21.53} & \coldRlim{-18.20} & \coldRlim{-13.67} & \coldRlim{-17.41} & \coldRlim{-16.18} & \coldRlim{-23.40} & \coldRlim{-16.27} & \coldRlim{-26.01} & \coldRlim{-27.86} \\ \hline\hline
\rule{0pt}{2.5ex} \multirow{2}*{\vbox{\setbox0\hbox{\strut (GHz)}\hbox to\wd0{\hss\strut $\nu$\hss}\copy0}} & LFT & \multicolumn{5}{c|}{MFT} & \multicolumn{5}{c|}{HFT} \\ \cline{2-12}
\rule{0pt}{2.5ex} & 140 & 100 & 119 & 140 & 166 & 195 & 195 & 235 & 280 & 337 & 402 \\ \hline
\rule{0pt}{2.5ex} $4^{\circ} <\theta< 8^{\circ}$ & \coldRlim{-23.23} & \coldRlim{-26.84} & \coldRlim{-29.71} & \coldRlim{-23.92} & \coldRlim{-34.64} & \coldRlim{-37.07} & \coldRlim{-33.00} & \coldRlim{-35.65} & \coldRlim{-32.52} & \coldRlim{-41.78} & \coldRlim{-38.03} \\ \hline
\rule{0pt}{2.5ex} $7^{\circ} <\theta< 12^{\circ}$ & \coldRlim{-23.88} & \coldRlim{-27.90} & \coldRlim{-30.71} & \coldRlim{-24.61} & \coldRlim{-35.63} & \coldRlim{-37.89} & \coldRlim{-34.17} & \coldRlim{-36.60} & \coldRlim{-33.34} & \coldRlim{-42.66} & \coldRlim{-38.68} \\ \hline
\rule{0pt}{2.5ex} $11^{\circ} <\theta$ & \coldRlim{-26.69} & \coldRlim{-25.74} & \coldRlim{-29.20} & \coldRlim{-29.83} & \coldRlim{-33.91} & \coldRlim{-34.30} & \coldRlim{-32.73} & \coldRlim{-34.38} & \coldRlim{-27.26} & \coldRlim{-37.79} & \coldRlim{-38.93} \\ \hline
\end{tabular}
}
\caption{Beam perturbation requirements for each frequency channel and each of the three angular windows of the beam perturbations giving $\delta r = 1.9\times 10^{-5}/66$, using unperturbed beams in the other frequency channels and angular windows. These results are expressed in terms of $\dRlim$, in dBi. The color of the cells correspond to a linear scale from green for the easiest requirements (on $\dRlim$) to red for the most challenging.}
\label{tab:dRlim requir}
\end{center}
\end{table}

We can see a remarkable fact from these requirements: they show little dependence on the angular range. In a given frequency channel, the difference between requirements in the three windows is at most of the order of a few dBis, much less than the variation between frequency channels, despite the large difference between the definitions of the windows. This means that the requirements on $\dRlim$ are quite independent of the beam shape and of the shape of the perturbation. So, the difference of power between the perturbed and unperturbed beams is the leading effect to the bias on $r$. And the relative differences in sensitivity between frequency channels come from their weight in the component separation, i.e. from the relative amplitude of Galactic foreground emissions in these channels. In particular, the requirements are the most stringent for the lowest frequency channel, important to clean for synchrotron, around the CMB frequencies and for the high frequency channels, important to clean for dust. Therefore, we believe that improving the beam modeling would not change significantly the requirements on $\dRlim$.
%So, they are a good starting point to quickly test the impact of design choices affecting beams on the feasability of required accuracy for observable quantities.

However, $\dRlim$ is not directly measurable because the total beam power, i.e. the integral of the beam appearing in the denominator of Eq.~\eqref{eq:dRlim}, is not precisely known. So, despite the apparent robustness of this physical quantity, we have to express requirements in terms of other related quantities.

\subsubsection{Average perturbation amplitude in the window}

Following the results of the previous subsection, we define $\dBlim$ which is closer to what can actually be used as a benchmark for calibration measurements (see also Eq.~\eqref{eq:sigma calib}):
\begin{equation}
    \dBlim^{\nu, W} \equiv \frac{\int B_{\rm lim}^{\nu, W} \left( \theta \right) W \left( \theta \right) d\Omega}{\int W \left( \theta \right) d\Omega} = \frac{\int \overline{B_{0}^{\nu}} \left( \theta \right) d\Omega}{\int W \left( \theta \right) d\Omega} \dRlim^{\nu, W}. \label{eq:dBlim}
\end{equation}

\begin{figure}[!htb]
\begin{center}
\includegraphics[width=\textwidth]{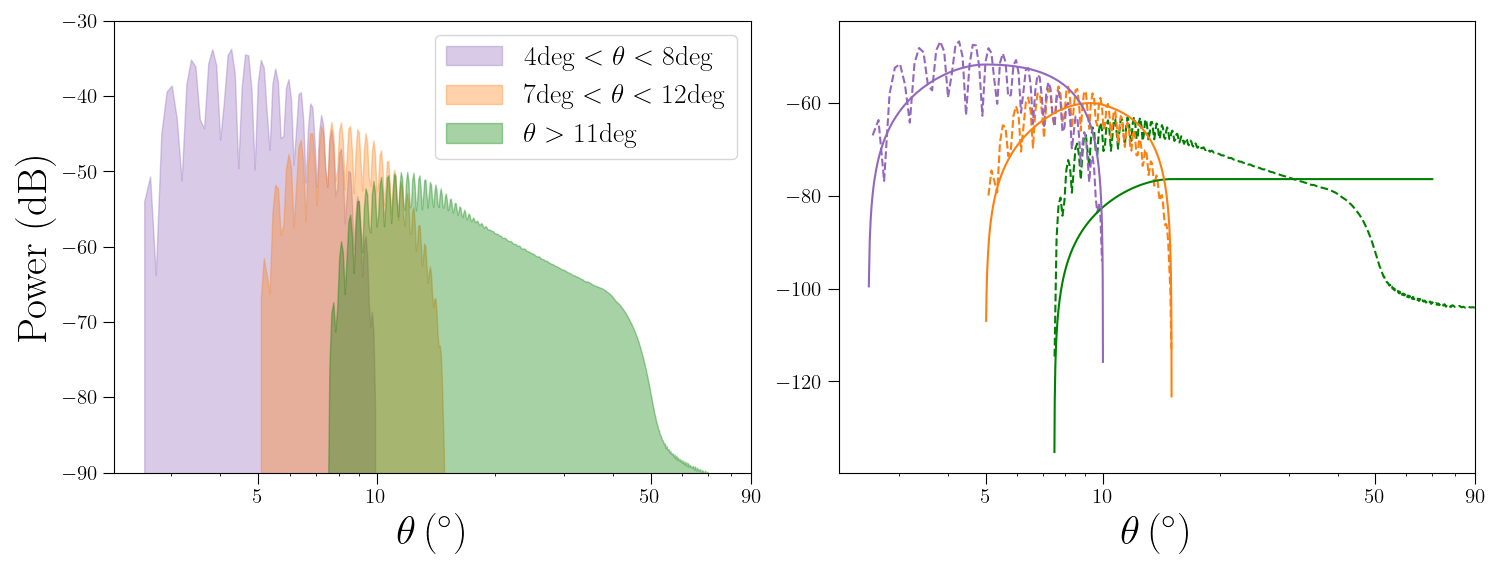}
\caption{(Left) Illustration of $\dRlim$ as the shaded area in the three angular windows in \textit{LiteBIRD}'s 100~GHz LFT channel, for a perturbation amplitude $\alpha=1$. (Right) Illustration of $\dBlim$ in the same channel for $\alpha=1$. The dashed lines correspond to the actual beam perturbations, while the solid lines represent what would be the perturbations if $\overline{B_{0}^{\nu}} \left( \theta \right)$ was constant in the window, equal to $\dBlim$.}
\label{fig:Rlim Blim}
\end{center}
\end{figure}

This quantity corresponds to the average amplitude, normalized to the peak, of the perturbation in the angular window and is illustrated in Figure~\ref{fig:Rlim Blim}. It is clear on one hand that $\dBlim$ is easier to measure than $\dRlim$ but on the other hand that the requirements will depend on the definition of the windows. In particular, the larger the window the more stringent the requirements on $\dBlim$. This is particularly relevant for the last angular window, which in principle ranges from $\sim 11^{\circ}$ up to $\sim 180^{\circ}$, combining regions with very different power levels. This would lead to unreasonably low requirements for $\dBlim$ in the last angular window. So, we have to make a choice of upper bound up $\theta_{\rm max}$ to which we should compute the integrals in the definition Eq.~\eqref{eq:dBlim}, based on beam dependent considerations. This makes the requirements on $\dBlim$ dependent on the beam modeling.

Because the amplitude of the beam models used for this analysis drops drastically at angles larger than $\sim 50^{\circ}$ (and are expected to decrease significantly at large angle thanks to the design of \textit{LiteBIRD}), this upper bound should not be too much higher than $\sim 50^{\circ}$. To stay conservative, we show the requirements on $\dBlim$ in Table~\ref{tab:dBlim requir} where we restrict the last window to angles smaller than $70^{\circ}$.

\begin{table}[!htb]
\begin{center}
{
\setlength{\tabcolsep}{0.5pt}
\begin{tabular}{|c|c|c|c|c|c|c|c|c|c|c|c|}
\hline
\multicolumn{12}{|c|}{\rule{0pt}{2.5ex} $\dBlim \left( \mathrm{dB} \right)$} \\ \hline\hline
\rule{0pt}{2.5ex} \multirow{2}*{\vbox{\setbox0\hbox{\strut (GHz)}\hbox to\wd0{\hss\strut $\nu$\hss}\copy0}} & \multicolumn{11}{c|}{LFT} \\ \cline{2-12}
\rule{0pt}{2.5ex} & 40 & 50 & 60 & \multicolumn{2}{c|}{68} & \multicolumn{2}{c|}{78} & \multicolumn{2}{c|}{89} & 100 & 119 \\ \hline
\rule{0pt}{2.5ex} $4^{\circ} <\theta< 8^{\circ}$ & \coldBlim{-42.55} & \coldBlim{-34.09} & \coldBlim{-39.46} & \coldBlim{-36.70} & \coldBlim{-30.45} & \coldBlim{-42.82} & \coldBlim{-39.42} & \coldBlim{-48.56} & \coldBlim{-38.84} & \coldBlim{-51.80} & \coldBlim{-54.82} \\ \hline
\rule{0pt}{2.5ex} $7^{\circ} <\theta< 12^{\circ}$ & \coldBlim{-46.62} & \coldBlim{-38.35} & \coldBlim{-43.67} & \coldBlim{-41.38} & \coldBlim{-35.62} & \coldBlim{-45.87} & \coldBlim{-42.48} & \coldBlim{-51.75} & \coldBlim{-42.04} & \coldBlim{-54.98} & \coldBlim{-57.91} \\ \hline
\rule{0pt}{2.5ex} $11^{\circ} <\theta< 70^{\circ}$ & \coldBlim{-66.40} & \coldBlim{-57.98} & \coldBlim{-63.20} & \coldBlim{-61.52} & \coldBlim{-56.00} & \coldBlim{-61.76} & \coldBlim{-59.07} & \coldBlim{-68.70} & \coldBlim{-59.51} & \coldBlim{-72.09} & \coldBlim{-75.00} \\ \hline\hline
\rule{0pt}{2.5ex} \multirow{2}*{\vbox{\setbox0\hbox{\strut (GHz)}\hbox to\wd0{\hss\strut $\nu$\hss}\copy0}} & LFT & \multicolumn{5}{c|}{MFT} & \multicolumn{5}{c|}{HFT} \\ \cline{2-12}
\rule{0pt}{2.5ex} & 140 & 100 & 119 & 140 & 166 & 195 & 195 & 235 & 280 & 337 & 402 \\ \hline
\rule{0pt}{2.5ex} $4^{\circ} <\theta< 8^{\circ}$ & \coldBlim{-51.25} & \coldBlim{-50.65} & \coldBlim{-54.58} & \coldBlim{-49.55} & \coldBlim{-60.87} & \coldBlim{-63.56} & \coldBlim{-59.01} & \coldBlim{-62.92} & \coldBlim{-60.57} & \coldBlim{-70.48} & \coldBlim{-67.91} \\ \hline
\rule{0pt}{2.5ex} $7^{\circ} <\theta< 12^{\circ}$ & \coldBlim{-54.11} & \coldBlim{-53.91} & \coldBlim{-57.78} & \coldBlim{-52.45} & \coldBlim{-64.07} & \coldBlim{-66.58} & \coldBlim{-62.38} & \coldBlim{-66.07} & \coldBlim{-63.60} & \coldBlim{-73.57} & \coldBlim{-70.77} \\ \hline
\rule{0pt}{2.5ex} $11^{\circ} <\theta< 70^{\circ}$ & \coldBlim{-74.60} & \coldBlim{-69.44} & \coldBlim{-73.96} & \coldBlim{-75.36} & \coldBlim{-80.06} & \coldBlim{-80.69} & \coldBlim{-78.63} & \coldBlim{-81.55} & \coldBlim{-75.23} & \coldBlim{-86.41} & \coldBlim{-88.72} \\ \hline
\end{tabular}
}
\caption{Beam perturbation requirements for each frequency channel and each of the three angular windows of the beam perturbations giving $\delta r = 1.9\times 10^{-5}/66$, using unperturbed beams in the other frequency channels and angular windows. These results are expressed in terms of $\dBlim$, in dB, assuming the last angular window to range up to $70^{\circ}$. The color of the cells correspond to a linear scale from green for the easiest requirements (on $\dBlim$) to red for the most challenging.}
\label{tab:dBlim requir}
\end{center}
\end{table}

As expected, the most challenging window and frequency channels are the one at large angle covering the largest area and at high frequencies. This comes from two effects: the importance of high frequency channels to estimate with a good lever arm the dust signal in channels at CMB frequencies, and the small FWHM of high frequency channels leading to very low average amplitude at large angles. In this context, it is clear that assigning the same bias budget to all channels is not optimal in terms of requirements, and we can alleviate a little the requirements in the most sensitive channels by making those in the other channels more stringent.

\begin{figure}[!htb]
\begin{center}
\includegraphics[width=0.49\textwidth]{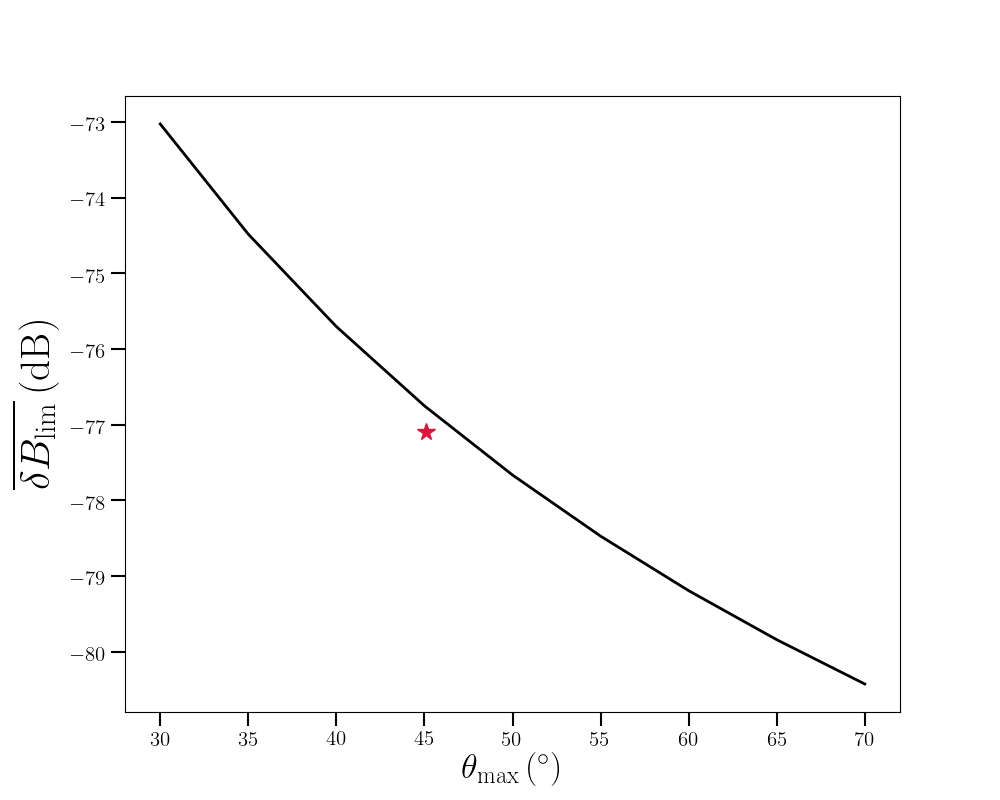}
\includegraphics[width=0.49\textwidth]{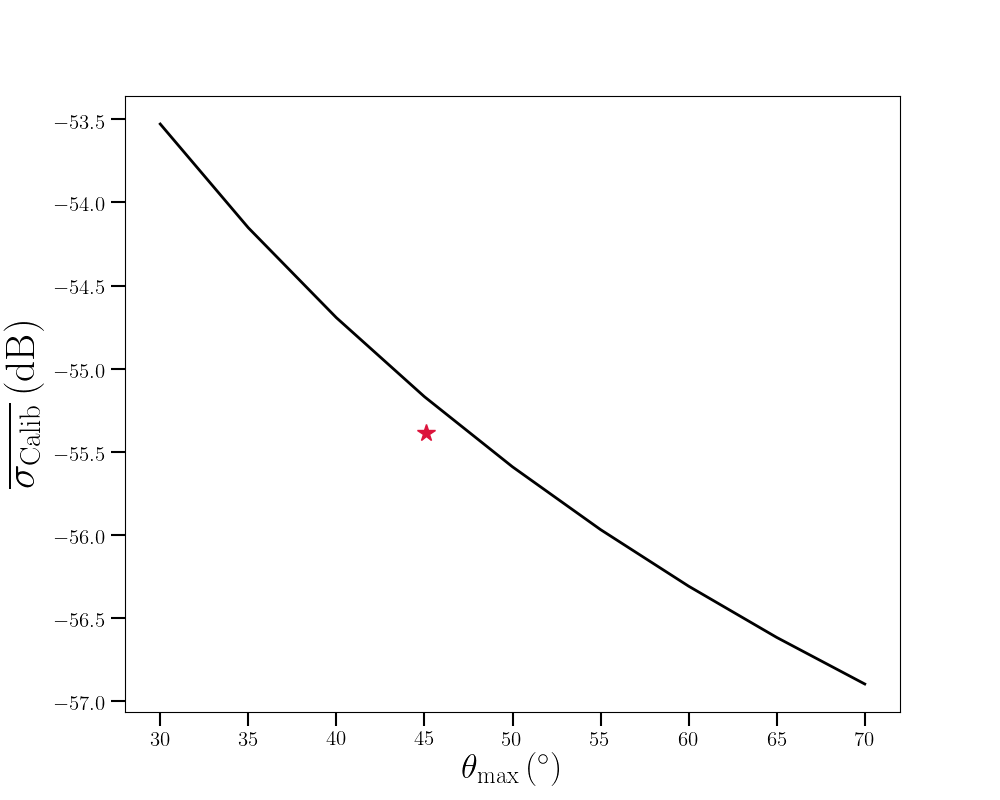}
\caption{(Left) Evolution of the common requirement $\overline{\dBlim}$ defined in Eq.~\eqref{eq:common dBlim} as a function of the upper bound angle $\theta_{max}$ after which the last angular window is $W \left( \theta > \theta_{\rm max} \right) = 0$. In the definition of $\dBlim$ Eq.~\eqref{eq:dBlim}, $\dRlim$ is kept constant as it is rather stable with respect to the definition of the windows. The red star corresponds to the value of $\overline{\dBlim}$ when $\theta_{\rm max}^{\nu}$ is obtained in each channel such that $\int^{\theta_{\rm max}^{\nu}} \overline{B_{0}^{\nu}} \left( \theta \right) W \left( \theta \right) d\Omega$ include 97.5\% of the theoretical perturbation power, at the mean of the $\theta_{\rm max}^{\nu}$. (Right) Evolution of the common requirement $\overline{\sigmaCal}$ defined in Eq.~\eqref{eq:sigma calib} as a function of $\theta_{\rm max}$ under the same assumptions.}
\label{fig:dBlim - sigCal vs theta_max}
\end{center}
\end{figure}

Having this in mind, we derived the requirements on $\dBlim$ assuming the whole frequency and angular range to share the same calibration accuracy. In this case, the requirement will be expressed as a single common $\overline{\dBlim}$, that we obtain by tuning the single channel bias budget $\delta r_{\rm lim}^{\nu, W}$ such that $\dBlim^{\nu, W} = \overline{\dBlim}$. It is possible to compute analytically an estimation of $\overline{\dBlim}$ using the previously found values of $\alpha_{lim}^{\nu, W}$ parameters with the scaling $\delta r^{\nu, W} \propto \left( \alpha^{\nu, W} \right)^{2}$. This leads to the following expression:
\begin{equation}
    \overline{\dBlim} = \sqrt{\frac{n_{\nu} \times n_{\theta}}{\sum\limits_{\nu, W} \left( \dBlim^{\nu, W} \right)^{-2}}}, \label{eq:common dBlim}
\end{equation}
where $n_{\theta}$ is the number of angular windows, $\dBlim^{\nu, \theta}$ are the requirements previously found. In the case where $\dBlim$ is defined up to $70^{\circ}$, these are taken from Table \ref{tab:dBlim requir}, and we find the common precision requirement to be $\overline{\dBlim} = -80.42 \mathrm{\ dB}$. In other words, under our assumptions the beam amplitude must be known in each frequency channel and each window with a precision of $\sim 10^{-8}$ with respect to the peak, including systematic and statistical sources of errors during calibration. Once again, this result depends on the definition of the windows, and in particular on the maximum angle $\theta_{\rm max}$ of the last window. If $\theta_{\rm max}$ is different, to reflect more accurately our knowledge of the optical characteristics of the telescope, or to investigate the potential impact of other design choices, this would have a significant impact on the value of $\overline{\dBlim}$. The left part of Figure~\ref{fig:dBlim - sigCal vs theta_max} shows how $\overline{\dBlim}$ varies as a function of $\theta_{\rm max}$, keeping $\dRlim$ constant as we saw that it should not depend on the window.

To put these requirements into perspective, one can compare with the level of accuracies that were reached for Planck during the ground calibration campaign. The difference between modeling and measurements was of the order of $6 \mathrm{\ dB}$ for iso-levels of $-90 \mathrm{\ dB}$ at $100 \mathrm{\ GHz}$, and $7 \mathrm{\ dB}$ at $-80 \mathrm{\ dB}$ at $320 \mathrm{\ GHz}$ \cite{Dubruel}.

\subsubsection{Noise limited calibration measurements}

If calibration measurements of the beams face systematic sources of errors, the requirements that apply to the accuracy of calibration measurements are the ones set for $\dBlim$ in the previous subsection. However, in the case of noise limited measurements, we do not need to go to such accuracy, because we can make several independent measurements of the same perturbation in many pixels on the sphere, distributed in the 2D angular ring of the perturbation. To make this more precise, the quantity effectively measured during ground beam calibration is the integrated power measured in a certain region of the sphere, normalized to the integrated power measured at the beam center, which can be modeled as:
\begin{equation}
    P_{\rm Calib}^{\nu} \left( \hat{r} \right) \equiv \int \overline{B_{0}^{\nu}} \left( \theta \right) \omega \left( \hat{r}' - \hat{r} \right) {\rm d}\Omega' \frac{1}{\int \overline{B_{0}^{\nu}} \left( \theta \right) \omega \left( \hat{r}' \right) {\rm d}\Omega'} + n_{\rm Calib},
\end{equation}
where $\omega(\hat{r})$ is a small integration window of the beam that can be interpreted as a pixel in the $\left( \theta, \phi \right)$ surface and $n_{\rm Calib}$ the noise in the beam calibration measurements.

\begin{figure}[!htb]
\begin{center}
\includegraphics[width=0.5\textwidth]{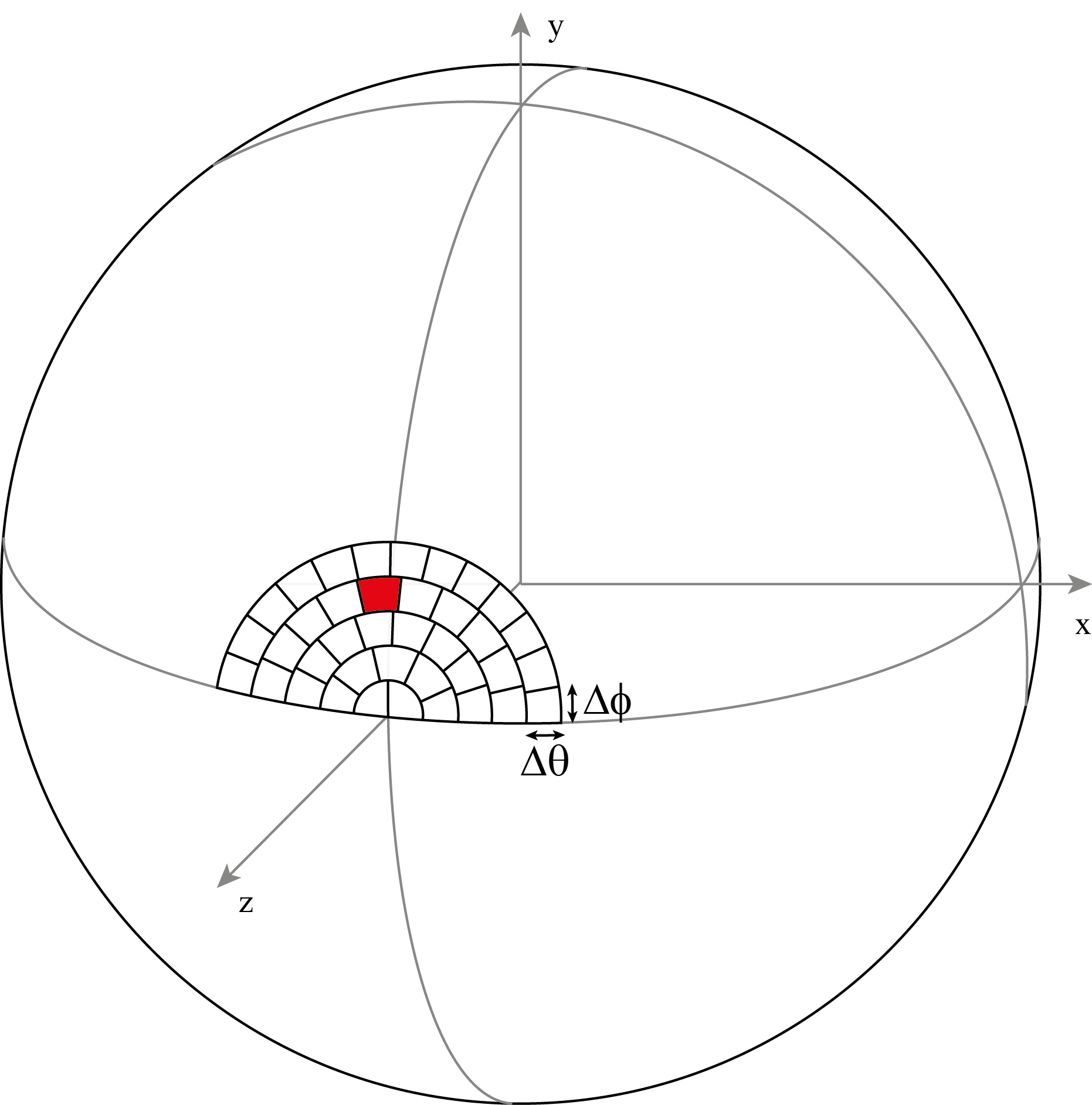}
\caption{Illustration of the grid of beam measurement pixels on the sphere. Each pixel, as the highlighted red one, is supposed to have the same area and are distributed in rings of constant $\Delta \theta$.}
\label{fig:Beam pixels}
\end{center}
\end{figure}

We estimate the precision required on the measured quantity $P_{\rm Calib}(\hat{r})$ assuming random uncorrelated errors in each measurement (and hence no systematic effects), and a grid of measurements at many angles as illustrated in Figure~\ref{fig:Beam pixels}. In this case the calibration measurement uncertainty $\sigmaCal = \sqrt{\langle n_{\rm Calib} n^{T}_{\rm Calib}\rangle}$ is related to the quantity $\dBlim$ by:
\begin{equation}
      \sigmaCal^{\nu, W} = \frac{\int W(\theta) {\rm d}\Omega}{\sqrt{\sum_{ij}W^2(\theta_{ij})} \Delta\Omega_{\rm pix}} \dBlim^{\nu, W} \equiv \sqrt{N_{eff}^{W}} \dBlim^{\nu, W}, \label{eq:sigma calib}
\end{equation}
with $\Delta\Omega_{\rm pix}$ the solid angle covered by one calibration measurement and $i,j$ the pixel number for a pixelized beam calibration map. The quantity $\sigma_{\rm Calib}$ can directly be interpreted as the accuracy of the beam measurements.
%during calibration.
In all of the following, we assume arbitrarily a constant pixel size $\Delta \Omega_{\rm pix} = 0.5^{\circ} \times 0.5^{\circ} = 0.25 \mathrm{\ deg}^{2}$, during beam ground calibration, but different values of $\Omega_{\rm pix}$ would lead to different requirements on $\sigmaCal$. As already mentioned, we assume the calibration beam measurements to be normalized to one at the peak. The factor $N_{eff}^{W}$ is the effective number of pixels in the region of the beam perturbation. As in the case of $\dBlim$, the requirements will depend on the definition of the windows. However, the dependency on the windows is not as straightforward because, as $\theta_{\rm max}$ in the last window is reduced, for instance, the effective number of pixel will decrease while $\dBlim$ will increase. The requirements for $\sigmaCal$ assuming $\theta_{\rm max} = 70^{\circ}$ are given in Table \ref{tab:sigmaCal requir}, where the effective numbers of pixels are $N_{eff} = 742.0, 1541.9, \mathrm{\ and\ } 52943.2$ in the three windows from small to large angles respectively.

In the same way as we derive a common requirement $\overline{\dBlim}$ using Eq.~\ref{eq:common dBlim}, we can also derive a single common $\overline{\sigmaCal}$. Under our general assumptions and assuming $\theta_{\rm max} = 70^{\circ}$, we find $\sigmaCal = -56.90 \mathrm{\ dB}$, and the $\theta_{\rm max}$ dependence of $\sigmaCal$ is shown in the right panel of Figure~\ref{fig:dBlim - sigCal vs theta_max}.

\begin{table}[!htb]
\begin{center}
{
\setlength{\tabcolsep}{0.5pt}
\begin{tabular}{|c|c|c|c|c|c|c|c|c|c|c|c|}
\hline
\multicolumn{12}{|c|}{\rule{0pt}{2.5ex} $\sigmaCal \left( \mathrm{dB} \right)$} \\ \hline\hline
\rule{0pt}{2.5ex} \multirow{2}*{\vbox{\setbox0\hbox{\strut (GHz)}\hbox to\wd0{\hss\strut $\nu$\hss}\copy0}} & \multicolumn{11}{c|}{LFT} \\ \cline{2-12}
\rule{0pt}{2.5ex} & 40 & 50 & 60 & \multicolumn{2}{c|}{68} & \multicolumn{2}{c|}{78} & \multicolumn{2}{c|}{89} & 100 & 119 \\ \hline
\rule{0pt}{2.5ex} $4^{\circ} <\theta< 8^{\circ}$ & \colsigCal{-28.20} & \colsigCal{-19.73} & \colsigCal{-25.11} & \colsigCal{-22.34} & \colsigCal{-16.10} & \colsigCal{-28.46} & \colsigCal{-25.07} & \colsigCal{-34.20} & \colsigCal{-24.49} & \colsigCal{-37.45} & \colsigCal{-40.47} \\ \hline
\rule{0pt}{2.5ex} $7^{\circ} <\theta< 12^{\circ}$ & \colsigCal{-30.68} & \colsigCal{-22.41} & \colsigCal{-27.73} & \colsigCal{-25.44} & \colsigCal{-19.68} & \colsigCal{-29.93} & \colsigCal{-26.53} & \colsigCal{-35.81} & \colsigCal{-26.10} & \colsigCal{-39.04} & \colsigCal{-41.97} \\ \hline
\rule{0pt}{2.5ex} $11^{\circ} <\theta< 70^{\circ}$ & \colsigCal{-42.79} & \colsigCal{-34.37} & \colsigCal{-39.60} & \colsigCal{-37.92} & \colsigCal{-32.39} & \colsigCal{-38.16} & \colsigCal{-35.46} & \colsigCal{-45.10} & \colsigCal{-35.91} & \colsigCal{-48.49} & \colsigCal{-51.39} \\ \hline\hline
% \end{tabular} \\
% \vspace{0.15cm}
% \begin{tabular}{|c|c|c|c|c|c|c|c|c|c|c|c|}
% \hline
\rule{0pt}{2.5ex} \multirow{2}*{\vbox{\setbox0\hbox{\strut (GHz)}\hbox to\wd0{\hss\strut $\nu$\hss}\copy0}} & LFT & \multicolumn{5}{c|}{MFT} & \multicolumn{5}{c|}{HFT} \\ \cline{2-12}
\rule{0pt}{2.5ex} & 140 & 100 & 119 & 140 & 166 & 195 & 195 & 235 & 280 & 337 & 402 \\ \hline
\rule{0pt}{2.5ex} $4^{\circ} <\theta< 8^{\circ}$ & \colsigCal{-36.90} & \colsigCal{-36.30} & \colsigCal{-40.23} & \colsigCal{-35.20} & \colsigCal{-46.52} & \colsigCal{-49.20} & \colsigCal{-44.65} & \colsigCal{-48.57} & \colsigCal{-46.22} & \colsigCal{-56.12} & \colsigCal{-53.56} \\ \hline
\rule{0pt}{2.5ex} $7^{\circ} <\theta< 12^{\circ}$ & \colsigCal{-38.17} & \colsigCal{-37.97} & \colsigCal{-41.84} & \colsigCal{-36.51} & \colsigCal{-48.13} & \colsigCal{-50.64} & \colsigCal{-46.44} & \colsigCal{-50.13} & \colsigCal{-47.66} & \colsigCal{-57.63} & \colsigCal{-54.83} \\ \hline
\rule{0pt}{2.5ex} $11^{\circ} <\theta< 70^{\circ}$ & \colsigCal{-51.00} & \colsigCal{-45.84} & \colsigCal{-50.36} & \colsigCal{-51.76} & \colsigCal{-56.45} & \colsigCal{-57.08} & \colsigCal{-55.03} & \colsigCal{-57.95} & \colsigCal{-51.63} & \colsigCal{-62.81} & \colsigCal{-65.12} \\ \hline
\end{tabular}
}
\caption{Beam perturbation requirements for each frequency channel and each of the three angular windows of the beam perturbations giving $\delta r = 1.9\times 10^{-5}/66$, using unperturbed beams in the other frequency channels and angular windows. These results are expressed in terms of $\sigmaCal$, in dB, assuming the last angular window to range up to $70^{\circ}$ and a constant pixel size $\Delta \Omega_{\rm pix} = 0.5^{\circ} \times 0.5^{\circ}$. The color of the cells correspond to a linear scale from green for the easiest requirements (on $\sigmaCal$) to red for the most challenging.}
\label{tab:sigmaCal requir}
\end{center}
\end{table}

\subsection{Requirements for the Modeling Case}
\label{subsection:Requirements for the Modeling Case}

\subsubsection{Comparison of the Detailed Method and the Axisymmetric Method}
\label{subsubsection:Comparison of the Detailed Method and the Axisymmetric Method}

The Axisymmetric Method, based on the axisymmetric beam approximation we detailed in Section~\ref{subsubsection:Modeling Case}, needs to be validated before being used for the purpose of investigating modeling uncertainties. This is done by comparing the results described in the previous section obtained using the Detailed Method with results obtained under the axisymmetric approximation. We use the pipeline of the Perturbation Case described in Section~\ref{subsubsection:Perturbation Case}, but here the input $4\pi$ maps and side-lobes maps are calculated with the averaged and symmetrized \texttt{GRASP} beams following the description of Section~\ref{subsubsection:Modeling Case}. The corresponding perturbed maps are then produced using Eq.~\ref{eq:map combination}. To perform a meaningful comparison, we perturb the beam in the same three angular windows and keep the same bias budget in each channel $\delta r_{\rm lim}^{\nu, W} = \left( 1.9 / 66 \right) \times 10^{-5}$. The results of the comparison between the two methods are given in Table \ref{tab:beamFSLcompat}, expressed as $\dBlim^{1} - \dBlim^{2}$ where $\dBlim^{i}$ is obtained using Method $i$. The values of $\dBlim^{2}$ in each frequency channel show a good agreement, with a difference with $\dBlim^{2}$ of a few dBs, up to a little more than $\sim 3 \mathrm{\ dB}$ in a few channels. We also note that the Axisymmetric Method seems to systematically underestimate the bias on $r$ which leads to a systematically higher $\dBlim$. Although the origin of such a factor of 2  of difference in some channels needs to be further understood, given that the results span almost six orders of magnitude (60~dB) throughout the different windows and frequency channels, this is still an impressive agreement. This implies that the asymmetries of the beam do not have a significant impact on our analysis, possibly thanks to the scanning strategy of \textit{LiteBIRD} which is efficient in symmetrizing the given \texttt{GRASP} beam. Note that the \textit{LiteBIRD} \texttt{GRASP} beams in this study do not have significant asymmetric features as explained in Section~\ref{subsubsection:Perturbation Case}. The case of more realistic beam asymmetries will be studied in a future work.

\begin{table}[!htb]
\begin{center}
\setlength{\tabcolsep}{5.5pt}
\begin{tabular}{|c|c|c|c|c|c|c|c|c|c|c|c|}
\hline
\multicolumn{12}{|c|}{\rule[-1.ex]{0pt}{3.5ex} $\dBlim^{1} \left( \mathrm{dB} \right) - \dBlim^{2} \left( \mathrm{dB} \right)$} \\ \hline\hline
\rule{0pt}{2.5ex} \multirow{2}*{\vbox{\setbox0\hbox{\strut (GHz)}\hbox to\wd0{\hss\strut $\nu$\hss}\copy0}} & \multicolumn{11}{c|}{LFT} \\ \cline{2-12}
\rule{0pt}{2.5ex} & 40 & 50 & 60 & \multicolumn{2}{c|}{68} & \multicolumn{2}{c|}{78} & \multicolumn{2}{c|}{89} & 100 & 119 \\ \hline
\rule{0pt}{2.5ex} $4^{\circ} <\theta< 8^{\circ}$ & -1.63 & -1.88 & -1.56 & -1.45 & -0.31 & -1.62 & -1.60 & -1.93 & -1.79 & -2.19 & -2.59 \\ \hline
\rule{0pt}{2.5ex} $7^{\circ} <\theta< 12^{\circ}$ & -1.91 & -1.48 & -1.63 & -1.59 & -0.96 & -2.24 & -2.18 & -2.46 & -2.29 & -2.63 & -2.99 \\ \hline
\rule{0pt}{2.5ex} $11^{\circ} <\theta< 70^{\circ}$ & -1.45 & -1.15 & -1.24 & -1.09 & -0.89 & -1.54 & -1.58 & -1.90 & -1.80 & -2.14 & -2.44 \\ \hline\hline
\rule{0pt}{2.5ex} \multirow{2}*{\vbox{\setbox0\hbox{\strut (GHz)}\hbox to\wd0{\hss\strut $\nu$\hss}\copy0}} & LFT & \multicolumn{5}{c|}{MFT} & \multicolumn{5}{c|}{HFT} \\ \cline{2-12}
\rule{0pt}{2.5ex} & 140 & 100 & 119 & 140 & 166 & 195 & 195 & 235 & 280 & 337 & 402 \\ \hline
\rule{0pt}{2.5ex} $4^{\circ} <\theta< 8^{\circ}$ & -3.91 & -1.76 & -2.23 & -3.63 & -1.64 & -1.92 & -1.19 & -1.67 & -2.66 & -2.82 & -3.11 \\ \hline
\rule{0pt}{2.5ex} $7^{\circ} <\theta< 12^{\circ}$ & -3.68 & -2.26 & -2.73 & -3.13 & -1.88 & -1.99 & -1.86 & -2.06 & -3.03 & -3.08 & -3.28 \\ \hline
\rule{0pt}{2.5ex} $11^{\circ} <\theta< 70^{\circ}$ & -2.33 & -1.94 & -2.22 & -2.18 & -2.08 & -2.08 & -1.82 & -1.76 & -2.46 & -2.97 & -3.46 \\ \hline
\end{tabular}
\caption{Difference in dB between $\dBlim$ obtained from measurement uncertainties approach (Section~\ref{subsubsection:Perturbation Case}) and the corresponding $\dBlim^{2}$ obtained using the axisymmetric approach (Section~\ref{subsubsection:Modeling Case}) for each frequency channel and each of the three angle ranges of the beam perturbations giving $\delta r_{\rm lim}^{\nu, W} = 1.9\times 10^{-5}/66$.}
\label{tab:beamFSLcompat}
\end{center}
\end{table}

In consideration of flexibility and less time consumption, we apply the new approach in the following study of modeling  uncertainties.

\subsubsection{Bias on the tensor-to-scalar ratio from beam mismodeling at large angle}
\label{subsubsection:Bias on the tensor-to-scalar ratio from beam mismodeling at large angle}

\begin{figure}[!htb]
\begin{center}
\begin{minipage}{0.49\textwidth}
\includegraphics[width=\textwidth]{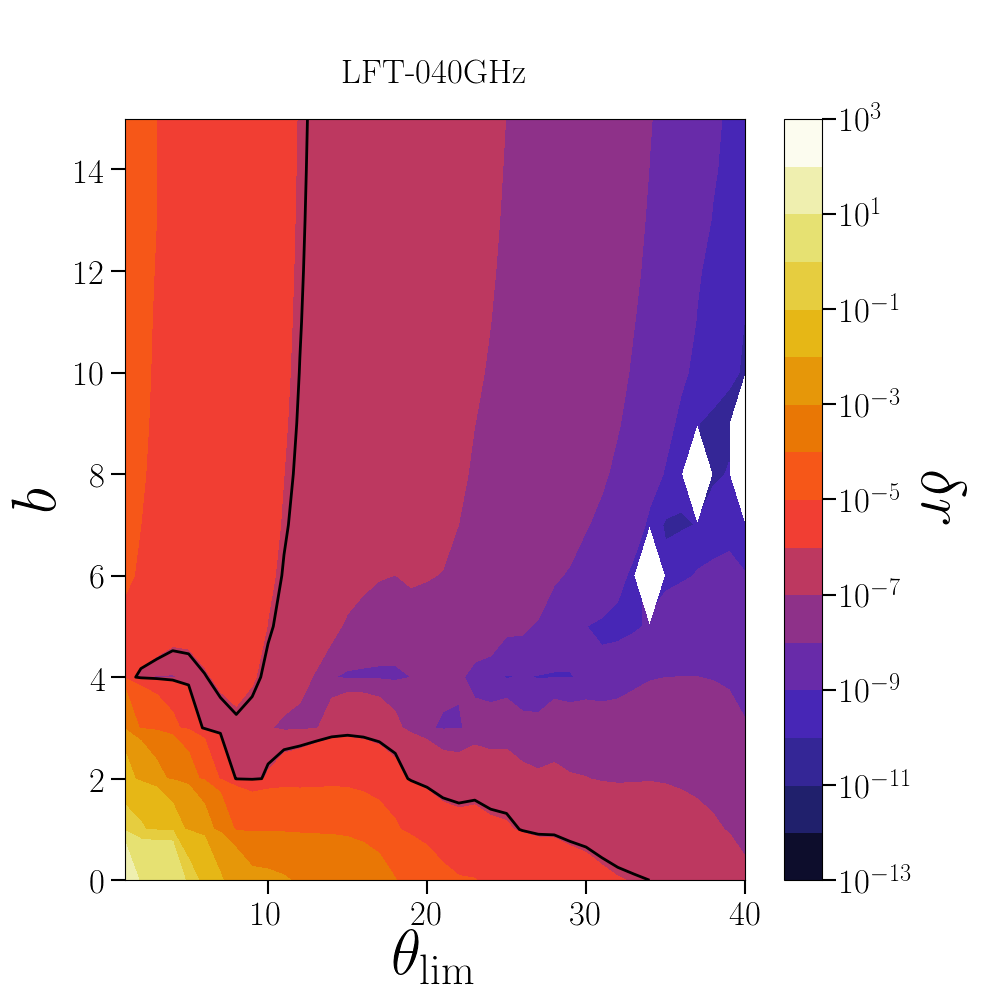}
\end{minipage}
\begin{minipage}{0.49\textwidth}
\includegraphics[width=\textwidth]{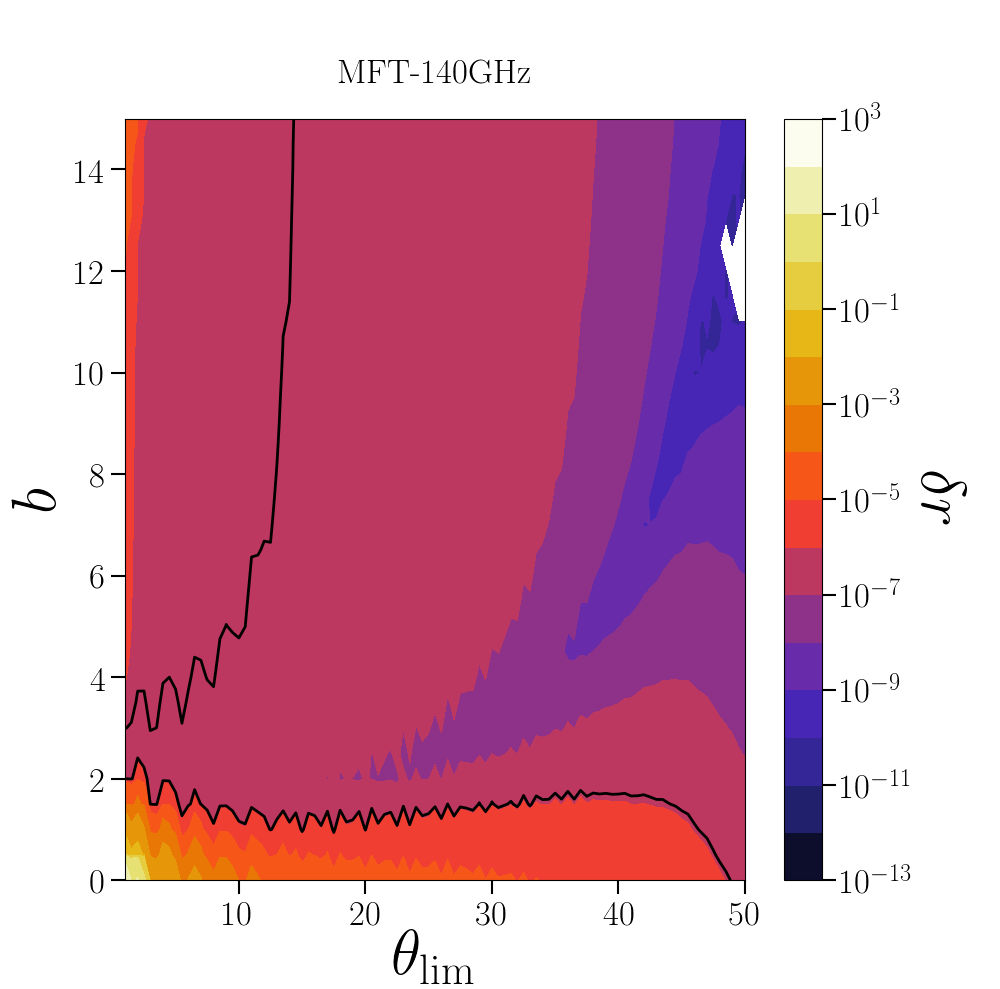}
\end{minipage}
\begin{minipage}{0.49\textwidth}
\includegraphics[width=\textwidth]{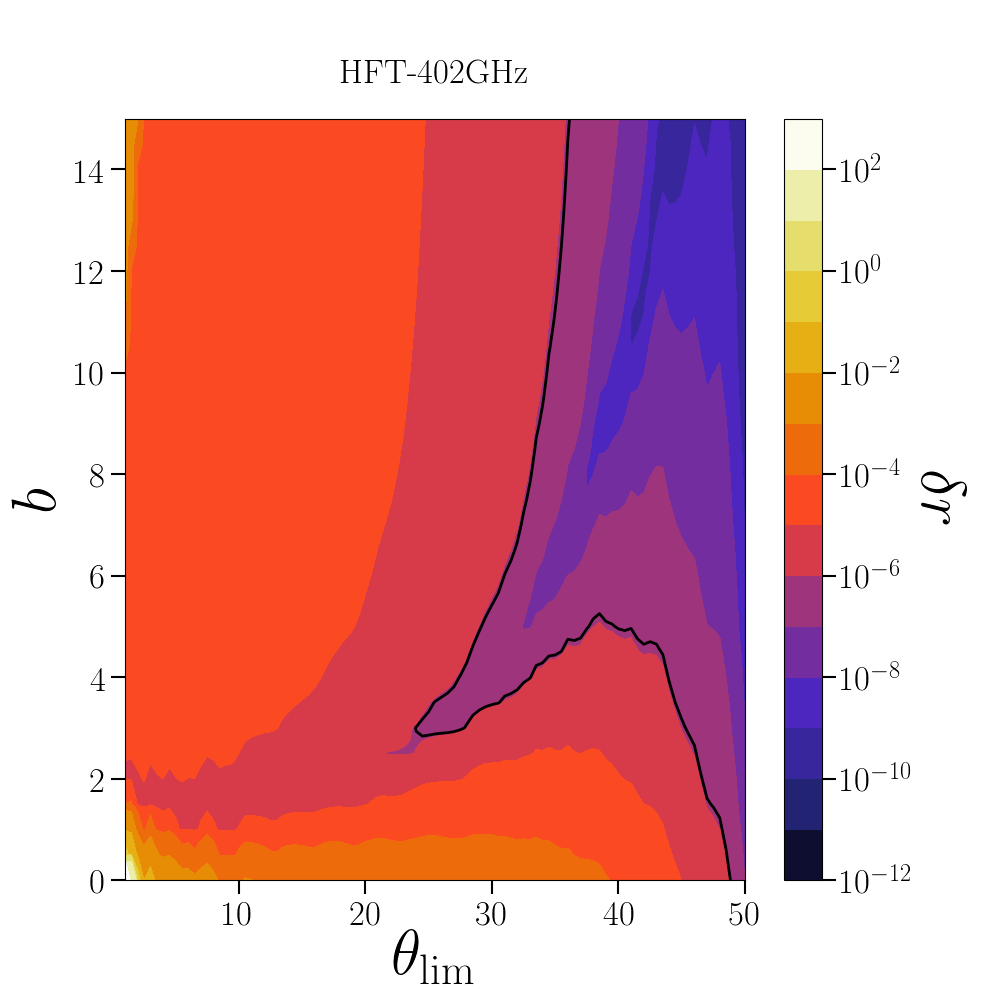}
\end{minipage}
\caption{Bias on $r$ in the 2D parameter space $\left( \theta_{\rm lim}, b \right)$ for three frequency channels, 40~GHz of the LFT (top left), 140~GHz of the MFT (top right) and 402~GHz of the HFT (bottom). The black contours correspond to the limit case where the bias is equal to the systematic error budget allocated to the individual frequency channels, $\delta r = \delta r_{\rm lim}^{\nu} = 1.9 \times 10^{-5}/22$.}
\label{fig: b-theta_lim}
\end{center}
\end{figure}

The main goal of this section is to find a way of figuring out the limit angle $\theta_{\rm lim}$ after which we can rely entirely on modeling to correct for beam effects. This angle will depend strongly on the ability of the beam model to reproduce key features of the true beam. Therefore, we have to explore the impact of both parameters of the beam model Eq.~\eqref{eq:beam model}, $\theta_{\rm lim}$ and $b$, at the same time. Following the same idea as in section \ref{subsection:Requirements for the Perturbation Case}, Figure~\ref{fig: b-theta_lim} shows the bias on $r$ on a grid of $\left( \theta_{\rm lim}, b \right)$ values when perturbing a single channel at a time keeping the other channels unperturbed, in three selected frequency channels: at the lowest frequency 40~GHz of the LFT, in one of the CMB frequency channels at 140GHz of the MFT and at the highest frequency 402~GHz of the HFT which we know from the previous section give the tightest constraints. In these figures, the region on the left of the black contour corresponds to sets of parameters for which $\delta r^{\nu} > \delta r_{\rm lim}^{\nu}$, the allocated budget in the given frequency channel, and the region on the right is where $\delta r^{\nu} < \delta r_{\rm lim}^{\nu}$. Following the spirit of the previous section when dealing with calibration uncertainties, we allocate the same budget to every frequency channel, namely $\delta r_{\rm lim}^{\nu} = \Delta r_{FSL} /n_{\nu} = 1.9 \times 10^{-5}/22$.

As we can see in each of these figures, neither of the flat beam model ($b=0$) and the cut beam model ($b \rightarrow \infty$), allow to reach $\theta_{\rm lim}$ close to the lowest possible values. In each frequency channel, there is a value of $b$ for which the limit bias is achieved for a lower value of $\theta_{\rm lim}$. This corresponds to the value of $b$ for which the power law beam model best fits the reference beam, leading to an induced bias compatible with the budget for lower values of $\theta_{\rm lim}$. However, the power law model is very simple and can only reproduce the most basic of properties from the reference beam. In other words, at least after some angle, the measurement of $r$ is not sensitive to the particular shape of the beam but to other features of the beam that are necessary global in nature such that they can be reproduced by a simple power law model.

\begin{figure}[!htb]
\begin{center}
\includegraphics[width=0.8\textwidth]{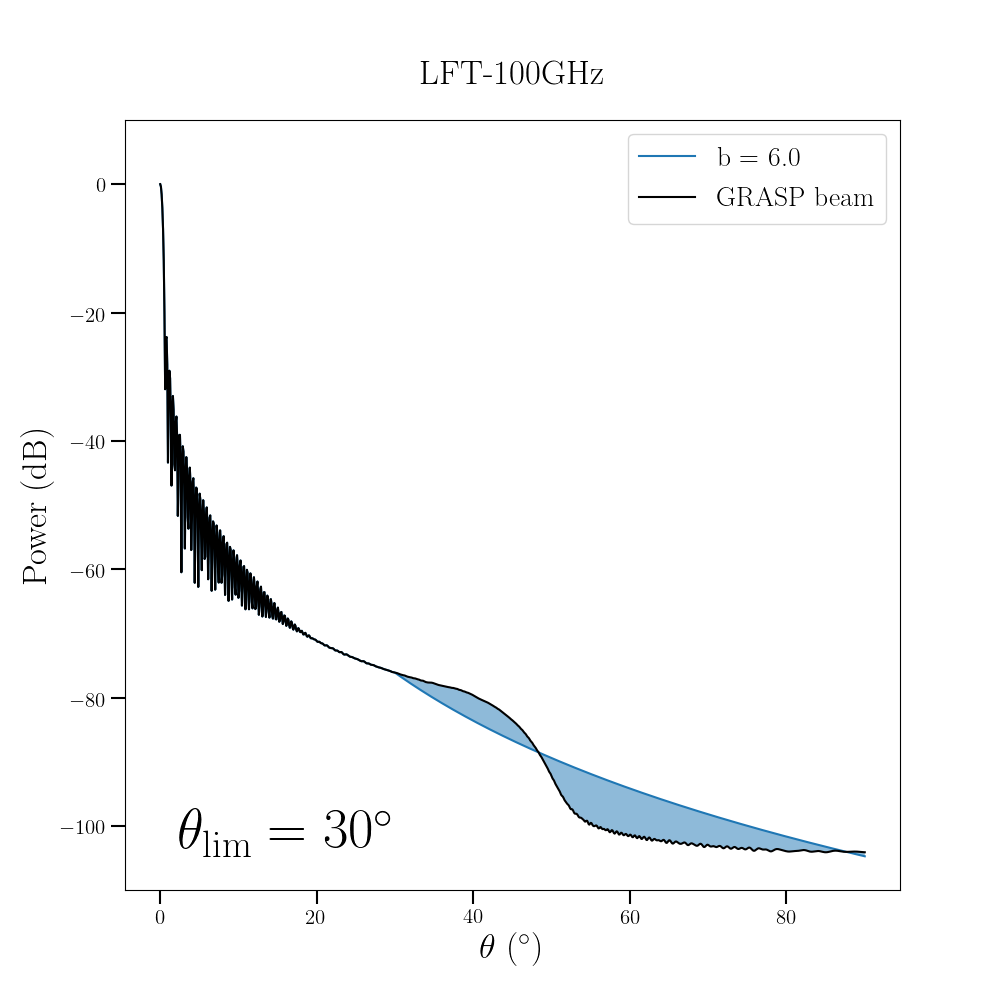}
\caption{Illustration of the residual power in the $\theta$ direction between the true beam and the beam model $\dKlim \left( \theta_{\rm lim}=30^{\circ}, b=6 \right)$ in the LFT 100~GHz frequency channel as the blue shaded area. The total residual power has to be integrated over $\phi$, which in the current context amounts only to a factor of $2\pi$.}
\label{fig:diff int}
\end{center}
\end{figure}

 Capitalizing on our results of the previous section, we investigate the link between $\delta r^{\nu}$ and the residual beam power between the true beam and the model:
\begin{equation}
    \dKlim^{\nu} \left( \theta_{\rm lim}, b \right) = \left| \int_{0}^{2\pi} \int_{\theta_{\rm lim}}^{180^{\circ}} \left[ B_{\rm model}^{\nu} \left( \theta ; \theta_{\rm lim}, b \right) - \overline{B_{0}^{\nu}} \left( \theta \right) \right] \mathrm{sin}\theta\, d\theta d\phi \right|.
\end{equation}

\begin{figure}[!htb]
\begin{center}
\begin{minipage}{0.49\textwidth}
\includegraphics[width=\textwidth]{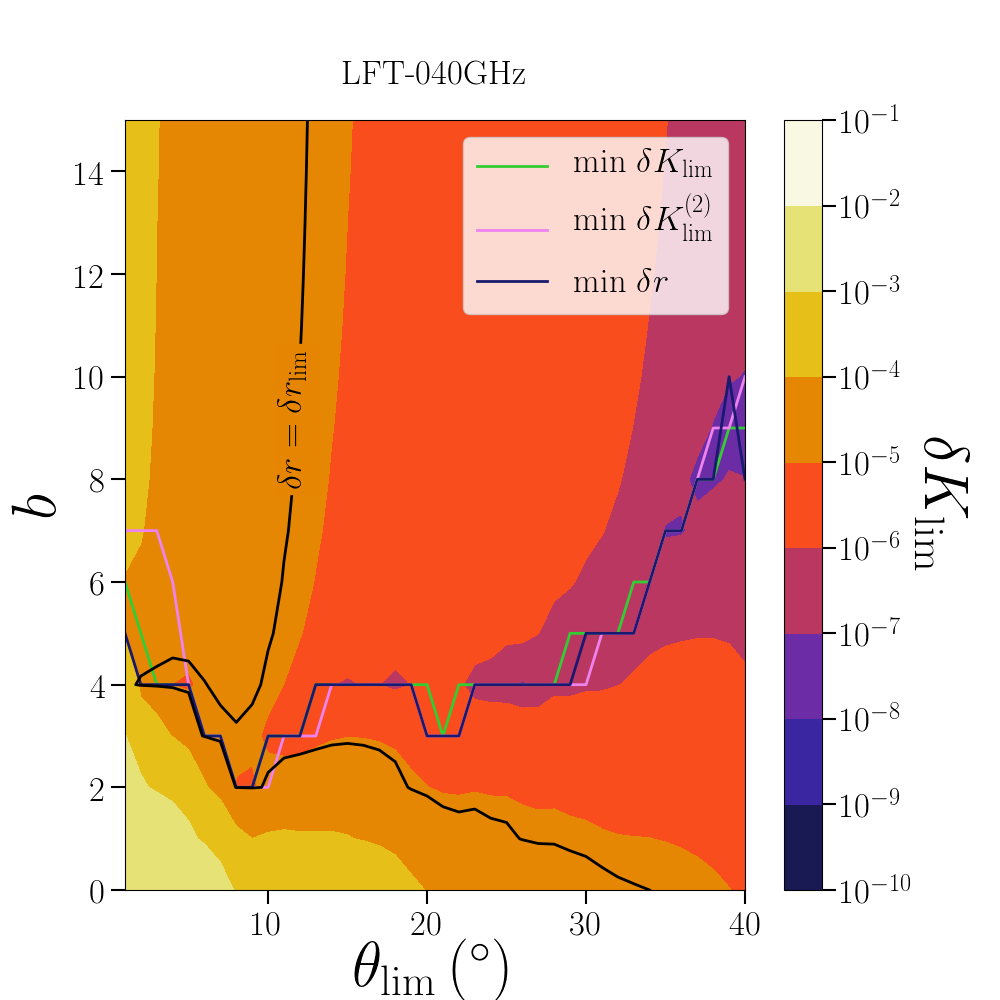}
\end{minipage}
\begin{minipage}{0.49\textwidth}
\includegraphics[width=\textwidth]{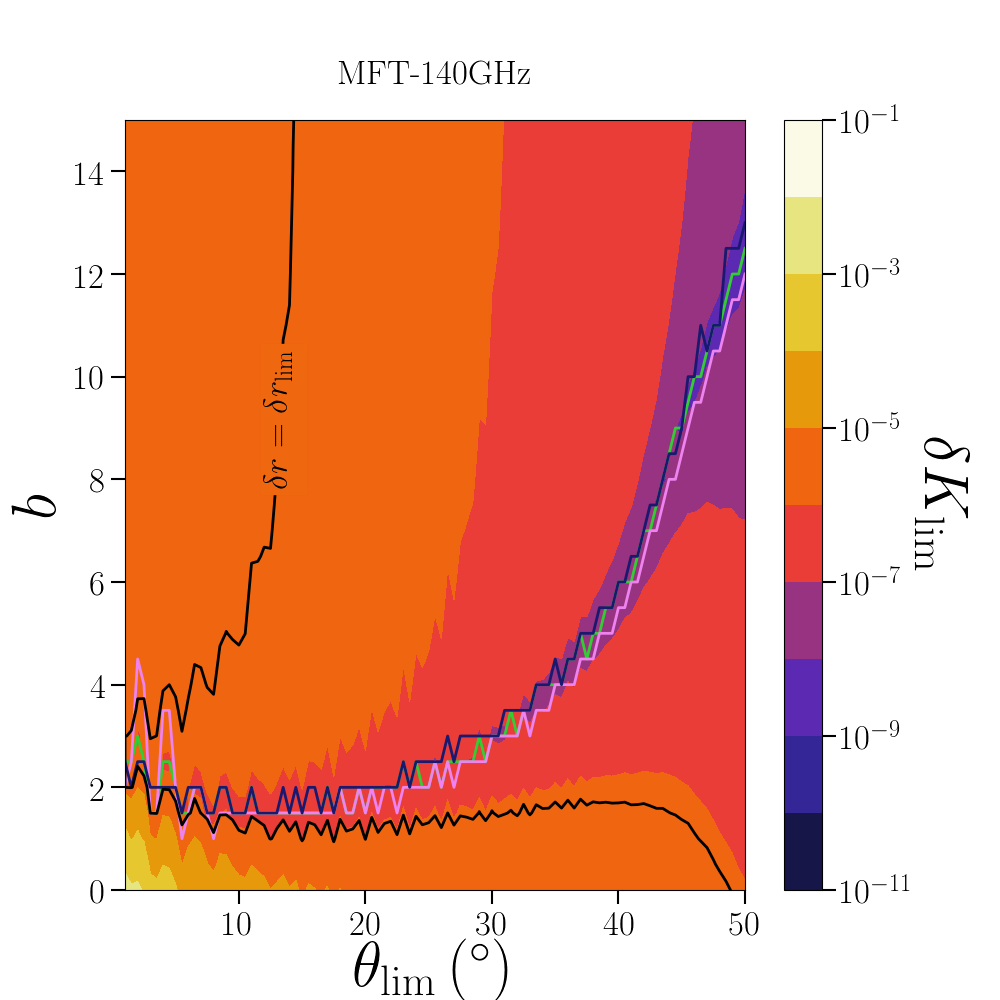}
\end{minipage}
\begin{minipage}{0.49\textwidth}
\includegraphics[width=\textwidth]{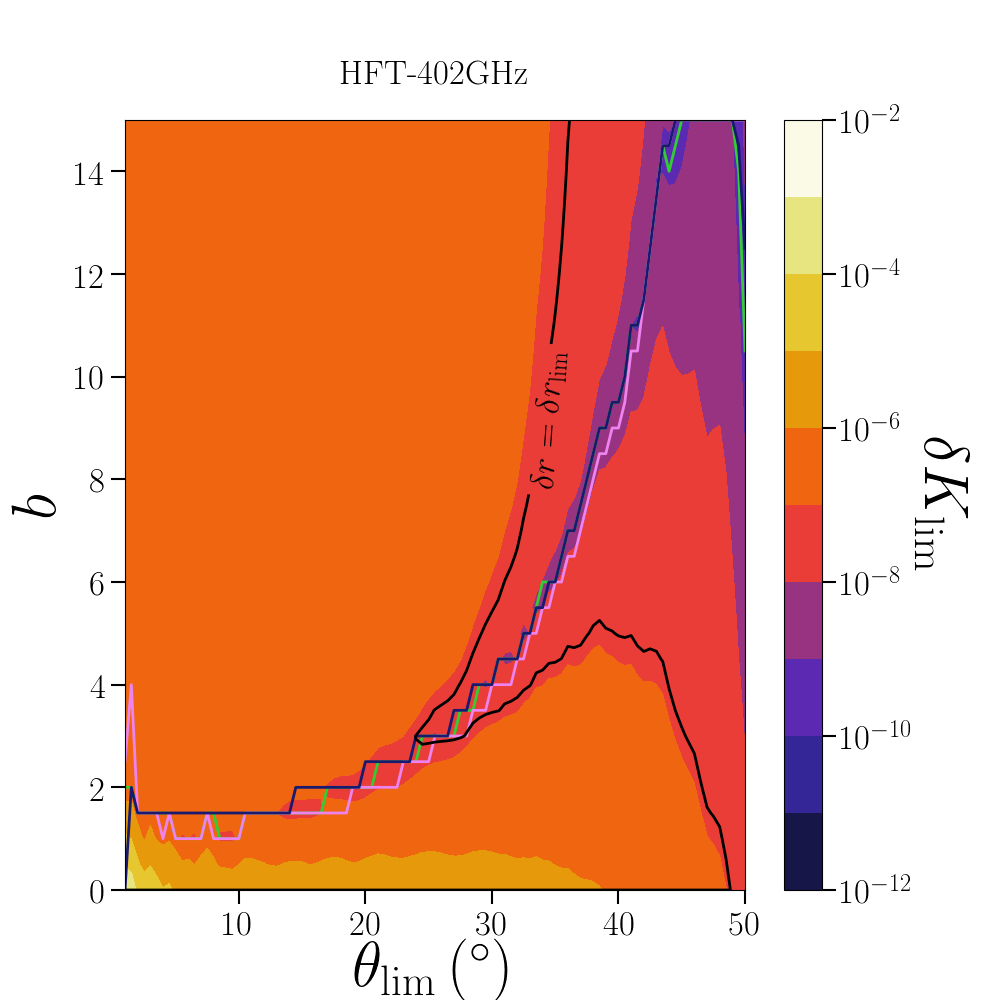}
\end{minipage}
\caption{Residual beam power $\dKlim$ in the 2D parameter space for three frequency channels, 40~GHz of the LFT (top left), 140~GHz of the MFT (top right) and 402~GHz of the HFT (bottom). The black contours are the same as the ones in Figure~\ref{fig: b-theta_lim}. The solid blue, green and pink lines correspond to the minimum value of $\delta r$, $\dKlim$ and $\dKlim^{\left( 2 \right)}$ respectively for a given value of $\theta_{\rm lim}$.}
\label{fig: theta_lim diffint}
\end{center}
\end{figure}

The meaning of the $\dKlim$ parameter is illustrated in Figure~\ref{fig:diff int} in the LFT 100~GHz channel and for $b=6$. If the beam profile is normalized by the integral of the beam power, this quantity is equivalent to the $\dRlim$ parameter in the previous section that we saw seemed to be a robust quantity across the angular range. We keep a different name for this quantity to emphasize the fact that $\dRlim$ is defined for a perturbation around a central beam with the same shape, while $\dKlim$ corresponds to a mismatch between a model and the true beam. To verify that $\delta K_{\rm lim}$ is indeed a relevant parameter, we computed its value on the same grid of $\left( \theta_{\lim}, b \right)$ as before and compared it to the corresponding values of $\delta r$. This is illustrated in Figure~\ref{fig: theta_lim diffint}.
The two parameters seem to have a similar behaviour in the 2D parameter plane. This is even more striking when looking at what happens for fixed values of $\theta_{\rm lim}$, i.e. for vertical slices. Indeed, in this context, we see that the value of $b$ that corresponds to the minimum of $\delta r$ is very close to the minimum of $\delta K_{\rm lim}$. Up to a very good accuracy, it appears that the power law that induces the least bias is the one that minimizes the residual power after $\theta_{\rm lim}$, regardless of the actual beam shape. For comparison, we also show the minimum of another quantity, $\delta K_{\rm lim}^{\left( 2 \right)}$, which corresponds to a $\chi^{2}$ between the reference beam and the beam model, defined as follow:
\begin{equation}
    \delta K_{\rm lim}^{\left( 2 \right)} \left( \nu, \theta_{\rm lim}, b \right) = \int_{0}^{2\pi} \int_{\theta_{\rm lim}}^{180^{\circ}} \left[ B_{\rm model}^{\nu} \left( \theta ; \theta_{\rm lim}, b \right) - \overline{B_{0}^{\nu}} \left( \theta \right) \right]^{2} \mathrm{sin}\theta\, d\theta d\phi.
\end{equation}
Although it would be an intuitive parameter to characterize how close the beam model is to the reference model as its minimum is the least square estimation, it is slightly off especially at high frequency. This gives further confidence in the particular relevance of the $\delta K_{\rm lim}$ parameter, and of the relative lack of importance of the shape of the beam profile.

\begin{figure}[!htb]
\begin{center}
\includegraphics[width=\textwidth]{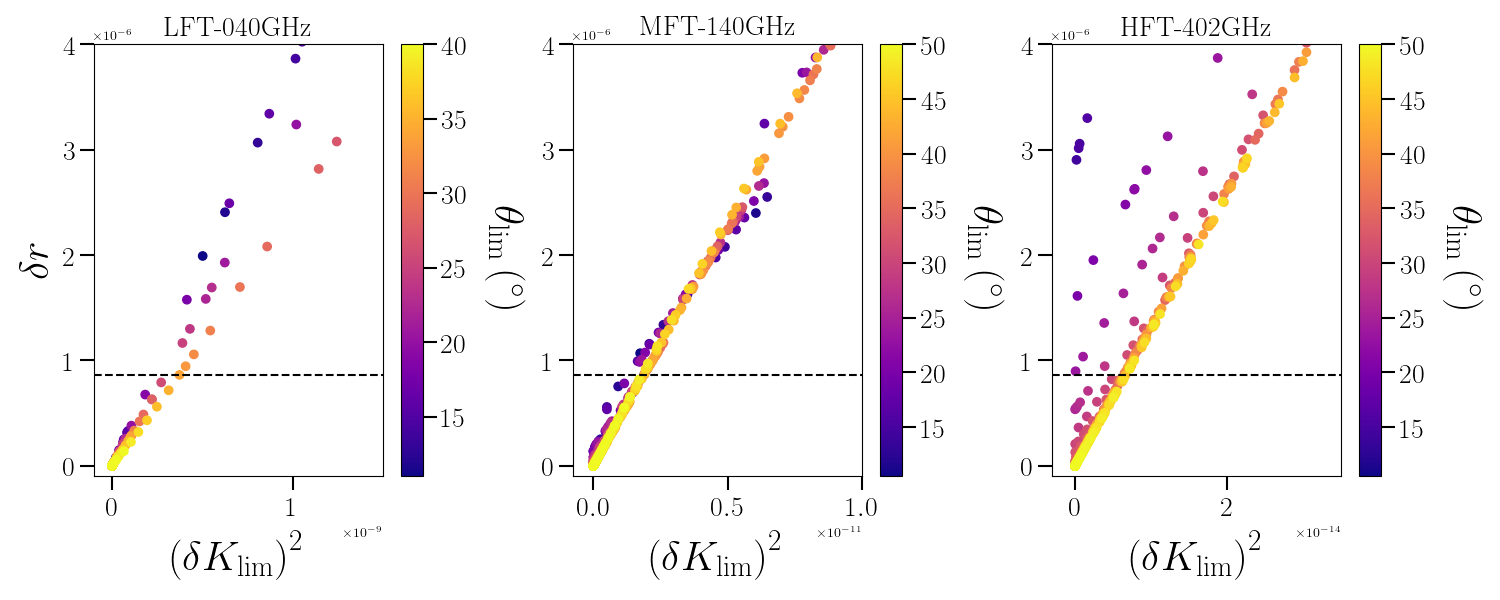}
\caption{Scatter plot in the 2D $\left( \delta r, \left( \delta K_{\rm lim} \right)^{2} \right)$ plane in the grid of $\theta_{\rm lim}$ and $b$ values used to produce figures \ref{fig: b-theta_lim} and \ref{fig: theta_lim diffint}, in the same three frequency channels. Because at very low $\theta_{\rm lim}$, the impact of imperfect beam knowledge is important and cannot be completely grasped by $\delta K_{\rm lim}$, we only show the points for which $\theta_{\rm lim} > 10^{\circ}$. In addition, the integral in the definition of $\delta K_{\rm lim}$ is bounded from below by the negative of the reference beam's integral so only the points where this integral is positive are plotted to facilitate the interpretation. We checked that the high level of correlation between the two parameters still hold when this integral is negative.}
\label{fig: deltar diffint}
\end{center}
\end{figure}

\begin{figure}[!htb]
\begin{center}
\includegraphics[width=\textwidth]{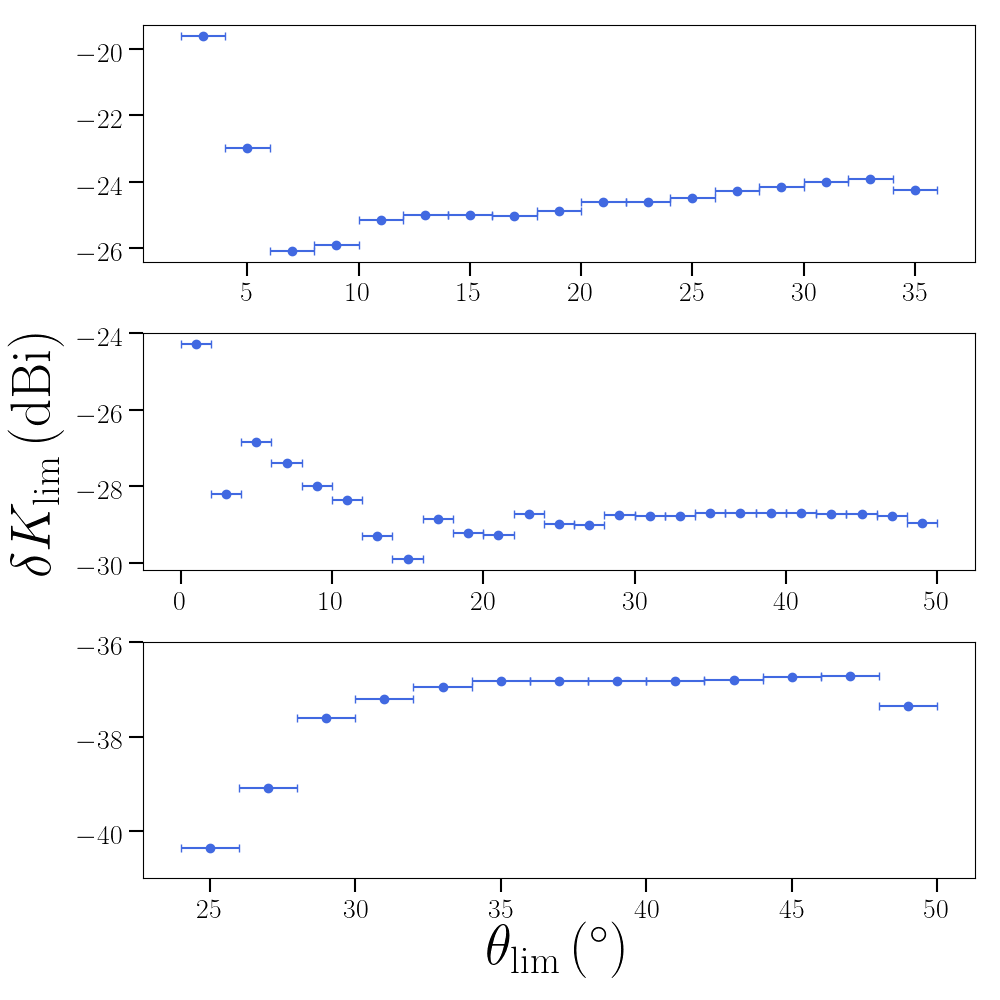}
\caption{Values of $\delta K_{\rm lim}$ for which $\delta r = \delta r_{\rm lim}$ as a function of $\theta_{\rm lim}$, in units of the total beam integral, at 40~GHz (top), 140~GHz (middle) and 402~GHz (low). These are computed for values of $b$ such that the integral in $\delta K_{\rm lim}$ is positive and for values of $\theta_{\rm lim} \leq \theta_{\rm lim} \left( \delta r = \delta r_{\rm lim}, b=0 \right)$.}
\label{fig: diff int}
\end{center}
\end{figure}

Figure~\ref{fig: deltar diffint} shows scatter plots in the $\left( \delta r, \left( \delta K_{\rm lim} \right)^{2} \right)$ plane with the same grid points used to produce previous plots. We see that the two parameters are tightly correlated. In fact, the bias is proportional to the residuals power spectrum, i.e. to the correlation of harmonic coefficients, so is proportional to the square of the perturbation at the map level. Therefore, $\delta r \propto \left( \delta K_{\rm lim} \right)^{2}$ as was already pointed out in Section \ref{subsection:Requirements for the Perturbation Case}. The very high correlation between the two parameters means that, to a given requirement on $\delta r$ would correspond a constant $\delta K_{\rm lim}$ for all values of $\theta_{\rm lim}$. To further check the correspondence between these two parameters and also to see when it breaks down, we computed the value of $\delta K_{\rm lim}$ (in units of the total beam integral) for which $\delta r = \delta r_{\rm lim}$. To compute it, we find the value of $b$ for which $\delta r = \delta r_{\rm lim}$, for a fixed value of $\theta_{\rm lim}$ and assuming that $\int_{\theta_{\rm lim}}^{180^{\circ}} \left[ B_{\rm model}^{\nu} \left( \theta ; \theta_{\rm lim}, b \right) - \overline{B_{0}^{\nu}} \left( \theta \right) \right] d\Omega > 0$ to avoid issues coming from the lower bound of this integral corresponding to the case when $b\rightarrow \infty$. The results are shown in Figure~\ref{fig: diff int} in the same three frequency channels as before. We see that, in each channel, there is a $\theta_{\rm lim}$ after which we reach a regime where $\delta K_{\rm lim}$ is approximately constant, as expected.
%, reaching a value very close to the requirements on $\dRlim$ in the corresponding channel (see Table \ref{tab:dRlim requir}). 
For lower values of $\theta_{\rm lim}$, a simple power law is not enough anymore as the actual shape of the beam becomes more important.

Therefore, as we conjectured in the previous section, there is a direct link between the bias on $r$ from the far side-lobes and the relative power difference between the true beam that convolves the sky maps and the beam model used to correct it. This relation can be used to derive meaningful requirements on the needed precision of the modeling: if we have an estimation of the residual power in the far side-lobes region at $\theta > \theta_{\rm lim}$, $\delta K_{\rm lim}$ can be interpreted as the required accuracy of this estimation. Alternatively, this relation opens new possibilities to mitigate the effect of the beam far side-lobes by calibrating for $\delta K_{\rm lim}$. It can also be used to determine the value of the angle $\theta_{\rm lim}$ after which we can rely on the beam model, as the angle at which the relation between $\delta r$ and $\dKlim$ breaks down. In the case of the 3 frequency channels we explored in this section, these angles would be $\sim 10^{\circ}$ for LFT 40 GHz,  $\sim 10^{\circ}$ for MFT 140 GHz and  $\sim 30^{\circ}$ for HFT 402 GHz.

\section{Discussions}
\label{section:Discussions}

Characterizing the beam properties has always been important for CMB space experiments to properly measure CMB intensity and $E$ modes of polarization. The knowledge of the instrument's beam characteristics is especially crucial to reach the exquisite level of sensitivity required to detect large-scale CMB $B$ modes. In particular, the beam far side-lobes uncertainty is expected to be one of the main sources of systematic error for \textit{LiteBIRD} \cite{LiteBIRD:2022cnt}. As previously stated, this study is the first to propagate beam systematic effects all the way from their accurate simulation to their impact on cosmological results. We first discuss the requirements for both cases under the assumptions we made in this paper. Then, we further address the limitations and assumptions of this study that leave room to define more meaningful and reliable requirements in future studies. Finally, we describe how the framework developed in this paper could be used to study realistic physical effects with the example of Ruze's lobes, or to define smooth systematic error curves that are easier to exploit for calibration measurements.

\subsection{Requirements on calibration}

First of all, the comparison between the Detailed Method and the Axisymmetric Method which showed that the results are similar is very important for future studies that will be able to capitalize on the simplicity and flexibility of the axisymmetric approach. In particular, in the following we use this method to understand and interpret the results we obtained.

As explained above, we tried to understand to what extent these results are independent of the particular beams used to convolve the simulated sky maps. Because we are always comparing the perturbed cases with the reference case of convolution with the full unperturbed beam, the analysis itself is only sensitive to the shape of the beam perturbations. In other words, if the perturbation vanishes, so does the bias on $r$, whatever the beam profile is. In principle in our case, this introduces a dependence on the beam shape because the perturbation is taken to have the same shape as the beam in the angular window of perturbation. In order to quantify how dependent our results are in the beam shape, we followed the Axisymmetric Method using perturbations with a different shape. In analogy with the perturbed beam defined in Eq.~\eqref{eq:perturbed beam}, we used perturbations of the form
\begin{equation}
    B_{\rm pert}^{\nu}(\theta) = \mu_{\nu} \left( B_0^{\nu} \left( \theta \right) + W \left( \theta \right) \delta B \right), \label{eq:const perturbed beam}
\end{equation}
where $\delta B$ is a constant parameter that drives the amplitude of the perturbation. The results of the comparison between the constant perturbation Eq.~\eqref{eq:const perturbed beam} and the perturbation defined in Eq.~\eqref{eq:perturbed beam} are detailed in Table \ref{tab:beamFSLconstB}, expressed in terms of $\dBlim - \dBlim^{\rm const}$ where $\dBlim$ is given in Table~\ref{tab:dBlim requir} and $\dBlim^{\rm const}$ is obtained using Eq.~\eqref{eq:const perturbed beam}. These results must be compared with results of Table \ref{tab:beamFSLcompat}. We see that the results are very similar, which supports the claim that our results depend little on the shape of the beam perturbation.

\begin{table}[!htb]
\begin{center}
\setlength{\tabcolsep}{5.5pt}
\begin{tabular}{|c|c|c|c|c|c|c|c|c|c|c|c|}
\hline
\multicolumn{12}{|c|}{\rule[-1.ex]{0pt}{3.5ex} $\dBlim \left( \mathrm{dB} \right) - \dBlim^{\rm const} \left( \mathrm{dB} \right)$} \\ \hline\hline
\rule{0pt}{2.5ex} \multirow{2}*{\vbox{\setbox0\hbox{\strut (GHz)}\hbox to\wd0{\hss\strut $\nu$\hss}\copy0}} & \multicolumn{11}{c|}{LFT} \\ \cline{2-12}
\rule{0pt}{2.5ex} & 40 & 50 & 60 & \multicolumn{2}{c|}{68} & \multicolumn{2}{c|}{78} & \multicolumn{2}{c|}{89} & 100 & 119 \\ \hline
\rule{0pt}{2.5ex} $4^{\circ} <\theta< 8^{\circ}$ & -1.25 & -1.23 & -1.06 & -0.97 & 0.11 & -1.86 & -1.84 & -2.08 & -1.96 & -2.34 & -2.75 \\ \hline
\rule{0pt}{2.5ex} $7^{\circ} <\theta< 12^{\circ}$ & -1.47 & -1.11 & -1.26 & -1.22 & -0.56 & -2.10 & -2.08 & -2.30 & -2.16 & -2.53 & -2.90 \\ \hline
\rule{0pt}{2.5ex} $11^{\circ} <\theta< 70^{\circ}$ & -0.57 & 2.46 & -0.53 & 0.15 & 0.66 & -2.53 & -2.95 & -4.07 & -4.32 & -3.22 & -0.35 \\ \hline\hline
\rule{0pt}{2.5ex} \multirow{2}*{\vbox{\setbox0\hbox{\strut (GHz)}\hbox to\wd0{\hss\strut $\nu$\hss}\copy0}} & LFT & \multicolumn{5}{c|}{MFT} & \multicolumn{5}{c|}{HFT} \\ \cline{2-12}
\rule{0pt}{2.5ex} & 140 & 100 & 119 & 140 & 166 & 195 & 195 & 235 & 280 & 337 & 402 \\ \hline
\rule{0pt}{2.5ex} $4^{\circ} <\theta< 8^{\circ}$ & -3.97 & -1.93 & -2.36 & -3.55 & -1.65 & -1.93 & -1.34 & -1.82 & -2.76 & -2.85 & -3.14 \\ \hline
\rule{0pt}{2.5ex} $7^{\circ} <\theta< 12^{\circ}$ & -3.32 & -2.18 & -2.61 & -2.93 & -1.84 & -1.97 & -1.72 & -2.00 & -2.97 & -3.03 & -3.25 \\ \hline
\rule{0pt}{2.5ex} $11^{\circ} <\theta< 70^{\circ}$ & 3.22 & -1.21 & 0.95 & 1.22 & 0.19 & -0.61 & -2.50 & -3.26 & 3.29 & -1.22 & -0.07 \\ \hline
\end{tabular}
\caption{$\dBlim$ using perturbations with constant $\delta B$ and with the fiducial shape in standard method for each frequency channel and each of the three angle ranges of the beam perturbations giving $\Delta r_{\rm FSL} = 1.9\times 10^{-5}/66$.}
\label{tab:beamFSLconstB}
\end{center}
\end{table}

Since the results depend on the perturbation windows, their choice is important and needs to be motivated. The choice of the three windows we made here (see Figure~\ref{fig:ring window functions}) is somewhat arbitrary and probably not optimal. Nevertheless, it divides the beam profile into a region close to the main beam, an intermediate region, and a region very far from the beam axis, with little overlap between the regions, which physically makes sense. This can, in particular, be used to plan the ground calibration strategy by dividing the beam into regions that can be calibrated with different accuracies. The definition of the windows lead to an apparent inconsistency where two very different requirements are defined at a same angle in the overlap region of two windows. This can be solved by defining a smooth error curve based on the requirements, see Section~\ref{subsubsection:Linking requirements to reconstructed beam error bars}. In addition, it will be possible to tune the accuracy in the different regions by tuning the resolution of the measurements, contrary to what we did here assuming constant resolution of $0.25 \mathrm{\ deg^{2}}$. Perturbations in more angular windows will be simulated in future studies, allowing to define requirements with a better angular resolution as well as a more thorough inspection of the effect of these perturbation windows.

In addition, these windows are completely axisymmetric, as well as the quantities $\delta \bar{B}_{lim}$, $\delta \bar{R}_{lim}$ and $\sigma_{\rm calib}$, which could be the source of an inconsistency in the results because the beams used to produce the simulations include some asymmetries, even if not at a realistic level. However, the comparison performed in Section~\ref{subsubsection:Comparison of the Detailed Method and the Axisymmetric Method} teaches us that the results are, to some extent, robust. This correspondence between a fully symmetric case and a case that includes some level of asymmetries could come from the combination of many symmetrically distributed detectors on the focal plane and from the symmetrizing effect of the scanning strategy, optimized to observe the same pixels from many different angles.

\subsection{Requirements on modelisation}

As explained in Section~\ref{subsubsection:Modeling Case}, we chose the power law beam model to reproduce the main falling tail feature of beam profiles while staying rather general and simple. From its generality, we expect our results to be rather independent of this particular modeling. Indeed, we saw that we could describe the bias on $r$ using a simple physical characteristic of the imperfect beam knowledge, $\delta K_{\rm lim}$, which is completely independent both of the reference \texttt{GRASP} beams and of the power law model. Therefore, we are confident that the results from this section are general and apply to a large variety of cases.

%The fact that the cosmological results was so weakly dependent on the actual shape of the reference beam and beam model actually came as a surprise at first. It is astonishing how the analysis can be completely oblivious to features of the beam at large angles, except for the total miscorrected power. As we understand it now, 
After some sufficiently large angle $\theta_{\rm lim}$, depending on the frequency channel, the effect of beam mismodeling corresponds to a modification of the signal at large scales. 
%Because Galactic foreground signal is also significant at large scales, this 
After calibrating on the dipole, this will have a net effect of scaling the normalisation of the multipoles $ell \geq 2$, introducing a bias proportional to the square of this change of normalisation, i.e. to $\left( \delta K_{\rm lim} \right)^{2}$. This regime only occurs for angles larger than some transition $\theta_{\rm lim}$ because for lower angles, the effect of the beam miscorrection is quite different. It acts at smaller scales which, after dipole, calibration will affect higher multipole. This is a potentially promising finding. Indeed, if we manage to measure the residual power in the far side-lobes region at $\theta > \theta_{\rm lim}$ with a precision better than $\delta K_{\rm lim}$, then we would be able to mitigate completely the systematic effect from beam far side-lobes. However, it does not look straightforward to measure the integrated power of the beam on such a large angular range, so this will require some sophisticated methods to calibrate this parameter in the context of the \textit{LiteBIRD} mission, possibly at the data analysis level including the $\delta K_{\rm lim}$ parameter in the component separation method.

\subsection{Limitations, assumptions and future improvements}

Because of the complexity of the present work, putting together multiple steps of the forecasting pipeline together, all undergoing active research, we had to face technical limitations as well as make key methodological assumptions to obtain the results described in Section~\ref{section:Results}. In particular, by refining the optical modeling or and the sky convolution, or by changing the data analysis pipeline (in particular the component separation stage), the requirements on the level of accuracy of beam measurements can potentially change by orders of magnitude and should, thus, be understood as the set of requirements obtained in the given context described here. For completeness, we detail in the following some of our assumptions that will need to be improved or further explored in future studies.

The beam models used for the above far side-lobes study are obtained from a \texttt{GRASP} simulation assuming ideal optical systems. In particular, the external satellite geometry such as the presence of fore-baffles and V-grooves for which we expect reflection and diffraction to occur is not taken into account, significantly modifying the reference beam shapes. This would have a positive effect on the requirements by lowering the amplitude of the far side-lobes, possibly reducing the relevant size of the last angular window (see Figure~\ref{fig:dBlim - sigCal vs theta_max}). In the optical system itself, no cross-polar contribution to the beam is taken into account in this study. We expect the cross-polar contribution to be reduced by the HWP, but it was shown in \cite{duivenvordeen2021} that high level of cross-polar contributions in the presence of a HWP are possible, and no effect from the HWP beyond the modulation of the polarization signal is included in the simulations of LiteBIRD either. Thus, the impact of the interactions between cross-polar beams and the HWP will need to be addressed in the future. As far as only the error on the transfer function is concerned, as we shall see in the next sections, it appears that our analyses are fairly independent of the particular beam shapes. Thus, the results of the previous sections would still be valid even when more realistic beam simulations are available, given the stability of $\dRlim$ and $\dKlim$.

Furthermore, because the simulated optical system is incomplete, the beam asymmetries are underestimated, especially for detectors on the edge of the focal planes. Enhanced asymmetries could lead to an additional contribution to the bias. The induced leakage from $E$ modes to $B$ modes should, nevertheless, still be mitigated by the HWP in absence of instrumental polarization, generated by diffraction on the V-grooves for instance. Therefore, all our results assume implicitly that the averaging over detectors in a given frequency channel and the scanning strategy of \textit{LiteBIRD} induce enough symmetrization of the effective beams, in contrast with the very asymmetric beams of the Planck satellite \cite{Planck-Beams}. The comparison between our two methods detailed in Section~\ref{subsubsection:Comparison of the Detailed Method and the Axisymmetric Method}, where one includes the current level of asymmetries but the other does not, tends to show that this is indeed the case for the incomplete beam simulations we have at hands. 

%In addition,
Finally, throughout this work, our goal was to study and isolate the effect of beam far side-lobes mismatch on the cosmological results. Therefore, we did not take into account other potential sources of errors nor their interplay with far side-lobes systematic effects. In particular, we used a very simple spatially homogeneous foreground model that may not be realistic since evidence for spatial variations of the SEDs was found in Planck data \cite{Planck:2018yye}. However, first of all, we expect our analysis to be reliable despite the simple foreground model because we always compare the residuals including effects from imperfect beam knowledge with residuals from a reference case (see equation Eq.~\eqref{eq:residuals}). However, it is expected that the results depend significantly on the particular component separation method that we used here. Because our method of foreground cleaning is parametric without any mitigation procedure and the impact of the beam far side-lobes is to modify the polarization frequency maps normalization factor from the dipole calibration procedure and effectively change the observed SEDs of the polarized foreground components,
%effective SEDs of the foreground components,
our approach will be particularly sensitive to this effect. Other methods, such as blind methods, are in principle more flexible and should be able to compensate this effect, to some extent. Therefore, we can expect the requirements to be significantly relaxed by using a blind component separation method. An equivalent approach to relax the requirements presented in this paper would be to modify the parametric component separation approach with a mitigation procedure of the far side-lobes effect, taking into account the error on the dipole calibration factor due to the far side-lobe mismatch, which is left for future work.
%Secondly, in the presence of a cocktail effect between complicated foreground emissions and imperfect beam knowledge, we also expect the foreground cleaning procedures to become more sophisticated than the one used here, allowing a better inclusion of beam effects and a better description of the complexity of foreground emissions. 
%Finally, another crucial effect is the dipole calibration step. 
As we see, considering the dipole calibration step is crucial to understand the impact of far side-lobes. Here, we assumed it to be perfect, however in a more realistic setting there will be errors and miscalibration having an unknown impact on beam far side-lobes systematic effects. The study of these interplays is, again, left for future work.

\subsection{Application to a realistic physical effect: Ruze's lobes}

As an example of how the derivation of requirements described in this work can be used in the presence of a realistic physical effect, we consider the case of Ruze's lobes. Note that, we do not present here a realistic study of the impact of the Ruze's lobes effect, but show how the formalism developed in this work could be used for such a study.

As described in his seminal paper \cite{Ruze-1966}, the effect of irregular optical surface (e.g. reflectors, lenses, etc.) on the beam profile can be modeled by introducing a random phase to the signal. It leads to an exponentially suppressed redistribution of the power from the main beam towards regions at higher angle, following:
\begin{equation}
	B^{\rm Ruze} \left( \theta \right) = e^{-\overline{\delta^{2}}} \left[ \overline{B_{0}} \left( \theta \right) + \left( \frac{2\pi c}{\lambda} \right)^{2} \sum_{n=1}^{+\infty} \frac{\left( \overline{\delta^{2}} \right)^{n}}{n \cdot n!} e^{- \left( \pi c \mathrm{sin} \left( \theta \right) /\lambda \right)^{2}/n} \right],
\end{equation}
where $c$ is the defect correlation length, i.e. radius of the typical defect, $\lambda$ the wavelength and $\overline{\delta^{2}}$ the phase front variance at the origin of the effect. This variance can be expressed in terms of the RMS surface error $\epsilon$, i.e. the size of the typical surface irregularities:
\begin{equation}
    \overline{\delta^{2}} = \left( \frac{4\pi\epsilon}{\lambda} \right)^{2}.
\end{equation}

\begin{figure}[!htb]
\begin{center}
\includegraphics[width=0.8\textwidth]{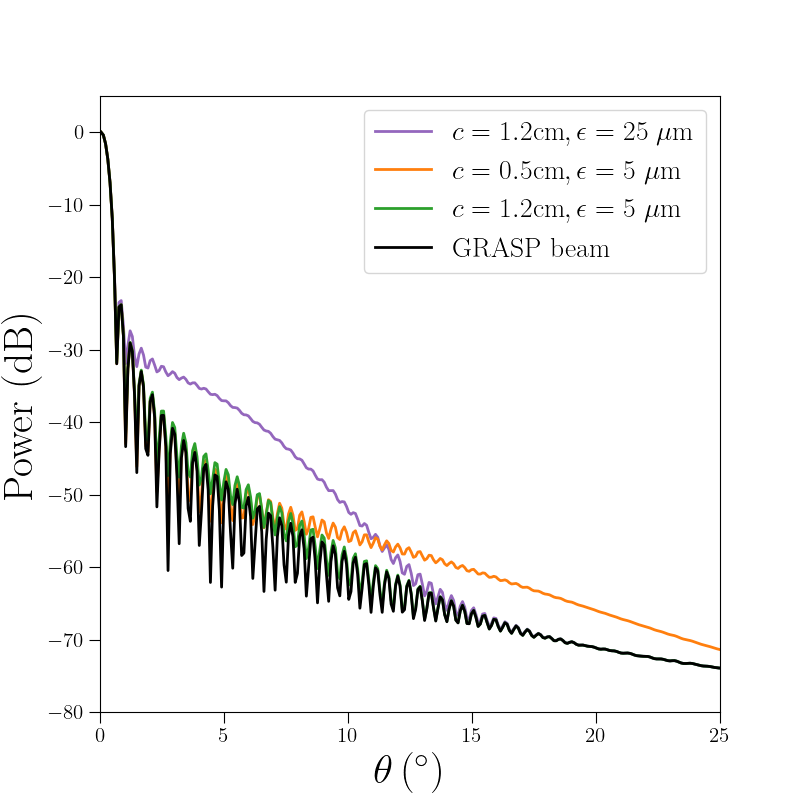}
\caption{Illustration of the impact of surface irregularities in the optical system on the shape of the beam profile, following the model Eq.~\eqref{eq:Ruze} for different set of parameters, in \textit{LiteBIRD}'s 100~GHz LFT channel.}
\label{fig:Ruze's lobes}
\end{center}
\end{figure}

In this exercise, we truncate the infinite sum, keeping only the terms for $n \leq 10$, and for a given frequency we are left with the set of two free parameters $\left( c, \epsilon \right)$. Figure \ref{fig:Ruze's lobes} illustrates the impact on the beam profile for multiple values of the parameters in the \textit{LiteBIRD}'s 100~GHz LFT channel. In order to be comparable with the types of perturbations considered in this paper, we choose to investigate the case where the parameters are $\left( c=1.2\mathrm{\ cm}, \epsilon=5\ \mu\mathrm{m} \right)$, which affects the angular range spanned by the two first windows. Note, however, that this model also induces a deformation of the main beam and near side-lobes, which we neglect here for the purpose of the exercise, as we are illustrating how the formalism developed in the context of the far side-lobes can be used. We know that the important parameter to  consider is $\dRlim$ which in this case is:
\begin{equation}
	\dRlim^{\rm Ruze} = \frac{\int \left[ B^{\rm Ruze} \left( \theta \right) - \overline{B_{0}}  \left( \theta \right) \right] W \left( \theta \right) d\Omega}{\int \overline{B_{0}} \left( \theta \right) d\Omega}. \label{eq:Ruze}
\end{equation}

We find that $\dRlim^{\rm Ruze} = -27.68 \mathrm{\ dB}$ in the first window and $\dRlim^{\rm Ruze} = -34.06 \mathrm{\ dB}$ in the second, to be compared with the requirements in Table \ref{tab:dRlim requir}. Being below the requirements in the two windows, we would conclude that such characteristics for the surface irregularities are compatible with the requirements on the beam far side-lobes to achieve the scientific goal of the mission.

\subsection{Linking requirements to reconstructed beam error bars}
\label{subsubsection:Linking requirements to reconstructed beam error bars}

\begin{figure}[!htb]
\begin{center}
\includegraphics[width=0.49\textwidth]{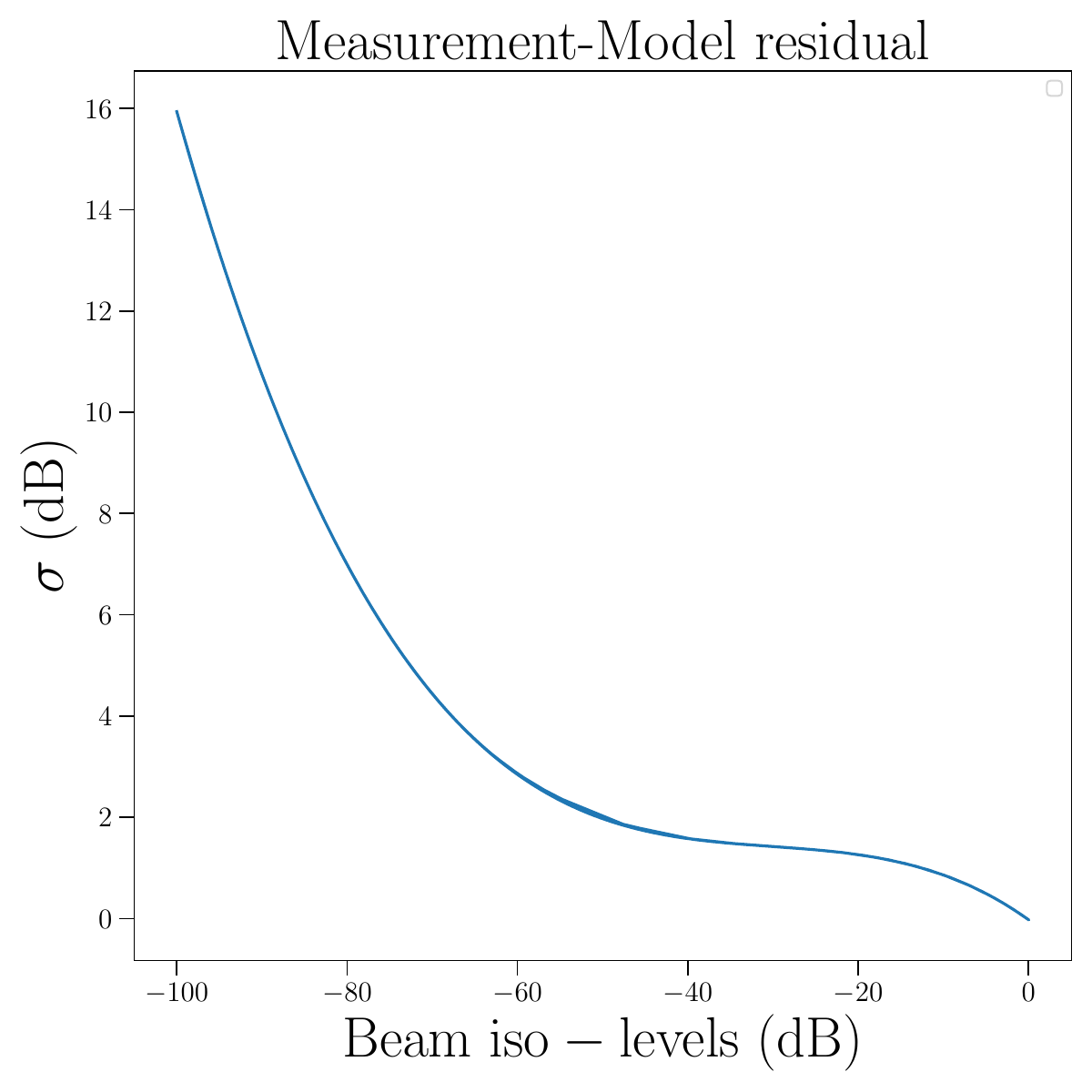}
\includegraphics[width=0.49\textwidth]{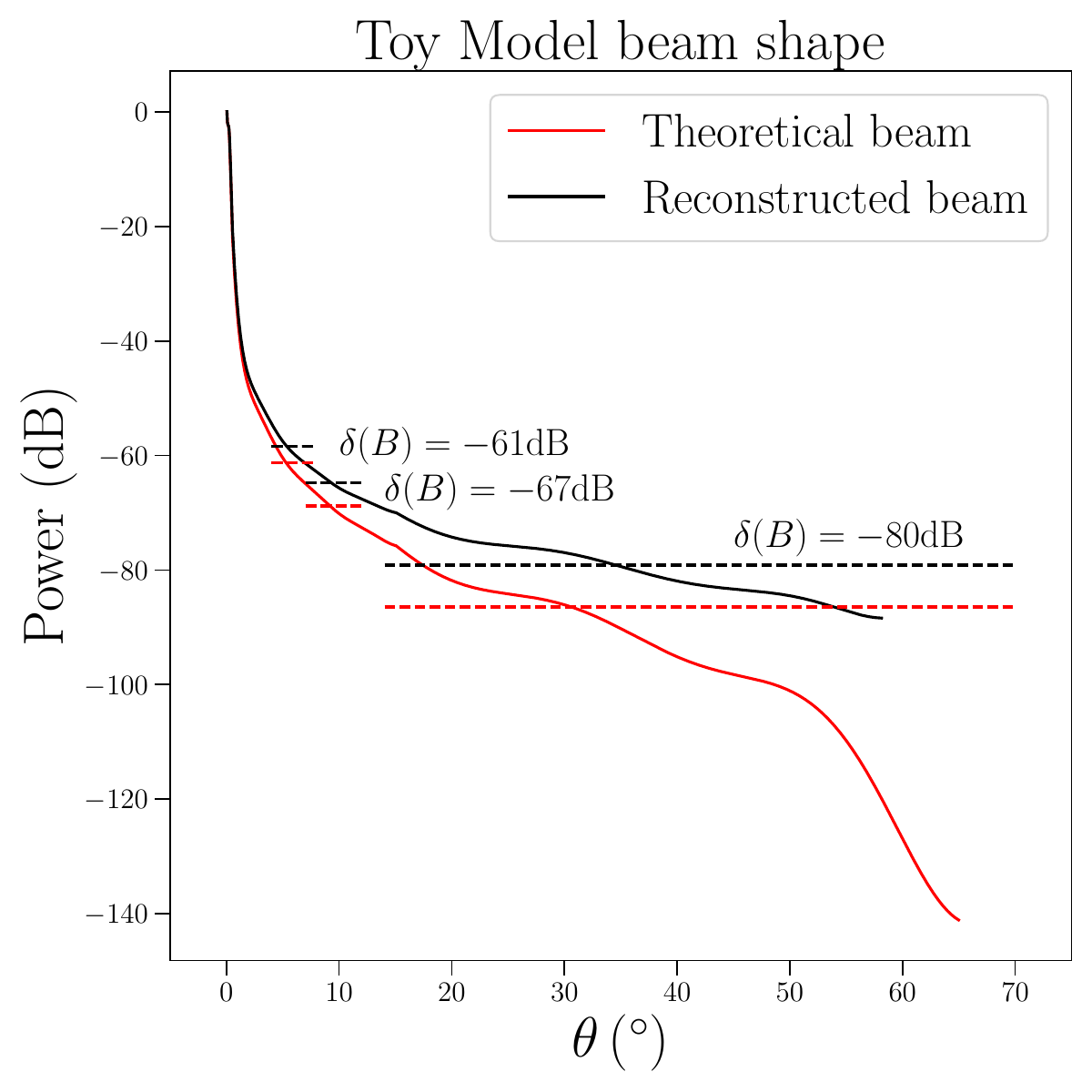}
\caption{Left: Illustration of assumed systematic residuals between model and measurements found by trial-and-error to match approximately the requirements of Table~\ref{tab:dBlim requir} for the MFT 166~GHz channel. Right: Illustration of an arbitrary theoretical beam shape (in red) together with the reconstructed one (in black) assuming the systematic residuals between this model and measurements given by the figure of the left panel.}
\label{fig:theo_achie}
\end{center}
\end{figure}

In this section, we explore how the requirements on $\dBlim$ defined in Table~\ref{tab:dBlim requir} can be translated in terms of error bars on the beam reconstruction. For illustration purposes, we start from an arbitrary symmetrized ``theoretical beam'', whose profile is shown in red in the upper panel of Fig.~\ref{fig:theo_achie}, for $\theta>0$. For concreteness, we assume that the beams will be reconstructed by a combination of measurements (ground calibration campaign) and optical modeling. As both the measurements and the simulations are subject to systematic errors (for example quality of the quiet zone on one side, and the precision of the optical model used as inputs for \texttt{GRASP} simulations on the other), we consider that the reconstructed beam will come together with a level of systematic error. This error typically comes from a residuals between model and measurements. We consider that the requirements apply to this systematic error and depend on the beam amplitude.

In Fig.~\ref{fig:theo_achie}, the black curve is the reconstructed beam assuming the systematic residual between the measurement and the theoretical beam which is illustrated on the figure of the left panel. We also indicate the $\dBlim^{\nu, W}$ values corresponding to the difference between the black and the red curves (the reconstructed beam amplitude is assumed to be zero above 60 degrees). The curve of systematic residuals is found by a trial-and-error approach to match approximately the corresponding $\dBlim$ values to those of the Table~\ref{tab:dBlim requir}, for the MFT 166~GHz channel. This is achieved for a residual error of $\simeq\ 3$ dB at -60 dB and 7 dB at -80 dB, ie the order of magnitude that has been achieved on the Planck RFQM at 100 GHz. Contrary to the requirements on $\dBlim$ that depend on the beam shape only through the size of the angular window, this smooth error curve dramatically depends on the theoretical beam shape.

%In future studies, when final requirements in more angular windows will be available, such requirements on a curve of systematic residuals can be constructed more naturally by interpolating between the requirements in each window, that correspond to some Beam iso-level. However, as we expect the average $\dBlim$ to be comparable in the same angular ranges as the three current windows, we expect the final requirement on the curve of systematic residuals to what can be achieved from the method detailed in this Section.

\section{Conclusion}
\label{section:Conclusion}

Instrument calibration is a critical step to mitigate systematic effects and define the instrument model necessary to perform a correct data analysis. We have studied the impact of beam far side-lobes through a simple procedure of beam perturbation in simulated observations. As a demonstration of the reliability of the framework, we set requirements on the calibration of beam amplitudes, under a set of assumption on the instrument, the analysis methods and mission choices, in three angular windows ranging from regions near the main beam up to $70^{\circ}$ which we related to physical quantities. We compared a procedure including a simulation of the focal plane and of the scanning strategy of \textit{LiteBIRD} with a simpler method including only axisymmetric beams and direct convolution of the sky signal and found the results to be comparable which allowed us to use the latter, a much simpler and faster method. Far from the main beam, we have found that the relevant parameter to be constrained is the difference of power in the far side-lobes between the model and the actual beam, regardless of the beam shape. Provided this parameter can be measured this could open a window to mitigate the beam far side-lobes systematic effect without having to rely too much on modeling in regions very far from the beam axis.

\section{Acknowledgements}
\label{section:Acknowledgements}

\textit{LiteBIRD} (phase A) activities are supported by the following funding agencies: ISAS/JAXA, MEXT, JSPS, KEK (Japan); CSA (Canada); CNES, CNRS, CEA (France); DFG (Germany); ASI, INFN, INAF (Italy); RCN (Norway); AEI (Spain); SNSA, SRC (Sweden); NASA, DOE (USA).
JE acknowledges the SciPol project funded by the European Research Council (ERC) under the European Union’s Horizon 2020 research and innovation program (Grant agreement No. 101044073). JEG acknowledge support from the Swedish National Space Agency (SNSA/Rymdstyrelsen) and the Swedish Research Council (Reg.\ no.\ 2019-03959). JEG also acknowledges support from the European Union (ERC, CMBeam, 101040169). The work of TM was supported by JSPS KAKENHI Grant Number 23H00107.
%\clearpage
\bibliographystyle{JHEP}
\bibliography{bibliography}

\appendix

\section{Correction by the effective beam}
\label{appendix:Correction by the effective beam}

Individual effective beams are produced in each frequency channel and for each Stokes parameter. We do not use the simulated beams at the detector level to produce the effective beams because these are transformed in a highly non-trivial way by the scanning strategy and map-making. So, we compute the effective beams by directly comparing the power spectra from the band-pass integrated \texttt{PySM} sky emission maps as reference and the unperturbed $\bold{m_{4\pi}^{\nu}}$ maps. The transfer functions in harmonic domain is, thus, given by:
\begin{equation}
    b_{\ell, \rm eff} = \sqrt{\frac{C_{\ell}^{\rm ref}}{C_{\ell}^{4\pi}}}, \label{eq:effective beam}
\end{equation}
that we decided to fit \eqref{eq:effective beam} with an empirical function $\beta_{\ell} \left( \lambda, \mu_{i} \right) = P_{\ell}^{3} \left( \mu_{i} \right) e^{-\ell^{2}/\lambda}$, where $P_{\ell}^{3}$ is a third order polynomial in $\ell$ with parameters $\mu_{i}$.

We apply beam deconvolution per frequency by translating input maps into harmonic domain, using their $\bold{a}_{\ell m}$ coefficients instead of pixel amplitudes. The beam corrected input signal is therefore:
\begin{equation}
    \bold{a}_{\ell m, \rm corr}^{\nu} = \frac{\bold{a}_{\ell m, \rm pert}^{\nu}}{b_{\ell, \rm eff}^{\nu}}.
\end{equation}

\end{document}